\documentclass[twocolumn]{aastex702}

\usepackage{xspace}
\usepackage{amsmath}
\usepackage{textgreek}

\usepackage{tabularx}
\usepackage{booktabs}
\usepackage{footnotehyper}
\usepackage{tikz}

\newcommand{\Msol}{\ensuremath{M_{\odot}}\xspace}
\newcommand{\Mfivehundred}{\ensuremath{M_{500,c}}\xspace}
\newcommand{\arcminsq}{\ensuremath{\mathrm{arcmin}^2}\xspace}
\newcommand{\LCDM}{\textLambda CDM\xspace}
\newcommand{\ALP}{\texttt{AstroLensPy}\xspace}
\newcommand{\ZA}{\texttt{ZitrinAnalytic}\xspace}
\newcommand{\Lenstool}{\texttt{Lenstool}\xspace}
\newcommand{\glafic}{\texttt{glafic}\xspace}
\newcommand{\LTM}{\texttt{LTM}\xspace}
\newcommand{\wslap}{\texttt{WSLAP+}\xspace}
\newcommand{\piemdenfw}{\texttt{PIEMDeNFW}\xspace}
\newcommand{\eazy}{\texttt{eazy}\xspace}
\newcommand{\Oguri}{\cite{Oguri2026_VENUS}\xspace}
\newcommand{\ncl}{57\xspace} 
\newcommand{\rms}{\ensuremath{\Delta_{\mathrm{RMS}}}\xspace}
\newcommand{\um}{\textmu m\xspace}
\newcommand{\Lya}{Lyman-\textalpha\xspace}
\newcommand{\norm}[1]{\left\lVert#1\right\rVert}
\newcommand{\totmis}{1733\xspace} 
\newcommand{\totmulim}{5013\xspace}
\newcommand{\totmisspec}{912\xspace}
\newcommand{\totnewmis}{227\xspace}
\newcommand{\avrms}{\ensuremath{0.54''}\xspace}

\newcommand{\panel}[3]{%
  \begin{tikzpicture}[baseline=(img.south west)]
    \node[anchor=south west, inner sep=0] (img)
      {\includegraphics[width=#3]{#2}};
    \node[anchor=south east, inner sep=1.5mm,
          yshift=-1.5mm,   
          font=\fontsize{10}{12}\selectfont]
      at (img.south east) {\textit{#1}};
  \end{tikzpicture}}

\usepackage{comment}

\newif\ifhideblock
\hideblockfalse      

\ifhideblock
  \excludecomment{toggleblock}
\else
  \includecomment{toggleblock}
\fi

\shorttitle{VENUS: A Census of Strong Lensing in a 60-Cluster JWST Survey}
\shortauthors{J. Allingham et al.}

\tabletypesize{\footnotesize}
\begin{document}

\title{VENUS: A Census of Strong Lensing in a 60-Cluster JWST Survey}


\correspondingauthor{Joseph Allingham}
\email{allingha@post.bgu.ac.il}

\author[0000-0003-2718-8640]{Joseph F. V. Allingham}
\affiliation{Department of Physics, Ben-Gurion University of the Negev, P.O. Box 653, Be'er-Sheva 84105, Israel}
\email{allingha@post.bgu.ac.il}

\author[0000-0001-6278-032X]{Lukas J. Furtak}
\affiliation{Cosmic Frontier Center, The University of Texas at Austin, Austin, TX 78712, USA} 
\affiliation{Department of Astronomy, The University of Texas at Austin, Austin, TX 78712, USA}
\email{furtak@utexas.edu}

\author[0000-0001-9411-3484]{Miriam Golubchik}
\affiliation{Department of Physics, Ben-Gurion University of the Negev, P.O. Box 653, Be'er-Sheva 84105, Israel}
\email[]{golubmir@post.bgu.ac.il}

\author[0000-0002-7876-4321]{Ashish K. Meena}
\email{ashishmeena766@gmail.com}
\affiliation{Department of Physics, Indian Institute of Science, Bengaluru 560012, India}

\author[0000-0002-0350-4488]{Adi Zitrin}
\affiliation{Department of Physics, Ben-Gurion University of the Negev, P.O. Box 653, Be'er-Sheva 84105, Israel}
\email{zitrin@bgu.ac.il}

\author[0000-0002-0302-2577]{John Chisholm}
\affiliation{Cosmic Frontier Center, The University of Texas at Austin, Austin, TX 78712, USA}
\affiliation{Department of Astronomy, The University of Texas at Austin, Austin, TX 78712, USA}
\email[]{chisholm@austin.utexas.edu}

\author[0000-0001-7201-5066]{Seiji Fujimoto}
\affiliation{David A. Dunlap Department of Astronomy and Astrophysics, University of Toronto, 50 St. George Street, Toronto, Ontario, M5S 3H4, Canada}
\affiliation{Dunlap Institute for Astronomy and Astrophysics, 50 St. George Street, Toronto, Ontario, M5S 3H4, Canada}
\email[]{seiji.fujimoto@utoronto.ca}

\author[0000-0002-5588-9156]{Vasily Kokorev}
\affiliation{Cosmic Frontier Center, The University of Texas at Austin, Austin, TX 78712, USA}
\affiliation{Department of Astronomy, The University of Texas at Austin, Austin, TX 78712, USA}
\email[]{vkokorev@utexas.edu}

\author[0000-0002-5258-8761]{Abdurro'uf}
\affiliation{Department of Astronomy, Indiana University, 727 East Third Street, Bloomington, IN 47405, USA}
\email{fnuabdur@iu.edu}

\author[0000-0002-8686-8737]{Franz E. Bauer}
\affiliation{Instituto de Alta Investigaci{\'{o}}n, Universidad de Tarapac{\'{a}}, Casilla 7D, Arica, 1010069, Chile}
\email{fbauer@academicos.uta.cl}

\author[0000-0001-5984-0395]{Maru\v{s}a Brada{\v c}}
\affiliation{University of Ljubljana, Faculty of Mathematics and Physics, Jadranska ulica 19, SI-1000 Ljubljana, Slovenia}
\affiliation{Department of Physics and Astronomy, University of California Davis, 1 Shields Avenue, Davis, CA 95616, USA}
\email[]{marusa.bradac@fmf.uni-lj.si}

\author[0000-0003-2680-005X]{Gabriel Brammer}
\affiliation{Cosmic Dawn Center (DAWN), Jagtvej 128, DK2200 Copenhagen N, Denmark}
\affiliation{Niels Bohr Institute, University of Copenhagen, Jagtvej 128, 2200 Copenhagen N, Denmark}
\email[]{gabriel.brammer@nbi.ku.dk}

\author[0000-0001-7410-7669]{Dan Coe}
\affiliation{Space Telescope Science Institute, 3700 San Martin Drive, Baltimore, MD 21218, USA}
\affiliation{Association of Universities for Research in Astronomy (AURA), Inc.~for the European Space Agency (ESA)}
\affiliation{Center for Astrophysical Sciences, Department of Physics and Astronomy, The Johns Hopkins University, 3400 N Charles St. Baltimore, MD 21218, USA}
\email[]{dcoe@stsci.edu}

\author[0000-0001-9065-3926]{Jose M. Diego}
\affiliation{Instituto de F\'isica de Cantabria,(CSIC-UC), Avda. Los Castros s/n. 39005, Santander, Spain}
\email[]{jdiego@ifca.unican.es}

\author[0000-0001-7113-2738]{Henry C. Ferguson} 
\affiliation{Space Telescope Science Institute, 3700 San Martin Drive, Baltimore, MD 21218, USA}
\email[]{ferguson@stsci.edu}

\author[0000-0001-8519-1130]{Steven L. Finkelstein}
\affiliation{Cosmic Frontier Center, The University of Texas at Austin, Austin, TX 78712, USA}
\affiliation{Department of Astronomy, The University of Texas at Austin, Austin, TX 78712, USA}
\email[]{stevenf@astro.as.utexas.edu}

\author[0000-0003-4512-8705]{Tiger Yu-Yang Hsiao}
\affiliation{Cosmic Frontier Center, The University of Texas at Austin, Austin, TX 78712, USA}
\affiliation{Department of Astronomy, The University of Texas at Austin, Austin, TX 78712, USA}
\email[]{tiger.hsiao@utexas.edu}

\author[0000-0002-6090-2853]{Yolanda Jim\'enez-Teja}
\affiliation{Instituto de Astrof\'isica de Andaluc\'ia--CSIC, Glorieta de la Astronom\'ia s/n, E--18008 Granada, Spain}
\affiliation{Observat\'orio Nacional, Rua General Jos\'e Cristino, 77 - Bairro Imperial de S\~ao Crist\'ov\~ao, Rio de Janeiro, 20921-400, Brazil}
\email[]{yojite@iaa.es}

\author[0000-0003-0942-817X]{Jose M. Palencia}
\affiliation{David A. Dunlap Department of Astronomy and Astrophysics, University of Toronto, 50 St. George Street, Toronto, Ontario, M5S 3H4, Canada}
\email[]{jose.palenciasainz@utoronto.ca}

\author[orcid=0009-0009-4388-898X]{Gregor Rihtar\v{s}i{\v c}}
\affiliation{University of Ljubljana, Faculty of Mathematics and Physics, Jadranska ulica 19, SI-1000 Ljubljana, Slovenia}
\email{gregor.rihtarsic@fmf.uni-lj.si}

\author[0000-0002-4201-7367]{Chris Willott}
\affiliation{NRC Herzberg, 5071 West Saanich Rd, Victoria, BC V9E 2E7, Canada}
\email[]{chris.willott@nrc.ca}

\author[0000-0001-8156-6281]{Rogier A. Windhorst}
\affiliation{School of Earth and Space Exploration, Arizona State University, Tempe, AZ 85287-6004, USA} 
\email{Rogier.Windhorst@asu.edu}

\author[0000-0003-3596-8794]{Hollis B. Akins}
\affiliation{Department of Astronomy, The University of Texas at Austin, Austin, TX 78712, USA}
\email[]{hollis.akins@gmail.com}

\author[0000-0001-5758-1000]{Ricardo O. Amor\'{i}n} 
\affiliation{Instituto de Astrof\'isica de Andaluc\'ia--CSIC, Glorieta de la Astronom\'ia s/n, E--18008 Granada, Spain}
\email[]{amorin@iaa.es}

\author[0000-0002-7908-9284]{Larry D. Bradley}
\affiliation{Space Telescope Science Institute, 3700 San Martin Drive, Baltimore, MD 21218, USA}
\email[]{lbradley@stsci.edu}

\author[0000-0003-2251-9164]{Chloe M. Cheng}
\affiliation{David A. Dunlap Department of Astronomy and Astrophysics, University of Toronto, 50 St. George Street, Toronto, Ontario, M5S 3H4, Canada}
\email[]{chloe.cheng@astro.utoronto.ca}

\author[0000-0003-1949-7638]{Christopher J. Conselice}
\affiliation{Jodrell Bank Centre for Astrophysics, Alan Turing Building, University of Manchester, Oxford Road, Manchester M13 9PL, UK}
\email[]{conselice@manchester.ac.uk}

\author[0000-0003-4263-2228]{David A. Coulter}
\affiliation{William H. Miller III Department of Physics and Astronomy, Johns Hopkins University, 3400 North Charles Street, Baltimore, MD 21218, USA}
\affiliation{Space Telescope Science Institute, 3700 San Martin Drive, Baltimore, MD 21218, USA}
\email{dcoulter@stsci.edu}

\author[0000-0001-8460-1564]{Pratika Dayal}
\affiliation{Canadian Institute for Theoretical Astrophysics, 60 St George St, University of Toronto, Toronto, ON M5S 3H8, Canada}
\affiliation{David A. Dunlap Department of Astronomy and Astrophysics, University of Toronto, 50 St. George Street, Toronto, Ontario, M5S 3H4, Canada}
\affiliation{Department of Physics, 60 St George St, University of Toronto, Toronto, ON M5S 3H8, Canada}
\email[]{pratika.dayal@utoronto.ca}

\author[0000-0001-8325-1742]{Guillaume Desprez}
\affiliation{Kapteyn Astronomical Institute, University of Groningen, P.O. Box 800, 9700AV Groningen, The Netherlands}
\email{g.p.a.desprez@rug.nl}

\author[0000-0003-0348-2917]{Miroslava Dessauges-Zavadsky}
\affiliation{Department of Astronomy, University of Geneva, Chemin Pegasi 51, 1290 Versoix, Switzerland}
\email[]{miroslava.dessauges@unige.ch}

\author[0000-0002-9382-9832]{Andreas L. Faisst}
\affiliation{IPAC, California Institute of Technology, 1200 E. California Blvd. Pasadena, CA 91125, USA}
\email{afaisst@caltech.edu}

\author[0000-0001-7232-5355]{Qinyue Fei}
\affiliation{David A. Dunlap Department of Astronomy and Astrophysics, University of Toronto, 50 St. George Street, Toronto, Ontario, M5S 3H4, Canada}
\email[]{qy.fei@utoronto.ca}

\author[0000-0003-1625-8009]{Brenda Frye}
\affiliation{Department of Astronomy/Steward Observatory, University of Arizona, 933 N. Cherry Ave, Tucson, AZ 85716}
\email{bfrye@arizona.edu}

\author[0000-0002-6047-430X]{Yuichi Harikane}
\affiliation{Department of Astronomy, Graduate School of Science, Kyoto University, Kitashirakawa-Oiwake-cho, Sakyo-ku, Kyoto 606-8502, Japan}
\email{hari@kusastro.kyoto-u.ac.jp}

\author[0000-0001-9187-3605]{Jeyhan S. Kartaltepe}
\affiliation{Laboratory for Multiwavelength Astrophysics, School of Physics and Astronomy, Rochester Institute of Technology, 84 Lomb Memorial Drive, Rochester, NY 14623, USA}
\email[]{jeyhan@astro.rit.edu}

\author[0009-0004-4332-9225]{Tomokazu Kiyota}
\affiliation{Astronomical Science Program, Graduate Institute for Advanced Studies, SOKENDAI, 2-21-1 Osawa, Mitaka, Tokyo 181-8588, Japan}
\affiliation{National Astronomical Observatory of Japan, 2-21-1 Osawa, Mitaka, Tokyo 181-8588, Japan}
\email[]{tomokazu.kiyota@grad.nao.ac.jp}

\author[0000-0003-2037-4619]{Conor Larison}
\affiliation{Space Telescope Science Institute, 3700 San Martin Drive, Baltimore, MD 21218, USA}
\email{clarison@stsci.edu}

\author[0000-0003-2540-7424]{Paulo A. A. Lopes}
\affiliation{Observatório do Valongo, Universidade Federal do Rio de Janeiro, Ladeira do Pedro Antônio 43, Rio de Janeiro, RJ, 20080-090, Brazil}
\email[]{plopes@ov.ufrj.br}

\author[0000-0003-1581-7825]{Ray A. Lucas}
\affiliation{Space Telescope Science Institute, 3700 San Martin Drive, Baltimore, MD 21218, USA}
\email{lucas@stsci.edu}

\author[0000-0002-4872-2294]{Georgios E. Magdis}
\affiliation{Cosmic Dawn Center (DAWN), Jagtvej 128, DK2200 Copenhagen N, Denmark}
\affiliation{DTU-Space, Technical University of Denmark, Elektrovej 327, 2800, Kgs. Lyngby, Denmark}
\email[]{geoma@space.dtu.dk}

\author[orcid=0000-0003-3243-9969]{Nicholas S. Martis}
\affiliation{University of Ljubljana, Faculty of Mathematics and Physics, Jadranska ulica 19, SI-1000 Ljubljana, Slovenia}
\email[]{nicholas.martis@fmf.uni-lj.si}

\author[0000-0002-1049-6658]{Masami Ouchi}
\affiliation{National Astronomical Observatory of Japan, 2-21-1 Osawa, Mitaka, Tokyo 181-8588, Japan}
\affiliation{Institute for Cosmic Ray Research, The University of Tokyo, 5-1-5 Kashiwanoha, Kashiwa, Chiba 277-8582, Japan}
\affiliation{Department of Astronomical Science, SOKENDAI (The Graduate University for Advanced Studies), 2-21-1 Osawa, Mitaka, Tokyo, 181-8588, Japan}
\affiliation{Kavli Institute for the Physics and Mathematics of the Universe (WPI), The University of Tokyo, 5-1-5 Kashiwanoha, Kashiwa, Chiba 277-8583, Japan}
\email[]{ouchims@icrr.u-tokyo.ac.jp}

\author[0000-0002-9651-5716]{Richard Pan}
\affiliation{Department of Physics and Astronomy, Tufts University, 574 Boston Avenue, Suite 304, Medford, MA 02155, USA}
\email[]{richard.pan@tufts.edu}

\author[0000-0001-7503-8482]{Casey Papovich}
\affiliation{Department of Physics and Astronomy, Texas A\&M University, College Station, TX, 77843-4242 USA}
\affiliation{George P. and Cynthia Woods Mitchell Institute for Fundamental Physics and Astronomy, Texas A\&M University, College Station, TX, 77843-4242 USA}
\email[]{papovich@tamu.edu}

\author[0000-0002-2282-8795]{Massimo Pascale}
\affiliation{Department of Physics \& Astronomy, University of California, Los Angeles, 430 Portola Plaza, Los Angeles, CA 90095, USA}
\email{mpascale@astro.ucla.edu}

\author[0000-0002-2361-7201]{Justin D. R. Pierel}
\affiliation{Space Telescope Science Institute, 3700 San Martin Drive, Baltimore, MD 21218, USA}
\email{jpierel@stsci.edu}

\author[0009-0007-0522-7326]{Tom Resseguier}
\affiliation{Center for Astrophysical Sciences, Department of Physics and Astronomy, The Johns Hopkins University, 3400 N Charles St. Baltimore, MD 21218, USA}
\affiliation{Space Telescope Science Institute, 3700 San Martin Drive, Baltimore, MD 21218, USA}
\email[]{tresseg1@jhu.edu}

\author[0000-0001-5492-1049]{Johan Richard}
\affiliation{Univ Lyon, Univ Lyon1, ENS de Lyon, CNRS, Centre de Recherche Astrophysique de Lyon UMR5574, 69230 Saint-Genis-Laval, France}
\email[]{johan.richard@univ-lyon1.fr}

\author[0000-0003-4223-7324]{Massimo Ricotti}
\affiliation{Department of Astronomy, University of Maryland, College Park, 20742, USA}
\email[]{ricotti@umd.edu}

\author[0000-0002-6265-2675]{Luke Robbins}
\affiliation{Department of Physics and Astronomy, Tufts University, 574 Boston Avenue, Suite 304, Medford, MA 02155, USA}
\email[]{andrew.robbins@tufts.edu}

\author[0000-0003-1889-0227]{William Sheu}
\affiliation{Department of Physics \& Astronomy, University of California, Los Angeles, 430 Portola Plaza, Los Angeles, CA 90095, USA}
\email{wsheu@astro.ucla.edu}

\author[0000-0002-4622-6617]{Fengwu Sun}
\affiliation{Department of Astronomy, School of Science, Westlake University, Hangzhou, Zhejiang 310030, P.\ R.\ China}
\email[]{sunfengwu@westlake.edu.cn}

\author[0009-0004-7808-1979]{Felix Vecchi}
\affiliation{Laboratoire d’Astrophysique, EPFL, Observatoire de Sauverny, 1290 Versoix, Switzerland}
\email[]{felix.vecchi@epfl.ch}

\author[0000-0003-1815-0114]{Brian Welch}
\affiliation{International Space Science Institute, Hallerstrasse 6, 3012 Bern, Switzerland}
\email{bwelch.astro@gmail.com}

\author[orcid=0009-0006-6763-4245]{Hiroto Yanagisawa}
\affiliation{Institute for Cosmic Ray Research, The University of Tokyo, 5-1-5 Kashiwanoha, Kashiwa, Chiba 277-8582, Japan}
\affiliation{Department of Physics, Graduate School of Science, The University of Tokyo, 7-3-1 Hongo, Bunkyo, Tokyo 113-0033, Japan}
\email{yana@icrr.u-tokyo.ac.jp}


\begin{abstract}  
We present strong-lensing models for the 57 galaxy clusters observed in Cycle 4 out of the 60 targeted by the JWST Vast Exploration for Nascent, Unexplored Sources (VENUS) program.
Models for the remaining three clusters will be incorporated after their observations.
The dataset comprises \totmulim individual images, corresponding to \totmis multiply-imaged systems, of which \totmisspec have spectroscopic redshift measurements and \totnewmis are newly identified.
Using this new wealth of observations, we construct parametric lens models with our \ALP software, reproducing the observed image positions with an average image-plane precision \rms of \avrms.
With a total imaging area of $576$ arcmin$^2$, VENUS offers the largest homogeneous set of cluster strong-lensing models to date.
The clusters yield a combined area of $191$ arcmin$^{2}$ magnified by $|\mu|\geq3$ for a source $z_s=10$, for example, establishing the VENUS fields as a premier survey for high-redshift studies.
Among our sample, the clusters MACS J0717.5+3745, RXC J0600.1-2007 and PLCK G287.0+32.9 display the largest Einstein radii ($\geq 60''$) and magnified areas ($\mathcal{A}_{z_s=10}^{|\mu|\geq3} \geq 9$ arcmin$^2$) for a source $z_s=10$, and therefore constitute the most potent gravitational telescopes known to date.
Finally, leveraging the statistical power of the sample, we derive scaling relations between various lensing observables (Einstein radius, magnified area, projected mass) and the SZ/X-ray-inferred cluster mass, \Mfivehundred.
\end{abstract}

\keywords{gravitational lensing: Strong; galaxy clusters: general}


\section{Introduction}\label{sec:intro}

Galaxy clusters form at the nodes of the Cosmic Web, by gravitational infall of baryonic and dark matter \citep{Voit2005,Kravtsov_Borgani2012,Bower2013}. 
As massive bodies they warp space-time, and when their mass density is sufficiently high, they
\textit{strongly} lens background sources lying behind them, thus creating multiple observable images. 
The distribution of these images for a given strong-lensing (SL) cluster enables its mass modeling, i.e.\ mapping the gravitational potential of the total matter distribution in the lens \citep{Kneib2011review, Natarajan2024_SLreview, Zitrin2026_SLreview}.
Because lensing is sensitive to the total projected mass, regardless of its nature, it constitutes an excellent means for mapping the distribution of the unseen dark matter (DM). 
As such, SL significantly informs DM studies \citep{Clowe2006Bullet, Randall2008, Diego2018}, without assuming the DM halo (DMH) to be in hydrostatic equilibrium with the intra-cluster medium \citep[ICM;][]{Munoz2024_ICM_massbias}.
SL, in particular of large samples of clusters at a range of redshifts, offers valuable constraints on the formation of structures \citep{Meneghetti2013, Natarajan2017Substracture} and cosmology \citep{Refsdal1964MNRAS,Jullo2010,Bonvin2017_Cosmograil,Caminha2022}.

Beyond the study of cluster lenses themselves, lensing magnifies the background similarly to a \textit{magnifying lens}, thus enabling the observation of the faint and distant Universe, otherwise inaccessible. 
This is beneficial for observing distant sources ranging from high-redshift galaxies
\citep{Coe2012highz,Salmon2020HighzRelics,Hsiao2023,RobertsBorsani2023highz}, stars \citep{Kelly2018NatAsCCE, Welch2022EarendelHST,Meena2023_lensedstars,Fudamoto2025_40lensedstars, Palencia2026_CCE_Dragonarc}, supernovae \citep[SNe,][]{Kelly2015Sci,Rodney2016_SN_Refsdal,Frye2024SN_H0pe,Pierel2024SN_Encore,Suyu2025_SN_Encore,Pascale2025_H0pe}, star clusters \citep{Vanzella2023_Sunrise,Fudamoto2024_CosmicSpear,Adamo2024_CosmicGems}, active galactic nuclei \citep[AGN,][]{Furtak2023AGN,Napier2023ApJ...954L..38N,Cloonan2025_AGNcllensed,Allingham2025_SPTCLJ0546} to little red dots \citep[LRDs,][]{Furtak2023_LRD}.

In order to exploit these properties and study galaxy clusters as well as their magnified backgrounds, several cluster lensing surveys
have been conducted with the \textit{Hubble} Space Telescope (HST), such as the Cluster Lensing And Supernova Survey with Hubble \citep[CLASH,][]{PostmanCLASH2012}, Hubble Frontier Fields \citep[HFF,][]{Lotz2016HFF}, Reionization Lensing Cluster Survey \citep[RELICS,][]{Coe2019RELICS}, and with the \textit{James Webb} Space Telescope, including the CAnadian NIRISS Unbiased Cluster Survey \citep[CANUCS,][]{Sarrouh2026_CANUCS}, Ultradeep NIRSpec and NIRCam ObserVations before the Epoch of Reionization \citep[UNCOVER,][]{Bezanson2024_UNCOVER}, Prime Extragalactic Areas for Reionization and Lensing Science \citep[PEARLS,][]{Windhorst2023_PEARLS}, GLIMPSE \citep[][]{Atek2025_GLIMPSE} and Strong LensIng and Cluster Evolution \citep[SLICE,][Mahler et al.\ in prep.]{Cerny2026SLICE}.

One such large SL survey is the Vast Exploration for Nascent, Unexplored Sources \citep[VENUS,][]{fujimoto_venus_2025}, targeting 60 known SL clusters.
VENUS provides a comprehensive JWST SL cluster survey with 10-band NIRCam imaging -- amounting to 240 hours of observation, leveraging the aforementioned advantages of SL clusters, and leading to remarkable discoveries, from lensed LRDs \citep{Golubchik2025_LRD_VENUS,Zhang2025LRD, Yanagisawa2026_LRDs}, lensed SNe \citep{Coulter2026_SN_Eos, Fujimoto2026_SNHelios, Larison2026_SNAres} to high-redshift galaxies and star clusters \citep{Nakane2025_VENUS_highzgal, Abdurrouf2026_CosmicSpear, Asada2026_GalEos, Pan2026_CL0152}.
In this context, in the same fashion as previous large SL modeling cluster campaigns \citep[e.g.][]{Oguri2012_SGAS, Zitrin2015CLASH25}, systematic lens modeling is necessary for all VENUS clusters. 
As Cycle 4 came to an end, we present here the first catalog of strong-lensing models for all VENUS clusters observed to date as the first data release (DR1), amounting to \ncl cluster models. Three clusters have not yet been observed as part of Cycle 4 and are rescheduled for Cycle 5.
Our DR1 models are produced with our parametric method \ALP \citep{Allingham2026_M1931}, the \texttt{Python} implementation of the \ZA method \citep{Zitrin2015CLASH25}; the models themselves will be made publicly available in a forthcoming release.
A companion article similarly details VENUS clusters' lens models produced with \glafic \citep{Oguri2026_VENUS}.

After presenting the VENUS and other archival observations in Section \ref{sec:data}, the lensing analysis methodology, priors and biases are described in Section \ref{sec:methods}. The specific cluster models and results are presented in Section \ref{sec:Results}, before a discussion in Section \ref{sec:discussion} on the statistical inference from our lens models and their impact for SL studies in general. The article is finally summarized in Section \ref{sec:summary}. 
In this article, we use a standard flat \LCDM cosmological model with $H_0=70\,\mathrm{km}\,\mathrm{s}^{-1}\,\mathrm{Mpc}^{-1}$, $\Omega_{\Lambda}=0.7$, and $\Omega_\mathrm{m}=0.3$, the AB system for magnitudes \citep{Oke1983ABandStandards} and $1\sigma$ ranges for uncertainties, unless stated otherwise.

\section{Data}\label{sec:data}

\subsection{VENUS JWST NIRCam observations} \label{sec:VENUS_imaging}

The Treasury JWST VENUS program \citep[\href{https://www.stsci.edu/jwst-program-info/download/jwst/pdf/6882/}{PID: 6882-GO}, PI: Fujimoto, see][]{fujimoto_venus_2025} targets 60 SL galaxy clusters between $z=0.18$ and $0.97$ with 10 NIRCam bands during Cycle 4, and re-visits a selection of 10 clusters during Cycle 5 with 2 NIRCam filters for another epoch, and with NIRSpec multi-objects spectroscopy (MSA).
The primary objective of surveying $576$ arcmin$^2$ of SL clusters is the study of high-redshift stars, galaxies and black holes to understand the formation and evolution of structures \citep{Harikane2023_highz_JWST,Zackrisson2024_JWST_PopIII}.
Such science goals require a precise understanding of the lensing magnification effect, as it constitutes the most important systematic in the extraction of the intrinsic brightness of background sources.

In this VENUS SL DR1 article, we focus only on the VENUS data acquired as part of Cycle 4, between 2025-08-07 and 2026-08-08. The target clusters are summarized in Table \ref{tab:all_clusters_1}.
The 10 NIRCam bands acquired during the first VENUS epoch include 7 NIRCam wide pass-band filters (F090W, F115W, F150W and F200W in the short-wavelength channel, and F277W, F356W, and F444W in the long-wavelength channel), and three medium bands (F210M, F300M and F410M), amounting to 239.8 hours of NIRCam observation time, thus reaching at least 28 AB magnitude per band with $5 \sigma$ certainty. 
Due to delayed observations for four clusters, the total NIRCam observation time at completion of Cycle 4 was 224.7 hours, and three clusters were not observed, while one was not yet visited for a second time. Consequently, out of the 60 VENUS clusters, we hereafter present SL models for \ncl clusters. 
Moreover, the inclusion of cluster Abell 2744 in VENUS is archival, as UNCOVER observations \citep{Bezanson2024_UNCOVER} already satisfy VENUS depth and multi-band observation criteria.

\startlongtable          
\begin{deluxetable*}{lcccccccccccc}
\tabletypesize{\footnotesize} 
\tablewidth{0pt}
\setlength{\tabcolsep}{3pt} 
\tablecaption{Summary of 60 VENUS clusters' observations, lensing constraints and image reproduction accuracy. Columns are:
(i) Cluster name, ordered by observation date;
(ii) R.A.\ (degrees, J2000);
(iii) Dec.\ (degrees, J2000);
(iv) Cluster redshift;
(v) Observation date;
(vi) $M_{500,c}$ values\tablenotemark{a}, extracted, with descending priority, from ACT DR6 \citep{ACT_DR6}, SPT-3G \citep[last DR,][as well as all SPT surveys]{Bleem2026_SPT-3G}, eROSITA-DE \citep{Bulbul2024_eRASS-DE} and \textit{Planck} DR2 \citep{Planck2016_galaxycat};
(vii) Total number of multiply-imaged systems (MIS);
(viii) Total number of systems with a spectroscopic confirmation;
(ix) Total number of lensed images;
(x) Total number of newly identified systems;
(xi) Image-plane \rms;
(xii) Number of free parameters in the model (including free redshifts);
(xiii) Strong-lensing $\chi^2$.
Note that the $\chi^2$ was derived with a nominal, uniform $\sigma_{\rm pos} = 0.2''$ error across all images and clusters (see Sec.\,\ref{sec:lensmodelling}).
Additionally, note that, in this table, knots from a same source galaxy and MIS from different galaxies are all counted as different MIS identifications. However adding new knots for an already known MIS is not counted as a new system.
Finally, note that MIS first identified by \Oguri are not counted as `newly identified' here, but we count all identifications performed separately.
\label{tab:all_clusters_1}}
\tablehead{
 \colhead{Cluster name} & \colhead{$\alpha_c$} & \colhead{$\delta_c$} & \colhead{$z$} &
 \colhead{Obs. date} & \colhead{$M_{500,c}$} &
 \colhead{$N_{\rm Sys}$} & \colhead{$N_{\rm Spec}$} & \colhead{$N_{\rm Im}$} &
 \colhead{$N_{\rm New}$} & \colhead{\rms} & \colhead{$N_{\rm Free}$} & \colhead{$\chi^2$} \\
 \colhead{} & \colhead{[deg, J2000]} & \colhead{[deg, J2000]} & \colhead{} &
 \colhead{} & \colhead{[$10^{14}M_\odot$]} &
 \colhead{} & \colhead{} & \colhead{} & \colhead{} & \colhead{[arcsec]} & \colhead{} & \colhead{}
}
\startdata
Abell 2744 & 3.5883333 & -30.3972500 & 0.31 & 2022-11-05\tablenotemark{b} & 15.25 & 65 & 59 & 186 & 0 & 0.37 & 55 & 632.8 \\
RXCJ0018.5+1626 & 4.6416667 & 16.4333333 & 0.55 & 2026-07-03 & 11.89 & 18 & 3 & 47 & 3 & 0.37 & 38 & 160.9 \\
MACSJ0025.4-1222 & 6.3764850 & -12.3812301 & 0.58 & 2026-08-08 & 8.32 & 10 & 2 & 30 & 4 & 0.16 & 28 & 18.3 \\
RXCJ0032.1+1808 & 8.0462500 & 18.1302778 & 0.40 & 2025-11-28 & 11.01 & 55 & 4 & 162 & 0 & 0.66 & 80 & 1755.4 \\
MACSJ0035.4-2015 & 8.8625000 & -20.2611944 & 0.35 & 2025-11-14 & 10.98 & 27 & 12 & 74 & 12 & 0.50 & 30 & 454.3 \\
Abell 68 & 9.2758333 & 9.1591667 & 0.26 & 2025-11-21 & 9.92 & 18 & 7 & 49 & 4 & 0.35 & 39 & 152.6 \\
Abell 2813 & 10.8535511 & -20.6209901 & 0.29 & 2025-11-14 & 11.68 & 18 & 2 & 58 & 2 & 0.63 & 33 & 574.6 \\
El Gordo & 15.7278497 & -49.2605321 & 0.87 & 2026-08-05 & 17.28 & 76 & 44 & 207 & 0 & 0.97 & 60 & 4881.4 \\
Abell 209 & 22.9691667 & -13.6105556 & 0.21 & 2025-11-19 & 9.82 & 3 & 0 & 7 & 0 & 0.30 & 7 & 16.1 \\
WHLJ013719.8-082841 & 24.3540956 & -8.4564151 & 0.57 & 2025-08-07 & 9.89 & 11 & 7 & 30 & 2 & 0.89 & 16 & 589.5 \\
CLJ0152.7-1357 & 28.1776447 & -13.9591200 & 0.83 & 2025-09-03 & 5.88 & 24 & 1 & 66 & 13 & 0.84 & 41 & 1158.8 \\
MACSJ0159.8-0849 & 29.9558333 & -8.8333333 & 0.41 & 2025-12-01 & 10.46 & 15 & 0 & 33 & 6 & 0.27 & 24 & 58.9 \\
RXCJ0232.2-4420 & 38.0773899 & -44.3466885 & 0.28 & 2025-09-10 & 13.32 & 18 & 1 & 68 & 12 & 0.47 & 34 & 370.7 \\
Abell 370 & 39.9696787 & -1.5719570 & 0.38 & 2026-01-09 & 10.22 & 93 & 79 & 281 & 1 & 0.70 & 52 & 3421.2 \\
Abell S295 & 41.3790390 & -53.0403040 & 0.30 & 2025-11-02 & 11.81 & 11 & 4 & 32 & 1 & 0.54 & 22 & 233.0 \\
Abell 383 & 42.0141300 & -3.5292300 & 0.19 & 2025-09-12 & 5.84 & 14 & 8 & 47 & 3 & 0.73 & 18 & 632.0 \\
MACSJ0257.1-2325 & 44.2925000 & -23.4366111 & 0.51 & 2025-08-18 & 9.89 & 10 & 3 & 29 & 4 & 0.30 & 19 & 63.5 \\
MACSJ0257.6-2209 & 44.4220833 & -22.1540000 & 0.32 & 2025-09-13 & 9.28 & 25 & 25 & 85 & 0 & 0.71 & 25 & 1083.6 \\
MACSJ0308.9+2645 & 47.2320833 & 26.7602222 & 0.36 & 2025-08-23 & 10.76 & 21 & 2 & 49 & 5 & 0.46 & 34 & 263.5 \\
PLCKG171.9-40.7 & 48.2370833 & 8.3720000 & 0.27 & 2025-09-13 & 10.71 & 15 & 0 & 47 & 1 & 0.47 & 31 & 264.0 \\
MACSJ0329.7-0211 & 52.4237500 & -2.1965833 & 0.45 & 2025-08-19 & 8.87 & 10 & 8 & 26 & 1 & 0.42 & 21 & 113.4 \\
Abell 3192 & 59.7201370 & -29.9298850 & 0.43 & 2025-09-12 & 12.49 & 25 & 3 & 71 & 5 & 0.88 & 39 & 1387.2 \\
MACSJ0416.1-2403 & 64.0381013 & -24.0674860 & 0.40 & 2026-01-10 & 11.26 & 138 & 127 & 379 & 0 & 0.53 & 36 & 2709.0 \\
MACSJ0417.5-1154 & 64.3904167 & -11.9062778 & 0.44 & 2025-10-06 & 16.95 & 41 & 20 & 121 & 1 & 0.68 & 49 & 1390.1 \\
MACSJ0429.6-0253 & 67.4004167 & -2.8855556 & 0.40 & 2025-09-14 & 5.48 & 6 & 3 & 21 & 2 & 0.22 & 14 & 26.2 \\
MACSJ0451.9+0006 & 72.9777280 & 0.1050335 & 0.43 & 2025-09-11 & 8.51 & 34 & 25 & 92 & 4 & 0.75 & 26 & 1279.2 \\
Abell 521 & 73.5287500 & -10.2236056 & 0.25 & 2026-01-14 & 7.07 & 16 & 15 & 38 & 0 & 0.11 & 12 & 11.9 \\
MACSJ0454.1-0300 & 73.5451620 & -3.0143231 & 0.55 & 2025-09-12 & 12.50 & 29 & 14 & 76 & 8 & 0.40 & 29 & 304.4 \\
MACSJ0553.4-3342 & 88.3436859 & -33.7078056 & 0.41 & 2025-12-29 & 15.36 & 62 & 4 & 171 & 5 & 0.79 & 73 & 2639.9 \\
RXCJ0600.1-2007 & 90.0408333 & -20.1358056 & 0.46 & 2025-11-29 & 10.58 & 58 & 23 & 173 & 24 & 0.68 & 66 & 1982.6 \\
SPT-CLJ0615-5746 & 93.9656173 & -57.7801994 & 0.97 & 2026-03-03 & 11.43 & 23 & 12 & 70 & 3 & 0.39 & 31 & 262.9 \\
Bullet & 104.6118466 & -55.9480870 & 0.30 & 2026-01-18 & 21.44 & 72 & 52 & 196 & 0 & 0.62 & 48 & 1908.9 \\
MACSJ0717.5+3745 & 109.3818750 & 37.7551389 & 0.55 & 2025-11-01 & 15.30 & 81 & 7 & 237 & 0 & 1.04 & 118 & 6434.5 \\
SMACSJ0723.3-7327 & 110.8312500 & -73.4543333 & 0.39 & 2026-06-05 & 10.59 & 16 & 7 & 45 & 0 & 0.34 & 31 & 133.1 \\
RXCJ0911.1+1746 & 137.7975000 & 17.7759722 & 0.51 & 2026-05-01 & 6.40 & 6 & 0 & 13 & 1 & 0.22 & 13 & 15.5 \\
MACSJ0940.9+0744 & 145.2237500 & 7.7403889 & 0.34 & 2026-05-11 & 2.91 & 10 & 8 & 30 & 3 & 0.27 & 21 & 56.0 \\
RXCJ0949.8+1707 & 147.4620833 & 17.1209167 & 0.38 & -- & 10.77 & -- & -- & -- & -- & -- & -- & -- \\
MACSJ1115.8+0129 & 168.9670833 & 1.4990556 & 0.35 & 2026-06-05 & 10.38 & 7 & 1 & 22 & 1 & 0.85 & 25 & 400.2 \\
PLCKG165.7+67.0 & 171.8127695 & 42.4753151 & 0.35 & 2026-03-26 & 4.00\tablenotemark{c} & 41 & 25 & 109 & 0 & 0.11 & 35 & 34.7 \\
MACSJ1149.5+2223 & 177.3987405 & 22.3985246 & 0.54 & 2026-06-05 & 10.89 & 93 & 82 & 269 & 0 & 0.86 & 41 & 5028.9 \\
PLCKG287.0+32.9 & 177.7116667 & -28.0811667 & 0.39 & 2026-06-08 & 13.95 & 33 & 14 & 101 & 2 & 0.75 & 58 & 1421.4 \\
MACSJ1206.2-0847 & 181.5506024 & -8.8009455 & 0.44 & 2026-06-14 & 12.34 & 44 & 40 & 131 & 7 & 0.48 & 27 & 744.4 \\
MACSJ1311.0-0310 & 197.7570833 & -3.1776389 & 0.49 & 2026-05-29 & 5.59 & 4 & 1 & 13 & 0 & 0.23 & 19 & 17.3 \\
Abell 1689 & 197.8748344 & -1.3434880 & 0.18 & 2026-06-11 & 23.15 & 71 & 25 & 213 & 21 & 0.74 & 64 & 2911.7 \\
RXJ1347.5-1145 & 206.8775000 & -11.7528056 & 0.45 & -- & 13.42 & -- & -- & -- & -- & -- & -- & -- \\
Abell 1835 & 210.2587500 & 2.8786667 & 0.25 & -- & 17.02 & -- & -- & -- & -- & -- & -- & -- \\
MACSJ1423.8+2404 & 215.9491667 & 24.0779167 & 0.55 & 2026-03-05 & 5.10 & 9 & 7 & 32 & 1 & 0.41 & 26 & 137.4 \\
RXJ1532.9+3021 & 233.2240833 & 30.3496389 & 0.35 & 2026-05-17 & 4.86 & 6 & 0 & 16 & 1 & 0.26 & 12 & 28.1 \\
PLCKG004.5-19.5 & 289.2687500 & -33.5245833 & 0.54 & 2025-10-04 & 14.49 & 9 & 5 & 23 & 8 & 0.74 & 20 & 311.6 \\
MACSJ1931.8-2635 & 292.9570833 & -26.5761000 & 0.35 & 2025-09-01 & 7.27 & 19 & 9 & 51 & 10 & 0.65 & 21 & 545.9 \\
SMACSJ2031-4036 & 307.9719583 & -40.6252056 & 0.34 & 2025-09-06 & 10.94 & 20 & 17 & 68 & 1 & 0.32 & 25 & 175.3 \\
MACSJ2129.4-0741 & 322.3587949 & -7.6910441 & 0.59 & 2025-10-30 & 10.96 & 21 & 13 & 65 & 1 & 0.83 & 26 & 1125.6 \\
RXJ2129.7+0005 & 322.4165324 & 0.0892660 & 0.23 & 2025-10-29 & 7.49 & 14 & 11 & 38 & 0 & 0.18 & 15 & 30.7 \\
SMACSJ2131.1-4019 & 322.7704586 & -40.3225768 & 0.44 & 2025-10-29 & 9.09 & 9 & 8 & 25 & 1 & 0.55 & 20 & 191.0 \\
MACSJ2137.3-2353 & 325.0632500 & -23.6613056 & 0.31 & 2025-10-30 & 6.91 & 8 & 4 & 22 & 4 & 0.40 & 10 & 87.2 \\
RXCJ2211.7-0349 & 332.9412940 & -3.8289112 & 0.40 & 2025-10-16 & 17.44 & 46 & 3 & 151 & 3 & 0.85 & 57 & 2723.1 \\
MACSJ2214.9-1359 & 333.7387175 & -14.0035406 & 0.50 & 2025-10-16 & 11.42 & 23 & 18 & 67 & 5 & 0.62 & 29 & 639.8 \\
Abell S1063 & 342.1850000 & -44.5301389 & 0.35 & 2025-09-22 & 22.58 & 34 & 31 & 91 & 0 & 0.52 & 26 & 620.8 \\
Abell 2537 & 347.0925000 & -2.1923333 & 0.30 & 2025-11-01 & 9.92 & 15 & 1 & 57 & 7 & 0.65 & 33 & 606.2 \\
SPT-CL J2325-4111 & 351.3001190 & -41.1990680 & 0.36 & 2025-10-01 & 9.01 & 33 & 6 & 103 & 24 & 0.65 & 50 & 1097.3 \\
\hline
Total &  &  &  &  &  & 1733 & 912 & 5013 & 227 &  &  &  \\
Average &  &  & 0.42 &  &  & 30.4 & 16.0 & 87.9 & 4.0 & 0.54 & 34.2 & 988.0 \\
\hline
\enddata
\tablenotetext{a}{\Mfivehundred denotes the mass enclosed within a sphere of radius $R_{500,c}$, within which the mean density is 500 times the critical density of the Universe at the cluster redshift, $\rho_c(z)$.}
\tablenotetext{b}{UNCOVER and MegaScience observations were acquired on 2022 November 2--15 and 2023 November 11--20.}
\tablenotetext{c}{\Mfivehundred for PLCKG165 is not, despite its name, reported in the \textit{Planck} or any other catalog. We use the combination of dynamical and lensing mass from \cite{Frye2019} and \cite{Pascale2022_PLCKG165} to estimate $M_{500,c}$ here.}
\end{deluxetable*}
The VENUS sample was obtained by selecting SL clusters with already well-identified multiply-imaged systems (MIS), which have been observed with HST ACS/F814W and F606W or F625W, in order to detect $z \sim 6$ dropouts, which have VLT/MUSE coverage, to maximize spectroscopic anchoring of the mass models, and which have a large magnified area.
The above selection criteria suffer three exceptions: 
the galaxy clusters MACS J0717.5+3745 and PLCK G165.7+67.0, whose northern position forbids observation with MUSE, and the cluster SPT-CL J2325-4111, not yet targeted with MUSE. 
These exceptions were made for these clusters because they have a significant highly magnified area, and an alternative spectroscopic coverage.
In addition, during the Cycle 4 of the VENUS survey, 7 JWST discretionary director's time (DDT) proposals were accepted, amounting to 19.4 hours of telescope time (accounting only already executed DDTs). This additional NIRCam depth, as well as spectroscopic detections supplement the lens modeling effort.

The VENUS data processing will be described in a future overview article (Fujimoto et al.\ in prep., Kokorev et al.\ in prep.), and we here briefly summarize the process.
The JWST products are consisting of all the individual single-exposure images, already calibrated with stages 1 and 2 of the JWST Pipeline\footnote{\url{https://github.com/spacetelescope/jwst}} \citep{Bushouse2022JWSTPipeline}, which were further processed with the \texttt{grizli} pipeline \citep{brammer_grizli_2023}, as was done for the DAWN JWST Archive \citep[DJA,][]{valentino22}\footnote{\url{https://dawn-cph.github.io/dja/}}. 
The dataset was further post-processed, by including additional background, $1/f$ noise and diffraction spike subtraction procedures, both at the amplifier level for each filter, and then the final drizzled mosaic \citep[][]{endsley24,kokorev25a}. 
The final NIRCam mosaics are drizzled onto a common grid with pixel scales of $0.03''$\,pix$^{-1}$.

\subsection{Archival JWST and HST imaging} \label{sec:ancillary_imaging}

We supplement our JWST dataset with ancillary archival HST and JWST imaging from past observations, whether it is from past large surveys 
(such as CLASH, RELICS, HFF, CANUCS, UNCOVER, PEARLS, GLIMPSE or SLICE) 
which largely overlap with the VENUS cluster sample, or other targeted observations. 
We emphasize that HST/ACS/F606W or F625W and F814W observations exist for all clusters.
Previous HST observations were processed uniformly as part of the Complete Hubble Archive for Galaxy Evolution initiative (CHArGE), as described in \cite{kokorev22}, and pre-existing JWST data were processed similarly to VENUS observations, thus creating a uniform imaging dataset on which SL modeling can rely. These mosaics will be presented in a future article (Kokorev et al.\ in prep.) and released with the VENUS DR1.

In order to supplement spectroscopic redshifts, used to anchor lens models, photometric redshifts constitute a useful prior for systems otherwise lacking spectral detection.
In this article, we use the first internal version of the VENUS \eazy \citep{BammerEazy2008} photometric redshift catalog, which will be described in a future dedicated VENUS data release (Fujimoto et al.\ in prep.).
Such photometric measurements leverage all processed HST and JWST imaging bands available, for each detection, using \texttt{Photutils} \citep{Bradley2025photutils}.

\subsection{Archival spectroscopic observations} \label{sec:spectroscopy}

Due to the VENUS selection criteria, 57 out of 60 clusters have already been covered by the VLT/MUSE IFU \citep{Bacon2010MUSE}. The literature provides a wealth of analysis for these exposures, especially when large catalogs exist for MUSE observations, such as the MUSE atlas \citep{Richard2021_MUSEatlas}. 
Although we did not systematically analyze every single exposure, we examined the archival MUSE data when needed, to confirm the redshift of a newly identified lensed system or to identify interlopers along the line of sight.
Moreover, a thorough literature search for past spectroscopic data reveals a wealth of ground-based (GMOS, MOS/FIRE, etc.) and space-based (NIRSpec) spectral observations, exploited here.
We complemented this pure literature-based compilation with a NED\footnote{\url{https://ned.ipac.caltech.edu}} and DJA\footnote{\url{https://dawn-cph.github.io/dja/spectroscopy/nirspec/}} research for spectral detections,
used for cluster member and MIS identifications.

\subsection{Multiply-imaged systems} \label{sec:MIS}

Provided the VENUS survey selection criterion of pre-existing, well-identified multiple images catalogs, we first retrieve the existing MIS from literature, and notable compilations from previous surveys.
We then systematically realigned the multiple images to the JWST frame of reference, based on Gaia astrometry \citep{Gaia2023_DR3}. Every individual cluster, including previous models and past MIS discovery, is listed in Section \ref{sec:lens_models}. 

After assembling a first lens model, we used VENUS and ancillary imaging to search for more multiple images, considering the combination of their location, geometry and color, aided by VENUS photometric and ancillary spectroscopic redshift catalogs. 
After verifying the compatibility of a new candidate MIS, we use our first lens model to detect possible additional counter-images. The candidates are then added to the next model, and kept if they commensurate with the model.
We repeat this iterative process to assemble the most complete MIS catalog possible before finalizing the lens model.

\section{Methods}\label{sec:methods}

\subsection{Cluster member galaxies selection} \label{sec:gal_selection}

Using the VENUS imaging catalogs (processed as described Section \ref{sec:ancillary_imaging}) and spectroscopy (Section \ref{sec:spectroscopy}), we build the catalog of cluster member galaxies for each cluster using the red-sequence in two distinct spaces \citep{GladdersYee2000Finder}.
We use the NIRCam F090W-F150W and F356W-F444W sequences, which empirically provide excellent results, especially if guided by abundant spectroscopic detections. 
For some specific clusters, we use a different combination of bands, if e.g.\ one of the aforementioned filters does not cover the cluster field, in which case we specify the red sequence combination used for each individual cluster in the data release \texttt{Readme} files (to be published with DR1).
Unless they are spectroscopically confirmed, detections of F150W magnitude $> 23$ AB are removed.

\subsection{Lens modeling software: \ALP}
\label{sec:lensing_software}

In this work, we use the parametric lensing software \ALP to model the clusters' gravitational potentials in the thin lens approximation.
First introduced in \citet{Allingham2026_M1931}, \textbf{\ALP} is the \texttt{Python3} implementation of the \textbf{\ZA} lensing software, itself a revised version of the \texttt{PIEMDeNFW} parametric method which was used to model the CLASH cluster sample, as described in \cite{Zitrin2015CLASH25}. 
In preparation for JWST's increased resolution, the \ZA pipeline, written in \texttt{MATLAB}, was revised to reduce run-time, and to accommodate arbitrarily higher resolutions so that it is no longer coupled to a fixed grid. 
Indeed, the method was then successfully applied to a variety of clusters in JWST data \cite{Pascale2022SMACS0723,Furtak2022UNCOVER,Meena2023_lensedstars,Golubchik2025_LRD_VENUS}, where more details can be found. 
\ALP conserves these main characteristics of \ZA, follows the same architecture, and thus allows for a fluid compatibility between both software. The fast computation is here ensured by a \texttt{Numba} \citep{Numba_17874185} compilation of \texttt{Python} functions into optimized machine code.

The software employs large-scale, smooth DMHs with the Pseudo-Isothermal Elliptical Matter Distribution (PIEMD; \citealt{Jaffe1983, Keeton2001models}) radial density profile:
\begin{equation}
    \rho (r) = \frac{\rho_0}{1 + \left(r / a \right)^2} = \frac{\sigma_0^2}{2 \pi G} \frac{1}{a^2 + r^2},
    \label{eq:PIEMD_profile}
\end{equation}
where $\rho_0$ is the central 3D density, $\sigma_0$ the 1D velocity dispersion, $a$ the core radius and $G$ Newton's universal gravitational constant.
Individual galaxies are modeled with the dual Pseudo-Isothermal Elliptical profiles \citep[dPIE, or pseudo-Jaffe; see][]{KassiolaKovner1993, Eliasdottir2007arXiv0710.5636E}, i.e.\ a combination of two PIEMD profiles, effectively truncating the density profile:
\begin{equation}
\begin{split}
    \rho (r) &= \frac{\rho_0}{\left[1 + \left(r / a \right)^2\right] \left[1 + \left(r / s \right)^2\right]}\\
    &= \frac{\sigma_0^2}{2 \pi G} \left[ \left(a^2 + r^2\right)^{-1} - \left(s^2 + r^2\right)^{-1} \right],
    \label{eq:dPIE_profile}
\end{split}
\end{equation}
where $s$ is the cut radius.
All cluster-member galaxies have their positional parameters (position, ellipticity, position angle) fixed, while their density parameters (velocity dispersion, core and cut radii) are scaled together through the \citet[][FJ]{FaberJackson} relationship for red elliptical galaxies. For a galaxy $i$ of luminosity $L_i$:
\begin{equation}
    \begin{split}
    \sigma_{0, i} &= \sigma_{0, \star} \left( \frac{L_i}{L_{\star}} \right)^{\lambda},\\
    a_i &= a_{\star} \left( \frac{L_i}{L_{\star}} \right)^{\beta},\\
    s_i &= s_{\star} \left( \frac{L_i}{L_{\star}} \right)^{\alpha},
    \label{eq:Faber-Jackson}
    \end{split}
\end{equation}
where the pivot scale core-radius, $a_{\star}$, is generally fixed
and the pivot scale cut-radius $s_{\star}$ and velocity dispersion $\sigma_{\star}$ of a reference L$^{\star}$ galaxy are always free parameters. 
The FJ power-law indices are generally fixed to their default values $\lambda = 1/4$, $\beta = 1/2$ and $\alpha = 1/2$, although they can be optimized parameters.
Moreover, for some individual galaxies whose precise modeling matters, we let the luminosity $L_i$ free, with an associated weight $w_i$, simply multiplying $L_i$ such that $L_i \rightarrow w_i L_i$, effectively enabling one degree of freedom along the FJ scaling relation.
For the same reasons, for some cluster members, the galaxy's core radius may be optimized as well.
Moreover, the redshifts of MIS lacking spectroscopic measurements may be left free.
At last, adding a shear field to the lens model is possible.

In order to compute the source position, \ALP takes the full Jacobian matrix into account \citep[e.g.][]{Keeton2010}. 
This effectively constitutes a Newton-Raphson (NR) approximation of the image plane optimization, successfully used before also in e.g.\ \citet[][]{Oguri2010_S1014}. With:
\begin{equation}
    \mathcal{A}(\boldsymbol{\theta}) = \frac{\partial \boldsymbol{\beta}}{\partial \boldsymbol{\theta}},
\end{equation}
the amplification matrix, where $\boldsymbol{\theta}$ is the image position and $\boldsymbol{\beta}$ the source position, a local $\boldsymbol{\delta \beta_{ij}} = \boldsymbol{\beta_{ij}} - \boldsymbol{\bar{\beta_{i}}}$ perturbation distance of the barycentered source $i$ from the observed image $j$ is, at the first-order Taylor expansion around the image position:
\begin{equation}
    \boldsymbol{\delta \beta} \simeq \mathcal{A} \cdot \boldsymbol{\delta \theta},
    \label{eq:Taylor_dbeta}
\end{equation}
thus allowing to approximate the image predicted position to:
\begin{equation}
    \boldsymbol{\theta_{ij}^p} \simeq \boldsymbol{\theta_{ij}} - \mathcal{A}_{ij}^{-1} \cdot \left( \boldsymbol{\beta_{ij}} - \boldsymbol{\bar{\beta_{i}}} \right),
\end{equation}
where $\boldsymbol{\theta_{ij}^p}$ is the predicted location for image $j$, from which we can compute $\chi^2$ in the pseudo-image plane:
\begin{equation}
    \chi^2_{\rm SL, imgNR} = \sum_{i=1}^{N_{\rm Sys}} \sum_{j=1}^{N_i} \frac{\left( \boldsymbol{\theta_{ij} - \boldsymbol{\theta_{ij}^p}} \right)^2}{\sigma_{ij}^2},
    \label{eq:img-pl-NR_chi2}
\end{equation}
where $N_{\rm Sys}$ is the number of MIS.
Note that near critical curves, where $\det \mathcal{A} \to 0$, this approximation breaks down, thus constituting the limit of the NR approximation.

For minimizing the $\chi^2$, \ALP can use two different MCMC samplers:
either the \ZA MCMC algorithm \citep{Zitrin2015CLASH25}, a Metropolis-Hastings-based algorithm with effective temperature annealing,
or, alternatively, \texttt{emcee} \citep{emcee_10996751} to perform a MCMC-like algorithm for exploring the parameter space through affine-invariant ensemble sampling moves \citep[cf.][]{GoodmanWeare2010StretchMove}. 
In practice, all \ALP models here presented use the \texttt{emcee} algorithm, which is parallelized \citep{emcee-Foreman-Mackey-2013}. 

After optimizing the $\chi^2$ in the NR approximation of the image plane, the final values provided in our lens models are all computed in the image plane, such as the best-fit $\chi^2$ and the \rms, representing the average positional error of a final model:
\begin{equation}
    \rms = \sqrt{\frac{1}{N_{\rm img}} \sum_{n=1}^{N_{\rm img}} \norm{\boldsymbol{\theta_{n}} - \boldsymbol{\theta_{n}^p}}^2},
    \label{eq:RMS}
\end{equation}
where $N_{\rm img}$ is the total number of constraining multiply-lensed images, $\boldsymbol{\theta_n}$ is the $n$-th observed image position, and $\boldsymbol{\theta^p_n}$ its model-predicted position.
We note that lower \rms are not a convincing witness of a mass model's quality if the number of MIS is not large enough (i.e.\ less than $\sim 5-10$ systems).

\subsection{Lens modeling} \label{sec:lensmodelling}

Although priors applied to each individual cluster differ, we here summarize the common lens modeling assumptions.
The whole potential is the sum of DMHs modeled with PIEMD profiles, the cluster galaxies modeled with dPIE and a possible external shear.
By default, every DMH has six free parameters: velocity dispersion and core radius, and positional parameters  -- central coordinates offset, ellipticity and position angle; the galaxies are all linked by the three FJ parameters: velocity dispersion and cut-radius of a typical L$^{\star}$ galaxy, while the core-radius is generally fixed to 0.20 kpc \citep[following the typical values of elliptical galaxies and other lens modeling work, e.g.][]{Richard2014FF}, but can also be let free.
The FJ power indices (see eq.\,\ref{eq:Faber-Jackson}) are generally fixed but can be let free, for example if the model does not converge convincingly otherwise. 

The role of BCGs, often disjoint from the rest of the main FJ sequence, is uniquely important \citep[e.g.][]{2013ApJ...765...24N}, and thus one (or multiple) BCG is optimized separately from the main FJ distribution, by appointing it a weight, and leaving the core radius free.
Although the ellipticity parameters can be optimized as well, we here fix them throughout our analysis to their measured values.
Moreover galaxies close to lensed images used as constraints are also appointed a weight as a free parameter. 
In the case of multi-plane lensing, which is not yet integrated to our software suite, a deflector is omitted from the model completely when required (e.g.\ if it is located in the foreground at a very low redshift), but, if necessary, it is attached to the main FJ sequence and appointed a free weight. 

Redshifts are fixed to their spectroscopic values, while other systems' redshifts are left free in lens models, using photometric redshifts as a prior. For rare models entirely lacking spectroscopic identifications, high-quality photometric redshifts ($\delta z < 0.05$ with $5\sigma$ certainty) may be fixed.
For the $\chi^{2}$ minimization, we adopt a positional error of $\sigma_{\rm pos} = 0.2''$ per image. 
With this nominal error, in Table \ref{tab:all_clusters_1}, we provide the best image-plane SL $\chi^2$. 
In order to derive consistent statistical errors, the positional errors should be rescaled for individual clusters, bringing the reduced $\chi^2 / N_{\rm DoF}$ close to 1.
For specific, well-constrained systems whose reproduction is paramount to the lens modeling (usually point sources such as e.g.\ a transient, a lensed LRD, etc.), we further restrict the positional error to give it more weight in the minimization.

\section{Results} \label{sec:Results}

\subsection{Observables} \label{sec:observables}

Relevant lens modeling results are summarized in Table \ref{tab:all_clusters_3}. 
The critical area, $\mathcal{A}^{\rm crit}_{z_s}$, is defined as the cumulative area enclosed by the most external (tangential) critical lines for a given source redshift $z_s$ \citep{FoxMahlerSharon2022}. 
Given the tangential magnification, $\mu_t = \left( 1 - \kappa - \gamma \right)^{-1}$, where $\kappa$ and $\gamma$ are respectively the lensing convergence and shear, the critical area is defined as the area for which $1 - \kappa - \gamma < 0$ or for which $1 - \kappa + \gamma < 0$.
The effective Einstein radius for a source redshift $z_s$ is defined as $\theta_{E, z_s} = \sqrt{\mathcal{A}^{\rm crit}_{z_s} / \pi}$.
Table \ref{tab:all_clusters_3} also provides the cumulated magnified area, $\mathcal{A}^{|\mu|\geq \mu_T}_{z_s}$, in the image plane above a given magnification threshold, $\mu_T$ $\in \left\{3, 10\right\}$ here, for a given source redshift, $z_s \in \left\{2, 10\right\}$ here.
Finally, $M_{\rm 2D}(<R_T)$ is the projected mass within a $R_T$ radius around the cluster center ($=200$\,kpc here).
We note that, for this latter observable, the cluster center is always taken to be the main BCG, i.e.\ the brightest cluster galaxy at the center of the main, most massive DMH. Although this definition leads to straightforward measurements, it can sometimes neglect a large part of the cluster mass, especially in merger cases -- such as the Bullet cluster.

\subsection{Cluster lens models} \label{sec:lens_models}

We present our \ncl \ALP lens models, in the order of ascending RA, with a quick summary of the known cluster models and remarkable features -- if any.
For many of these clusters, the models presented here rely on the largest number of MIS used as constraints to date.
While most MIS identifications were carried out independently, a subset was adopted from our companion paper, \Oguri, as described hereafter.
The modeling characteristics of each cluster (numbers of MIS and images, image-plane \rms) and original MIS identifications are summarized in Table \ref{tab:all_clusters_1}.
\textbf{For each cluster, the lens model, including the deflection maps, magnification, etc., the cluster member galaxies and MIS catalogs will be publicly released on the VENUS site\footnote{\url{https://jwst-venus.github.io}} at a later date.}
Once the 3 remaining VENUS targets are observed, we will update our results for this article to model all 60 clusters.
\begin{longtable*}{l c cc cc cc cc}
	\caption{Lensing observables of our \ALP models for VENUS DR1 clusters. Columns: (i) Cluster name; (ii) Cylindrical 2D mass contained within 200 kpc from the center (defined as the main BCG); 
    (iii) Effective Einstein radius for a source at $z_s = 2$; (iv) Effective Einstein radius for a source at $z_s = 10$; (v) Projected mass contained within the area $\mathcal{A} = \pi \theta_{E}^2$ enclosed by the critical lines for a source at redshift $z_s = 2$; (vi) Projected mass contained within the area enclosed by the critical lines for a source at redshift $z_s = 10$; (vii) Area on the field in the image plane for which the magnification is $|\mu|\geq 3$, for a source redshift $z_s = 2$; (viii) Area on the field in the image plane for which the magnification is $|\mu|\geq 3$, for a source redshift $z_s = 10$; (ix) Area on the field in the image plane for which the magnification is $|\mu|\geq 10$, for a source redshift $z_s = 2$; (x) Area on the field in the image plane for which the magnification is $|\mu|\geq 10$, for a source redshift $z_s = 10$.
    For all statistics provided here, the systematics dominate statistical errors, and, based on previous studies \citep{Zitrin2015CLASH25}, we take the total error budget to be $10\%$ on the Einstein radii, $15\%$ on the critical area enclosed mass, $20\%$ on the magnified area above a threshold and $30\%$ on the integrated cylindrical mass.
    }
	\label{tab:all_clusters_3}\\
\hline\hline
Cluster name & $M_{\rm 2D}(<200\,\mathrm{kpc})$ & \multicolumn{2}{c}{$\theta_{E}$} & \multicolumn{2}{c}{$M_{\rm 2D}(\in \mathcal{A})$} & \multicolumn{2}{c}{$\mathcal{A}(|\mu|\ge 3)$} & \multicolumn{2}{c}{$\mathcal{A}(|\mu|\ge 10)$} \\
 & [$10^{14}\Msol$] & \multicolumn{2}{c}{[$''$]} & \multicolumn{2}{c}{[$10^{14}\Msol$]} & \multicolumn{4}{c}{[arcmin$^2$]}\\
\cmidrule(lr){3-4} \cmidrule(lr){5-6} \cmidrule(lr){7-8} \cmidrule(lr){9-10}
$z_s$ &  & $2$ & $10$ & $2$ & $10$ & $2$ & $10$ & $2$ & $10$ \\
\hline
\endfirsthead
\multicolumn{10}{c}{Table \thetable\ (continued)}\\
\hline\hline
Cluster name & $M_{\rm 2D}(<200\,\mathrm{kpc})$ & \multicolumn{2}{c}{$\theta_{E}$} & \multicolumn{2}{c}{$M_{\rm 2D}(\in \mathcal{A})$} & \multicolumn{2}{c}{$\mathcal{A}(|\mu|\ge 3)$} & \multicolumn{2}{c}{$\mathcal{A}(|\mu|\ge 10)$} \\
 & [$10^{14}\Msol$] & \multicolumn{2}{c}{[$''$]} & \multicolumn{2}{c}{[$10^{14}\Msol$]} & \multicolumn{4}{c}{[arcmin$^2$]}\\
\cmidrule(lr){3-4} \cmidrule(lr){5-6} \cmidrule(lr){7-8} \cmidrule(lr){9-10}
$z_s$ &  & $2$ & $10$ & $2$ & $10$ & $2$ & $10$ & $2$ & $10$ \\
\hline
\endhead
\hline
\endfoot
\hline
\hline
\endlastfoot
\hline\hline
Abell 2744 & 1.58 & 27.95 & 35.75 & 1.08 & 1.56 & 5.23 & 7.23 & 0.96 & 1.47 \\
RXCJ0018.5+1626 & 1.99 & 22.71 & 33.94 & 1.31 & 2.36 & 2.25 & 3.88 & 0.59 & 1.15 \\
MACSJ0025.4-1222 & 1.47 & 11.06 & 27.29 & 0.31 & 1.39 & 2.26 & 3.80 & 0.30 & 1.35 \\
RXCJ0032.1+1808 & 1.90 & 39.91 & 50.71 & 2.68 & 3.89 & 5.59 & 7.51 & 1.64 & 2.36 \\
MACSJ0035.4-2015 & 1.55 & 19.95 & 25.34 & 0.59 & 0.84 & 1.48 & 2.10 & 0.37 & 0.53 \\
Abell 68 & 1.66 & 20.04 & 26.47 & 0.43 & 0.67 & 2.45 & 3.10 & 0.71 & 0.96 \\
Abell 2813 & 1.95 & 26.19 & 31.27 & 0.95 & 1.23 & 2.42 & 2.95 & 0.77 & 1.02 \\
El Gordo & 1.90 & 16.68 & 45.45 & 1.04 & 5.37 & 2.17 & 3.98 & 0.58 & 1.34 \\
Abell 209 & 0.78 & 13.51 & 15.13 & 0.16 & 0.19 & 0.42 & 0.49 & 0.11 & 0.13 \\
WHLJ013719.8-082841 & 2.07 & 10.48 & 34.55 & 0.30 & 2.53 & 3.55 & 4.54 & 0.87 & 2.36 \\
CLJ0152.7-1357 & 1.33 & 13.10 & 25.36 & 0.63 & 1.56 & 0.61 & 1.62 & 0.13 & 0.46 \\
MACSJ0159.8-0849 & 1.92 & 24.62 & 31.49 & 1.19 & 1.62 & 1.70 & 2.37 & 0.47 & 0.63 \\
RXCJ0232.2-4420 & 2.17 & 25.81 & 32.08 & 0.91 & 1.26 & 3.56 & 4.49 & 0.98 & 1.41 \\
Abell 370 & 2.28 & 36.58 & 45.43 & 2.34 & 3.14 & 3.97 & 5.04 & 1.26 & 1.80 \\
Abell S295 & 1.29 & 25.99 & 36.35 & 0.82 & 1.35 & 3.57 & 4.81 & 0.85 & 1.35 \\
Abell 383 & 1.60 & 16.90 & 19.68 & 0.26 & 0.33 & 1.42 & 1.71 & 0.46 & 0.58 \\
MACSJ0257.1-2325 & 1.43 & 14.55 & 21.01 & 0.49 & 0.82 & 0.82 & 1.40 & 0.21 & 0.38 \\
MACSJ0257.6-2209 & 1.95 & 26.43 & 31.57 & 1.06 & 1.35 & 2.03 & 2.73 & 0.59 & 0.79 \\
MACSJ0308.9+2645 & 2.57 & 37.45 & 45.36 & 2.39 & 3.03 & 3.41 & 4.32 & 0.89 & 1.17 \\
PLCKG171.9-40.7 & 2.33 & 31.85 & 38.55 & 1.27 & 1.70 & 4.27 & 5.26 & 1.34 & 1.73 \\
MACSJ0329.7-0211 & 1.88 & 23.84 & 33.22 & 1.13 & 1.84 & 2.22 & 3.26 & 0.64 & 0.95 \\
Abell 3192 & 1.95 & 28.05 & 34.54 & 1.66 & 2.15 & 1.83 & 2.48 & 0.58 & 0.88 \\
MACSJ0416.1-2403 & 1.55 & 25.79 & 32.69 & 1.09 & 1.55 & 1.90 & 2.53 & 0.50 & 0.70 \\
MACSJ0417.5-1154 & 2.02 & 31.22 & 41.22 & 1.86 & 2.80 & 3.65 & 5.45 & 0.85 & 1.41 \\
MACSJ0429.6-0253 & 1.29 & 16.25 & 20.26 & 0.50 & 0.65 & 0.66 & 0.93 & 0.18 & 0.25 \\
MACSJ0451.9+0006 & 1.79 & 25.52 & 33.11 & 1.24 & 1.76 & 1.65 & 2.10 & 0.50 & 0.67 \\
Abell 521 & 0.75 & 8.23 & 9.74 & 0.07 & 0.10 & 0.26 & 0.33 & 0.07 & 0.09 \\
MACSJ0454.1-0300 & 1.87 & 21.81 & 30.96 & 1.19 & 1.94 & 1.61 & 2.61 & 0.44 & 0.81 \\
MACSJ0553.4-3342 & 1.78 & 35.75 & 45.06 & 2.16 & 3.00 & 3.15 & 3.96 & 0.87 & 1.21 \\
RXCJ0600.1-2007 & 2.05 & 46.33 & 63.70 & 3.86 & 6.30 & 7.46 & 9.65 & 2.54 & 3.37 \\
SPT-CLJ0615-5746 & 2.36 & 17.14 & 34.95 & 1.48 & 3.75 & 0.88 & 2.42 & 0.27 & 0.72 \\
Bullet & 1.62 & 26.52 & 33.64 & 0.87 & 1.25 & 3.55 & 5.14 & 0.83 & 1.17 \\
MACSJ0717.5+3745 & 1.45 & 49.18 & 67.35 & 5.57 & 9.00 & 6.46 & 9.55 & 2.55 & 3.12 \\
SMACSJ0723.3-7327 & 1.71 & 18.46 & 25.76 & 0.58 & 0.99 & 1.87 & 2.70 & 0.51 & 0.81 \\
RXCJ0911.1+1746 & 1.16 & 16.87 & 23.28 & 0.61 & 0.91 & 0.62 & 0.95 & 0.19 & 0.22 \\
MACSJ0940.9+0744 & 0.95 & 11.27 & 13.44 & 0.20 & 0.26 & 0.36 & 0.51 & 0.09 & 0.13 \\
RXCJ0949.8+1707 & -- & -- & -- & -- & -- & -- & -- & -- & -- \\
MACSJ1115.8+0129 & 1.24 & 21.92 & 25.55 & 0.81 & 0.95 & 0.74 & 0.91 & 0.19 & 0.26 \\
PLCKG165.7+67.0 & 0.91 & 12.79 & 15.69 & 0.24 & 0.32 & 0.46 & 0.61 & 0.14 & 0.16 \\
MACSJ1149.5+2223 & 1.78 & 22.41 & 32.96 & 1.15 & 2.06 & 1.97 & 3.20 & 0.54 & 0.91 \\
PLCKG287.0+32.9 & 2.67 & 45.40 & 60.52 & 3.64 & 5.50 & 6.89 & 9.01 & 2.45 & 3.30 \\
MACSJ1206.2-0847 & 2.08 & 27.19 & 36.28 & 1.50 & 2.24 & 2.41 & 3.44 & 0.67 & 1.00 \\
MACSJ1311.0-0310 & 1.41 & 15.53 & 22.16 & 0.57 & 0.89 & 0.77 & 1.22 & 0.19 & 0.35 \\
Abell 1689 & 3.00 & 50.35 & 54.73 & 2.27 & 2.51 & 5.36 & 6.07 & 1.49 & 1.68 \\
RXJ1347.5-1145 & -- & -- & -- & -- & -- & -- & -- & -- & -- \\
Abell 1835 & -- & -- & -- & -- & -- & -- & -- & -- & -- \\
MACSJ1423.8+2404 & 1.76 & 18.82 & 28.42 & 0.89 & 1.58 & 1.42 & 2.46 & 0.33 & 0.71 \\
RXJ1532.9+3021 & 1.24 & 15.30 & 18.78 & 0.37 & 0.49 & 0.70 & 0.97 & 0.18 & 0.25 \\
PLCKG004.5-19.5 & 2.51 & 38.24 & 50.88 & 3.86 & 5.56 & 3.66 & 5.18 & 1.15 & 1.78 \\
MACSJ1931.8-2635 & 1.60 & 22.35 & 26.74 & 0.79 & 1.00 & 1.29 & 1.57 & 0.34 & 0.46 \\
SMACSJ2031-4036 & 1.53 & 17.64 & 25.74 & 0.46 & 0.87 & 2.10 & 2.93 & 0.59 & 0.86 \\
MACSJ2129.4-0741 & 1.84 & 21.70 & 33.79 & 1.26 & 2.36 & 1.99 & 3.04 & 0.44 & 1.05 \\
RXJ2129.7+0005 & 1.20 & 13.94 & 16.61 & 0.20 & 0.26 & 0.82 & 1.04 & 0.20 & 0.27 \\
SMACSJ2131.1-4019 & 1.67 & 21.29 & 28.32 & 0.96 & 1.37 & 1.15 & 1.59 & 0.40 & 0.52 \\
MACSJ2137.3-2353 & 1.25 & 17.33 & 20.20 & 0.46 & 0.55 & 0.63 & 0.81 & 0.15 & 0.19 \\
RXCJ2211.7-0349 & 2.86 & 43.14 & 52.15 & 3.56 & 4.51 & 3.46 & 3.94 & 1.18 & 1.40 \\
MACSJ2214.9-1359 & 1.69 & 19.74 & 26.59 & 0.94 & 1.38 & 1.02 & 1.62 & 0.31 & 0.47 \\
Abell S1063 & 2.11 & 30.57 & 39.07 & 1.44 & 2.03 & 3.35 & 4.50 & 0.90 & 1.38 \\
Abell 2537 & 2.01 & 27.72 & 33.65 & 1.05 & 1.38 & 2.59 & 3.27 & 0.78 & 0.99 \\
SPT-CL J2325-4111 & 2.46 & 32.72 & 43.19 & 1.88 & 2.79 & 4.58 & 5.32 & 1.35 & 2.11 \\
\hline
Total & 100.74 &  &  & 72.57 & 115.05 & 138.70 & 190.61 & 39.66 & 59.58 \\
Average & 1.77 & 24.56 & 33.12 & 1.27 & 2.02 & 2.43 & 3.34 & 0.70 & 1.05 \\
Standard deviation & 0.48 & 10.11 & 12.28 & 1.08 & 1.69 & 1.71 & 2.22 & 0.57 & 0.76 \\
\hline
\end{longtable*}

\begin{toggleblock}
\begin{figure*}
    \centering
    \includegraphics[width=0.85\linewidth]{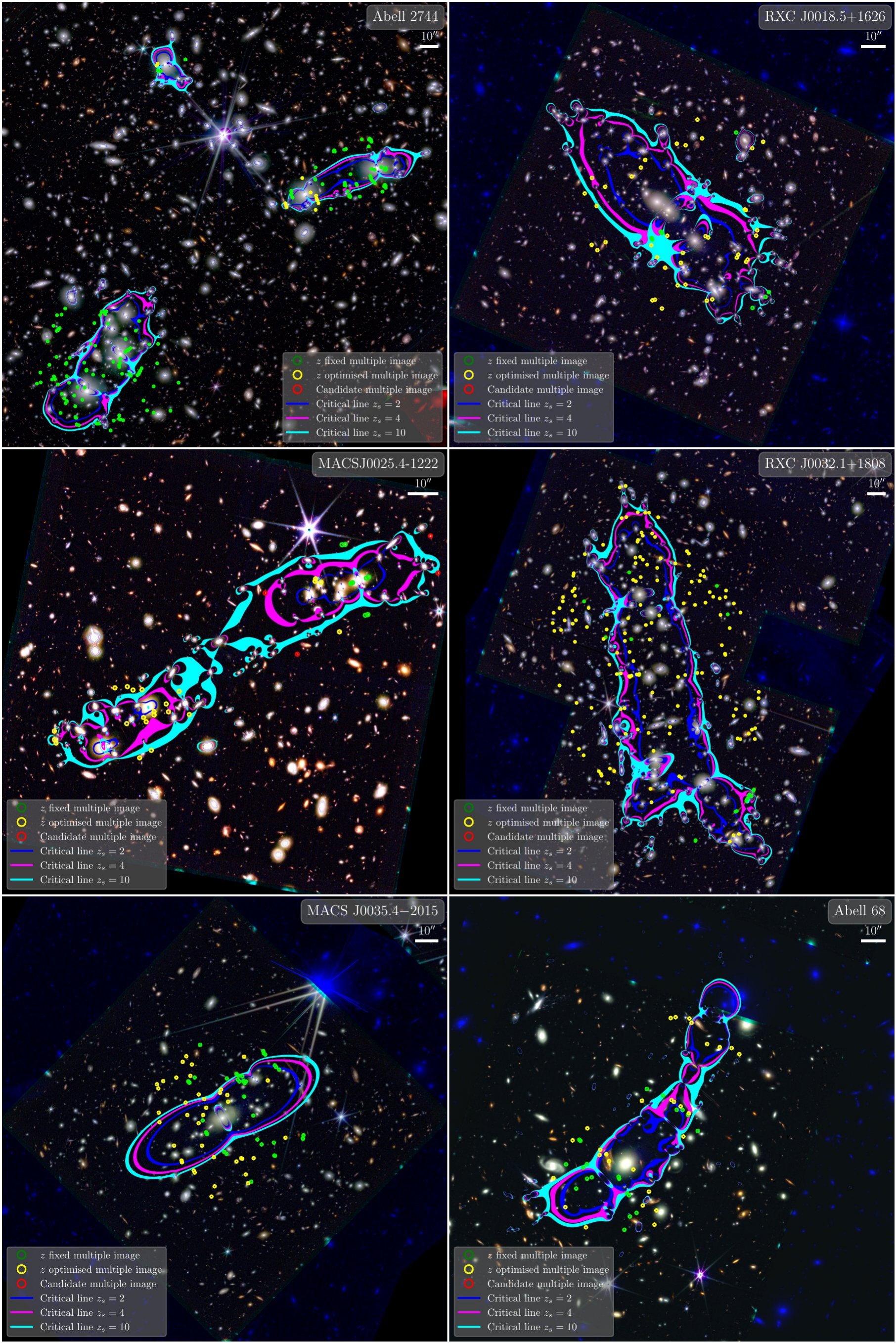}
    \caption{RGB false-color images for clusters A2744, R0018, M0025, R0032, M0035 and A68. 
    The RGB images are primarily obtained through the following coaddition: Blue: NIRCam/F090W+F115W+F150W; Green: F200W+F210M+F277W; Red: F300M+F356W+F410M+F444W. In some cases the combination may in practice differ (e.g.\ the A68 blue band makes use of HST/ACS to show out-of-NIRCam field structures). 
    \textit{Blue, magenta and cyan curves:} critical curves at various source redshifts, covering the $|\mu|\geq100$ regions. 
    \textit{Green circles:} fixed-$z$ multiple images (spec-$z$ or excellent photometric redshifts, see Sec.\,\ref{sec:lensmodelling}). \textit{Golden circles:} other MIS with a free redshift. \textit{Red circles:} candidate systems (when available), not included in modeling.}
    \label{fig:clusters_1}
\end{figure*}
\end{toggleblock}

\subsubsection{Abell 2744 (A2744)} \label{sec:a2744}

Abell~2744 ($z=0.31$) is perhaps the best-studied SL field to date. 
It was included in the HFF survey with HST and then became the second SL cluster to be observed with JWST in the framework of the GLASS-JWST \citep{treu22}, UNCOVER, PEARLS, and All the Little Things \citep[ALT,][]{Naidu2024_ALP} programs, as well as numerous JWST follow-up programs in subsequent cycles. As such, A2744 has an exceptionally long history of SL modeling and analysis. We refer the reader to recent dedicated A2744 SL modeling papers for a detailed history of this cluster, e.g.\ \citet{Furtak2022UNCOVER}, \citet{bergamini22}, or \citet{Patel2026_A2744}.

While not observed anew in VENUS, A2744 is included in the sample because it has sufficient JWST coverage to allow a similar analysis as for VENUS clusters (see Fujimoto et al.\ in prep.). We therefore here do not derive a new SL model of A2744, but rather use the \ZA UNCOVER model by \citet{Furtak2022UNCOVER} and \citet{Price2025_UNCOVER}, newly optimized with \ALP. This model parametrizes the cluster with 5 DMHs and is constrained with 186 multiple images of 65 sources, 59 of which have spectroscopic redshifts \citep{bergamini22,RobertsBorsani2023highz,Furtak2024_LRD,Naidu2024_ALP,Price2025_UNCOVER} and achieves a final lens-plane average image reproduction error of $\rms=0.37''$. We refer the reader to \citet{Price2025_UNCOVER} for all the details about this model.

\subsubsection{RXC J0018.5+1626 (R0018)} \label{sec:r0018}

RXC~J0018.5+1626 ($z=0.55$), also known as MACS~J0018.5+1626 or CL~J0016+1609, is a very X-ray luminous galaxy cluster that was included in the RELICS and SLICE surveys prior to VENUS. It most prominently holds a multiply-imaged blue-dominated double-peaked \Lya emitter at $z=3.218$ \citep{Furtak2022_Lya}, as well as the multiply-image SN Helios at $z=3.34$ \citep{Fujimoto2026_SNHelios}. The cluster was initially discovered as a strong lens by \citet{Zitrin2011_12macsclusters} with the \LTM method. This model was built on in \citet{Furtak2022_Lya}, which used five MIS, one of which is the $z=3.218$ \Lya emitter, to refine the \LTM model of R0018. Now, with the VENUS JWST data, we model the cluster with two DMHs. Our model is constrained with 47 multiple images of 18 sources, 3 of which have spectroscopic redshifts \citep{Mainali2019_PhD,Furtak2022_Lya}, and achieves a lens-plane $\rms=0.37''$. 
This model was presented in \citet{Fujimoto2026_SNHelios}, alongside four other VENUS and SLICE SL models derived with \ZA, \glafic, \Lenstool, and \wslap, where they are used to constrain SN Helios.
We refer the reader to our dedicated paper Furtak et al.\ (in prep.), which will follow-up on the $z=3.218$ lensed \Lya emitter, for more details on our SL model of R0018.

\subsubsection{MACS J0025.4-1222 (M0025)} \label{sec:m0025}

Cluster MACS J0025.4-1222 ($z=0.58$) corresponds to a major merger system, whose collisional profile, with an ICM peak located between two DMHs, makes it a perfect example of a `Bullet-like' cluster (see Sec.\,\ref{sec:cl0658}), and owes it the nickname `Baby Bullet'.
First imaged with HST, its mass map was modeled by \cite{Bradac2008macsbullet, Zitrin2011_12macsclusters}, and later by \cite{Cibirka2018}, with RELICS observations.
The cluster is composed of two disjoint DMHs, each constrained by SL arcs. Only the NW halo presents a spectroscopically-confirmed lensed source, leaving the SE clump with only photometric redshift constraints (in a fashion similar to AS295, see Section \ref{sec:as295}).
Supplementing six previously reported MIS (including \citealt{Oguri2026_VENUS}'s) with four new detections, we assemble a catalog of 10 multiply-lensed galaxies to constrain the lensing potential. The observed images are reproduced with a final precision \rms of $0.16''$.

\subsubsection{RXC J0032.1+1808 (R0032)} \label{sec:r0032}

RXC~J0032.1+1808 ($z=0.40$) is a rich, bi-modal major merging cluster that was included in the RELICS survey prior to VENUS. 
It displays a spectacular ALMA-detected sextuply-imaged dusty star-forming galaxy at $z=3.631$ \citep{Dessauges-Zavadsky2017_RXC0032,Dessauges2023_ALMA}, but otherwise crucially lacks spectroscopic redshifts. 
A comprehensive SL study of R0032 based on the RELICS HST data was presented in \citet{Acebron2019RXC0032}, including four SL models of the cluster made with \LTM, \piemdenfw, \Lenstool, and \glafic, which were constrained with 18 multiple images of 5 sources and all agreed on a very large Einstein radius of $\theta_{E, z_s=2}\sim40''$ for a source at $z=2$. 
Using the combination of SLICE and VENUS data, we model R0032 with three DMHs following \citet{Acebron2019RXC0032}. Our model is constrained with 162 multiple images of 55 sources, 4 of which have spectroscopic redshifts \citep{Dessauges-Zavadsky2017_RXC0032,Tsujita2025MACS0417}, and achieves a final lens-plane image reproduction error of $\rms=0.66''$. This JWST-based model confirms the cluster's large Einstein radius, of $\theta_{E, z_s=2}=39.9 \pm 4.0''$ (see Tab.~\ref{tab:all_clusters_3}).

\subsubsection{MACS J0035.4-2015 (M0035)}

The cluster MACS~J0035.4-2015 ($z=0.35$) was observed with HST in RELICS, prior to JWST in VENUS. It contains a rare multiply-imaged AGN at $z=2.06$, which is detected in X-rays, ALMA, and displays strong rest-frame UV variability \citep{Furtak2023AGN}. \citet{FoxMahlerSharon2022} list two publicly available RELICS models of M0035 constructed with \glafic and \Lenstool, and \citet{Furtak2023AGN} use an HST-based \ZA model in their analysis. An additional parametric SL model of M0035 was published in \citet{Somma2025_MACS0035}. For VENUS, we model the cluster based on JWST/NIRCam data with \ALP, using one DMH. The model is constrained with 74 multiple images of 27 sources, 12 of which have spectroscopic redshifts \citep{Furtak2023AGN,Somma2025_MACS0035}, and has a final lens-plane image reproduction error of $\rms=0.50''$.

\subsubsection{Abell 68 (A68)} \label{sec:a68}

Abell 68 ($z=0.25$) was previously observed with JWST as part of SLICE.
We complement these data with archival HST imaging (PID 11591). A68 is an X-ray luminous \citep{Ebeling1996}, dynamically unrelaxed cluster with a bimodal mass distribution \citep{smith2005,Richard2007,Cerny2026SLICE}.
We optimize the model using 18 multiply-imaged systems, comprising a total of 49 multiple images. Of these, 14 systems were previously known \citep{Richard2007,Cerny2026SLICE}, and 4 are newly identified in the VENUS data. Seven systems (1, 2, 3, 5, and 11, where systems 1 and 2 are both divided in two knots) have spectroscopic redshifts \citep{smith2005,Richard2007}. The mass distribution is described by 3 DMHs. 

Notably, system 3 is a lensed LRD candidate with a blue companion at $z=5.421$ \citep[mentioned in][discovered independently]{Baggen2026_LRD}, with a magnification of $\mu\sim11$ for each of its three images and predicted time delays of 0.40 and 0.11 years relative to the first-arriving image. The final model reproduces the image positions with a \rms of $0.35''$.

\subsubsection{Abell 2813 (A2813)} \label{sec:a2813}

Abell 2813 ($z=0.29$) is a major merger cluster, whose lensing profile was first studied by \cite{MayPE2013_PhD}, who identified a spectroscopic detection for the giant arc system 2, thus allowing to fix a redshift in the lens optimization. This same system's redshift was confirmed with VLT/XShooter by \cite{Jullo2026_XSHOOTER}, who also measured the redshift of another giant arc (system 1).
This cluster was included in the RELICS survey, \href{https://archive.stsci.edu/hlsps/relics/abell2813/models/lenstool/}{yielding an according, yet unpublished lens model}.
Its complex profile, at the crossing of several BCGs, magnifies two galaxies into two $17''$-long giant arcs studied by \cite{MayPE2013_PhD}, and a crescent-like MIS.
Rooting our SL model in the literature and \Oguri MIS identifications, we report two new systems, amounting to a total 18 MIS constraining two DMHs, and yielding a final \rms of $0.63''$.

\subsubsection{El Gordo (ACT-CL J0102.9-4915, ACTCL0102)} \label{sec:actcl0102}

Galaxy cluster `El Gordo' (ACT-CL J0102.9-4915, $z=0.87$) is an exceptionally massive cluster for its redshift, undergoing a major merger, with a highly elongated configuration. 
Imaged by the RELICS and PEARLS surveys, and widely covered by spectroscopic surveys, it has been largely SL-modeled already \citep{Zitrin2013_Gordo, Cerny2018, Caminha2023_ElGordo, Diego2020_Gordo, Diego2023_Gordo}. PEARLS, in particular, revealed a wealth of new MIS galaxies, precisely studied and reconstructed in their source plane, such  as `El Anzuelo' and `La Flaca' \citep{Frye2023_ElGordo, Diego2023_Gordo, Galan2024_Anzuelo}.
We keep the tessellation of these lensed systems into several knots for our lens model, amounting to a total 76 lensed systems.
The foreground galaxy group at $z=0.633$, whose lensing significance was underlined by \citet{Caminha2023_ElGordo}, was included in the main galaxy catalog, but the weight of these galaxies were left free.
With three free-centroid DMHs, we model the cluster with an image-plane precision $\rms = 0.97''$.

\subsubsection{Abell 209 (A209)} \label{sec:a209}

Abell 209 ($z=0.21$), included in the X-ray-selected CLASH sample, was classified as unrelaxed by \citet{smith2005, jimenez-Teja2018}. The only strong-lensing model of A209 to date was presented by \citet{Zitrin2015CLASH25}, who identified the first multiple images in this cluster and reported a modest effective Einstein radius of $\theta_E \simeq 9 \pm 1''$ for a source at $z_s=2$, with the images of two systems forming around a bright cluster member located $\sim18''$ from the BCG. 
We adopt the same 3 multiply-imaged systems (7 images) as in \citet{Zitrin2015CLASH25}. To our knowledge, none of the systems has a spectroscopic redshift, so we fix the redshift of system 1 to its photometric redshift $z_{\rm phot}=1.13$ and optimize the redshifts of the remaining systems. A single DMH reproduces the image positions with a \rms of $0.30''$.

\subsubsection{WHL J013719.8-082841 (W0137)}
\label{sec:w0137}

The cluster WHL J013719.8-082841 ($z=0.57$), initially observed by the RELICS survey, hosts the Sunrise arc (systems 1-4), a triply-imaged, highly magnified galaxy constraining the cluster mass \citep{Welch2022EarendelHST,Welch2022EarendelJWST}. 
This arc hosts Earendel, an exceptionally magnified source initially interpreted as an individual star. Later NIRSpec spectroscopy established a redshift of $z_{\rm spec}=5.926\pm0.013$ for the Sunrise galaxy, while the spectral properties of Earendel tend to classify it as a compact, metal-poor stellar cluster rather than an individual star \citep{Scofield2025_Earendel, Pascale2025_Earendel}.
Further photometric and spectroscopic analysis of the cluster field revealed a possible multiply-imaged, $z_{\rm spec} = 8.22$ galaxy \citep[system 10;][]{Bradley2023_w0137, Vikaeus2024}.

Supplementing the literature by the DJA NIRSpec archive \citep{brammer_2025_NIRSpec_DJA}, we compiled a total of 11 MIS (including 7 with spectral identification), two of which are newly discovered here, and one also identified by \Oguri. 
Using a centroid-free one-halo model to describe the cluster's relaxed state, our model converged to a \rms of $0.89''$.

\subsubsection{CL J0152.7-1357 (ACTCL0152)} \label{sec:actcl0152}

Cluster ACTCL0152 ($z=0.83$) was included in the RELICS and SLICE surveys, and consequently modeled by \cite{Acebron2019CL0152, Cerny2026SLICE} respectively. With only one spectroscopically confirmed redshift, the mass-sheet degeneracy is much more difficult to break than for most VENUS targets. 
The photometric wealth of VENUS observations allowed us to detect 13 new systems, thus more than doubling the total known to 24. On top of the main DMH, surrounding the main BCG and already identified in past models, we identified multiple images along the North-East -- South-West axis, thus requiring to model the cluster with a total of 3 DMHs, all with fixed centers coinciding with the main BCGs. Although constraints remain lackluster in these regions, mostly due to the lack of spectroscopy, free redshifts were constrained using the best-fit VENUS photometric redshifts, thus allowing to limit degeneracies.
System 11, i.e.\ the multiply-lensed VENUS-CLJ0152-QG1 (VCQG1) quiescent galaxy at $z_{\rm phot} = 5.18$ \citep{Pan2026_CL0152}, was here fixed to its photometric redshift, leveraging excellent photometry. Moreover, the fluxes detected for the three images of this source were used to constrain their magnification ratio.
Our \ALP optimization, already used in \cite{Pan2026_CL0152}, yields a $\rms = 0.84''$. This cluster is a clear major merger (supported by X-rays, weak lensing and intracluster light, see \citealt{Maughan2003_2cl, Jee2005_CL0152, JimenezTeja2024_ICL}), and may be Bullet-like (see Section \ref{sec:cl0658}), although the limited X-ray and spectroscopic observations do not allow to conclude.

\subsubsection{MACS J0159.8-0849 (M0159)} \label{sec:m0159}

MACS J0159.8-0849 ($z=0.41$) is dynamically relaxed and harbours a cool core \citep{Giacintucci2019}. The first published strong-lensing model was presented by \citet{Cibirka2018} as part of RELICS, who recovered a single, dominant mass concentration and a comparatively symmetric distribution, consistent with a largely unperturbed system.
In agreement with its relaxed state, a single DMH is sufficient to describe the mass distribution. 
Our model is constrained by 15 MIS with 33 images in total. These include the 4 systems of \citet{Cibirka2018}, their candidates c5 and c6, which are now confirmed by the JWST imaging, and 6 newly identified systems. Based on VENUS imaging, we also revise the identification of images 1.3 and 2.3 of \citet{Cibirka2018}, adopting new positions for these counter-images. Candidate c5 (system 2 in our model) is an arc, for which we use 4 distinct clumps to constrain the critical curve.
The model reaches a \rms of $0.27''$.

Interestingly, system 1 (system 3 in \citealt{Cibirka2018}) is a LRD candidate with a blue companion at $z_{\rm phot}=3.47$. Its two images are magnified by factors of $\sim3-5$, with a predicted time delay of $\sim60-70$ years between them.
We further report a second LRD candidate with a blue companion, which is not multiply imaged, located at ($\alpha$, $\delta$) = (29.94966, -8.822161)
(J2000) with $z_{\rm phot}\sim5$ and $\mu\sim2.5$.

\subsubsection{RXC J0232.2-4420 (SPTCL0232)} \label{sec:sptcl0232}

Galaxy cluster RXC J0232.2-4420 ($z=0.28$, also catalogued as SPT-CL J0232-4421) was imaged by RELICS and SLICE, and subsequently modeled by \cite{Mainali2019_PhD, Cerny2026SLICE} respectively, the former of which confirmed the only spec-$z$ system in this cluster. Our MUSE research did not supply more multiple images redshift; however the excellent VENUS photometry allowed to tightly constrain a few systems, thus reliably constraining the mass distribution. With the addition of 12 new systems, we tripled the number of MIS known, reaching 18. Using a DMH per BCG for a total of two, and letting the centroids free to evolve, \ALP constrained the lensing potential with a final $\rms = 0.47''$ precision, displaying a relaxed geometry. In the west, system 6 appears to be a multiply-lensed LRD, but it proved difficult to accurately reproduce its three images.

Another confirmed spec-$z$ system \citep{Mainali2019_PhD} exists in the eastern part of the cluster, but is lensed by a galaxy group in the background of the cluster, around $z \in [0.6, 0.7]$. It was therefore not included here.

\subsubsection{Abell 370 (A370)} \label{sec:a370}

Galaxy cluster Abell 370 ($z=0.38$) was the very first strong-lensing cluster to be identified, in \cite{Soucail1987_A370, Paczynski1987_Dragon, Soucail1988_A370}, with its giant, $30''$-extended Dragon arc ($z=0.7251$). Its many lensed systems, as well as its large magnified area have always made it a prime target of SL surveys, such as HFF, CANUCS and MAGNIF \citep[\href{https://www.stsci.edu/jwst-program-info/download/jwst/pdf/2883/}{GO-2883}][]{Sun2023_MAGNIF}. 
Subsequent lens models include \Lenstool \citep{Richard2010A370,Richard2014FF,Lagattuta2017_A370,Lagattuta2019_A370,Gledhill2024_A370,Limousin2025_A370}, \wslap \citep{Diego2016_A370, Diego2025_A370}, \glafic \citep{Kawamata2018_SL,Li2024_A370_Dragon_Flashlights} and \texttt{GRALE} \citep{Ghosh2021_A370}. 
This exceptional cluster was also object of considerable spectroscopic coverage \citep{Christensen2012Specs,Lagattuta2017_A370,Lagattuta2019_A370,Richard2021_MUSEatlas,Lagattuta2022_A370,Gledhill2024_A370,Sarrouh2026_CANUCS}, yielding precise SL constraints.

The high surface brightness and extreme magnification of the Dragon arc led to the discovery, comparing the CANUCS epoch with other programs such as GO-3538, of more than a hundred CCEs, i.e.\ individually lensed stars at redshift $z=0.7251$ \citep{Li2024_A370_Dragon_Flashlights, Fudamoto2025_40lensedstars, Palencia2026_CCE_Dragonarc}. 
For this reason, this cluster is regularly monitored by JWST (e.g.\ by the Dragon survey, \href{https://www.stsci.edu/jwst-program-info/download/jwst/pdf/7345/}{GO-7345} and \href{https://www.stsci.edu/jwst-program-info/download/jwst/pdf/12583/}{12583}, PI: Fudamoto), thus leading to the discovery of multiple transients, such as galaxy-galaxy strongly-lensed (GGSL) SN Typhon (Pierel et al.\ in prep.) and the exceptional cluster-lensed SN Hydra \citep[Larison et al.\ in prep.; see also the HST-era GGSL SN][]{Chen2022_lensedSN_A370}.
The Dragon arc CCE abundance has also been exploited to constrain DM models \citep{Broadhurst2025_DM_Dragon, Palencia2026_CCE_Dragonarc}.

The model here assembled inherits the same spectroscopic constraints and MIS as previously cited models. A lot of clumps from the Dragon arc, broken down into 11 MIS, were used as constraints, for a total of 93 MIS corresponding to 281 different images. Our model, consisting of 4 DMHs -- all with a free centroid, converged down to $\rms = 0.70''$, a good fit considering the difficulty to model the cluster.

\subsubsection{Abell S295 (AS295)} \label{sec:as295}

Galaxy cluster Abell S295 ($z=0.30$) was first noticed for its giant, $23''$-extended arc near the BCG, as part of the ROSAT survey \citep{Edge1994_AS295arc}. 
A later observation by the RELICS survey yielded a first lens model \citep{Cibirka2018}. Although it was part of the SLICE survey, no new model was established with JWST yet. 
Using the VENUS observations, we here report a new MIS, yielding a lens model based on a total 11 systems. 
Our model is based on two centroid-free DMHs to model the cluster's major merger geometry.
Despite the good final $\rms = 0.54''$, this target simply does not have enough lensing constraints in the center and south eastern regions, and therefore cannot be strongly constrained. Moreover, the spectroscopic constraints on 4 systems are also stemming from the same giant arc-lensed galaxy in the NE. In order to limit the degeneracy in the SW halo, two systems with excellent photometry are fixed to their photo-$z$. A lot of red dots are present across the field, but only spectroscopy can differentiate possible lensed LRDs from brown dwarves. 
In combination with a deeper study of the cluster's ICM, spectroscopic observations of AS295 may reveal if the cluster is a Bullet-like collision.

\subsubsection{Abell 383 (A383)} \label{sec:a383}

Cluster Abell 383 ($z=0.19$) is a relaxed, cool-core cluster, displaying giant arcs, the longest of which spans $\sim 20''$. It has therefore been included in CLASH and studied for its SL properties since (\citealt{Zitrin2011CLASH383, Richard2011A383highz, Zitrin2015CLASH25}, lensing features also discussed in \citealt{Geller2014_A383}), but no JWST observations has complemented it before VENUS. 
We replicate the main assumptions of these models, with one DMH whose centroid is free around the BCG, representing the cluster's relaxed geometry. A first VENUS model was presented in \cite{Golubchik2025_LRD_VENUS}, which presents the doubly-lensed LRDs and its blue companion, i.e.\ the `Red meets Blue' system (systems 5 and 15, $z=6.03$). We here update this model, using 14 MIS and yielding $\rms = 0.73''$.

\subsubsection{MACS J0257.1-2325 (M0257)} \label{sec:m0257}

MACS J0257.1-2325 ($z=0.51$) is one of the twelve MACS clusters at $z>0.5$ \citep{EbelingMacsCat2001, EbelingMacs12_2007}. Its first strong-lensing model was presented by \citet{Zitrin2011_12macsclusters}; subsequent imaging includes HST/RELICS \citep{Coe2019RELICS} and public JWST/NIRCam SLICE imaging. 
We here reproduce the model combining these data presented in \citet{Nakane2025_VENUS_highzgal}. We adopt previously identified systems whose internal knots are resolved by JWST and treated as independent constraints, together with additional systems identified in the SLICE and VENUS imaging. 
The most notable background source, identified with VENUS imaging as system 12, is the `Misty Moons', a galaxy at $z_{\rm phot}\sim11.3$. We fix its redshift in our lens optimization, alongside system 1 at $z_{\rm spec}=1.09$ (divided into three knots), confirmed by MUSE spectroscopy.
The remaining six source redshifts are optimized.
The mass model uses one cluster-scale halo near the BCG with a free centroid, and a second, substantially less massive halo located on the second BCG.
Our new \ALP model converged with $\rms = 0.30''$.

Our updated lens model predicts a magnification of $\mu\sim37$ and $\mu\sim25$ for the two brightest Misty Moons images, thus yielding a physical resolution into star-forming clumps of $\sim10$--$70\,\mathrm{pc}$ \citep[compatible with our previous \ZA model predictions, see][]{Nakane2025_VENUS_highzgal}.
It is intrinsically sub-L$^{\star}$ ($M_{\rm UV}\sim-18$), among the faintest galaxies known at $z>10$, and its clumps dominate its stellar-mass budget. Our model predicts five images: a candidate third image is detected in the VENUS imaging, while the two remaining counter-images fall below the detection limit. The predicted geometry resembles a hyperbolic umbilic configuration \citep[for a summary an exotic lensing configurations, see][]{OrbandeZivry2009_exoticlenses, Meena2023_HU}, confirming all five images would strongly constrain the radial mass profile.

\subsubsection{MACS J0257.6-2209 (A402)}

MACS J0257.6-2209 ($z=0.32$), also catalogued as Abell 402, was identified in the Abell survey and in the MAssive Cluster Survey \citep{Abell1989cat,EbelingMacsCat2001,ReppEbeling2018}. 
\citet{MannEbeling2012evolution} place this cluster among the relaxed systems on the basis of its X-ray and optical morphology, but its central entropy is high, and its member galaxies show signs of strong ram-pressure stripping, both of which argue against a fully settled system \citep{Kolokythas2025}. 
A cavity in its BCG was later discovered by the SLICE survey, caused by a supermassive black hole activity \citep{McDonald2026_A402}.
\citet{Richard2021_MUSEatlas} presented a strong-lensing model based on deep VLT/MUSE spectroscopy, constrained by 25 multiply-imaged systems and 85 multiple images, all spectroscopically confirmed and dominated by faint \Lya emitters revealed by MUSE. 
We adopt the MIS and spectroscopic redshifts of \citet{Richard2021_MUSEatlas}, re-optimizing the model with the VENUS imaging. Following their analysis, the mass distribution is described by two centroid-free cluster-scale halos, reaching a \rms of $0.71''$.

\subsubsection{MACS J0308.9+2645 (M0308)} \label{sec:m0308}

The cluster MACS J0308.9+2645 ($z=0.36$) was imaged with HST in the RELICS survey as well as with JWST (GO-5293) prior to VENUS. The cluster hosts the `Cosmic Spear', a multiply-imaged ALMA [C\,\textsc{ii}]$158$\um galaxy at $z_{\mathrm{spec}}=6.2078\pm0.0002$ \citep{Fudamoto2024_CosmicSpear} sheared into multiple clumps in the JWST imaging \citep[system 6 in our model;][]{Abdurrouf2026_CosmicSpear}, as well as the multiply-imaged SN Ares \citep[system~1, $z_{\mathrm{spec}}=1.276$;][]{Larison2026_SNAres}.

A first SL model of M0308, based on RELICS, was derived by \citet{Acebron2018_MACS0308_PLCKG171}, using the \LTM method and 3 MIS, none of which had spectroscopic redshifts. We model the cluster using 49 multiple images belonging to 21 sources, two of which have spectroscopic redshifts \citep[][Resseguier et al.\ in prep.]{Fudamoto2024_CosmicSpear,Larison2026_SNAres} which are fixed in the SL modeling. The cluster is parametrized with one smooth DMH, reaching a final lens-plane $\rms=0.46''$. In addition to the VENUS \ALP and \glafic models, a \Lenstool model using VENUS data will be published in Resseguier et al.\ (in prep).

\subsubsection{PLCK G171.9-40.7 (PLCKG171)} \label{sec:plckg171}

PLCK~G171.9-40.7 ($z=0.27$) is a rich and massive SZ-selected \textit{Planck} cluster that was included in the RELICS survey with HST, based on which a first \LTM SL model of PLCKG171 was derived by \citet{Acebron2018_MACS0308_PLCKG171} using 11 multiple images of 3 sources without spectroscopic redshifts. 
Our SL model of PLCKG171 parametrizes the cluster with one cluster-scale DMH. 
We use 47 multiple images, detected in NIRCam, of 15 sources as constraints, none of which have spectroscopic redshifts. This model achieves an average image reproduction error of $\rms=0.47''$, but we note that, in the complete absence of spectroscopic redshifts of any MIS, the SL model remains uncertain.

\subsubsection{MACS J0329.7-0211 (M0329)} \label{sec:m0329}

MACS J0329.7-0211 ($z=0.45$) was included within the CLASH survey as a dynamically relaxed cluster on the basis of its \textit{Chandra} X-ray morphology \citep{PostmanCLASH2012}, and hosts a cool core with a radio minihalo (\citealt{Giacintucci2019}, compatible with a relaxed state, confirmed by intracluster light observations; Jiménez-Teja et al., in prep). Its strong-lensing properties were first studied by \citet{Zitrin2012CLASH0329}, who identified six MIS and candidates, including a quadruply-imaged galaxy at $z\sim6$, three images of which were secure enough to constrain the mass model. 
\citet{Caminha2019} subsequently assembled nine spectroscopically confirmed MIS from MUSE spectroscopy, and \citet{Richard2021_MUSEatlas} presented a model from the same MUSE data, adding two multiply-imaged \Lya pairs as constraints.
Our model is constrained by the 24 multiple images of \citet{Richard2021_MUSEatlas} (across 9 MIS). We adopt their spectroscopic redshifts where available, and we leave the redshifts of system 3 and the newly identified system 10 free to be optimized by the model. The mass distribution is described by two smooth halos, the second halo being centered on the bright galaxy north-west of the cluster core; its presence has been taken as a possible sign of an ongoing merger, despite the relaxed classification of the cluster. 
The model reaches an \rms of $0.42''$.

\subsubsection{Abell 3192 (A3192)} \label{sec:a3192}

Cluster Abell 3192 is identified to cluster RXC J0358.8-2955 at redshift $z = 0.43$, in the background of a $z=0.17$ structure mentioned in \cite{Hamilton-Morris2012_A3192}.
\cite{Hsu2013_A3192} first investigated the complex merger geometry, and acquired more spectroscopic data, but the main lens models assembled for A3192 were established following the RELICS and SLICE observations \citep[][respectively]{Okabe2020_39SLcl, Cerny2026SLICE}. 
Compiling past models (including \citealt{Oguri2026_VENUS}), we found five additional MIS, including a dim giant arc, with a possible hyperbolic-umbilic geometry (system 25). With three DMHs free to move around the BCGs, our model, constrained by 25 MIS, converged with $\rms = 0.88''$.
The cluster's apparent merger geometry may be confirmed only through additional spectroscopic observations beyond the three MIS with confirmed spectroscopic redshifts. In particular, the `bulletticity' of the cluster, i.e.\ its Bullet-cluster-like character (a major merger with distinct, post-crossing, shocked ICM; see Section~\ref{sec:cl0658}), remains to be established.

\subsubsection{MACS J0416.1-2403 (M0416)} \label{sec:m0416}

Galaxy cluster MACS J0416.1-2403 ($z=0.40$) is a major merger, largely surveyed by CLASH, HFF, CANUCS and PEARLS, leading to numerous SL analyses \citep{Zitrin2013_M0416, Jauzac2015_M0416, Grillo2015_0416, Bonamigo2017_M0416, Kawamata2018_SL, Limousin2022, Bergamini2023_M0416, Diego2024_Spock, Rihtarsic2025_M0416, Limousin2025_M0416, Cha2026_M0416, Oguri2026_VENUS}. Additional MIS or spectroscopic detections without modeling were reported in \cite{Richard2011_clusters_spectro, Christensen2012Specs, Ebeling2014_HFFspectro, Hoag2016_M0416, Rodney2018_transient_M0416}. 
With the VENUS observations, the number of identified MIS was brought to 138 MIS, including 127 with confirmed spectroscopy, for a total 379 images.
We here follow classic M0416 modeling with 4 DMHs, one of which (DMH3) is centroid-fixed. Our reference model converges with $\rms = 0.53''$, while an alternative model, based on only 3 DMHs reached $\rms = 0.68''$.

\subsubsection{MACS J0417.5-1154 (M0417)} \label{sec:m0417}

The cluster MACS~J0417.5-1154 ($z=0.44$) was imaged with HST in the RELICS program, and later observed both in imaging and wide-field slit-less spectroscopy (WFSS) with JWST in CANUCS prior to VENUS. It contains a rare cluster-lensed AGN at $z=0.872$ with X-ray variability \citep{Pandge2019_M0417}, as well as a multiply-imaged interacting merging pair of spiral galaxies known as the `Question Mark Pair' \citep[][]{Jauzac2019MACS0417,Estrada-Carpenter2024MACS0417}, in which VENUS observations spotted SN Athena, whose next multiply-lensed image is predicted to reappear in early 2027 (Pascale et al.\ in prep.).

M0417 was modeled with \Lenstool using RELICS and MUSE data by \citet{Jauzac2019MACS0417} and \citet{Mahler2019MACS0417}. A CANUCS \Lenstool model with numerous new spectroscopic redshifts from JWST/NIRISS \citep{Sarrouh2026_CANUCS} will be published in Desprez et al.\ (in prep.). We here model M0417 for VENUS using two cluster-scale DMHs. Our model is constrained with 121 multiple images of 41 sources, 20 of which have spectroscopic redshifts \citep{Jauzac2019MACS0417,Mahler2019MACS0417,Tsujita2025MACS0417,Sarrouh2026_CANUCS}, and achieves a lens-plane average image reproduction error of $\rms=0.68''$.

\subsubsection{MACS J0429.6-0253 (M0429)} \label{sec:m0429}

The massive X-ray cluster MACS~J0429.6-0253 ($z=0.40$) was included in the CLASH survey with HST, after being modeled by \cite{Zitrin2011_12macsclusters}. As a CLASH cluster, M0429 has a long history of lensing analyses: CLASH-based SL \LTM, \piemdenfw and \glafic models include two identified MIS without spectroscopic redshifts \citep{Zitrin2015CLASH25, Umetsu2016}. 
Thanks to several MUSE pointings, \citet{Caminha2019} measured three MIS' redshifts and derived an HST and MUSE-based \Lenstool SL model of M0429, using these three MIS with 11 multiple images. We model M0429 using one DMH, and constrain our model with 21 multiple images belonging to 6 sources, 3 of which have spectroscopic redshifts \citep{Caminha2019}. Our model achieves a lens-plane image reproduction error of $\rms=0.22''$.

\subsubsection{MACS J0451.9+0006 (M0451)} \label{sec:m0451}

Cluster MACS J0451.9+0006 (M0451, $z=0.43$) was targeted by spectroscopic observations by \cite{Richard2011_clusters_spectro, Schaerer2015_spectroM0451} and the MUSE atlas \citep{Richard2021_MUSEatlas}, the latter of which assembled the only pre-existing lens model with \Lenstool, based on 25 spectroscopically-confirmed MIS. 
The spectroscopically confirmed systems assembled in these articles, and new multiple images detections by \Oguri amount to 30 lensed systems, which we supplement with 4 new MIS identifications.
The newly assembled model includes two DMHs, whose centers are let free. The final \rms reaches $0.75''$, with a relaxed geometry.

\subsubsection{Abell 521 (A521)} \label{sec:a521}

Abell 521 ($z=0.25$) is a dynamically disturbed merging cluster, hosting a spectacular multiply-imaged galaxy at $z=1.043$, first identified by \citet{Richard2010locuss20} with spectroscopy. 
Using MUSE observations, its morphology, star-forming clumps and lens model were later investigated and further refined by \cite{Patricio2018_maggals,Messa2022_A521}, with additional constraints from ALMA observations \citep{Dessauges2023_ALMA}.
On a different note, the cluster's ICM appears to be shocked, as a result of an ongoing merger \citep{Bourdin2013_A521}.

Following literature, we identify 15 knots in the lensed source galaxy, and add another system discovered in VENUS imaging. 
A thorough search for more MIS proved unfruitful, thus only constraining the region around the BCG, for a final model of image-plane $\rms = 0.11''$.
We note that the limited SL-constrained area does not witness the ongoing merger, but we attribute this to the limited Einstein radius of $\theta_{E,z_s=2} = 8.2 \pm 0.8''$, i.e.\ the smallest value across VENUS clusters.

\subsubsection{MACS J0454.1-0300 (M0454)} \label{sec:m0454}

The SL mass distribution of cluster MACS J0454.1-0300 ($z=0.55$, also identified as MACS J0451.6-0305 in the literature) was modeled by \cite{MacKenzie2014, Jauzac21_3SLcl}, with previous spectroscopic identifications including \cite{Chapman2002}.
JWST cycle-3 GO program \href{https://www.stsci.edu/jwst-program-info/download/jwst/pdf/5058/}{5058} (PI: Furtak \& Meena), reaching a depth of $29$ AB mag at $5 \sigma$, later deepened the cluster SL model using six NIRCam broadband filters, motivated by the possibility of observing lensed stars through caustic crossing events (CCEs). \citet{Meena2026_swallowtails} presents a comparable SL model to the one here presented, in order to study the rare swallowtail multiply-imaged configuration of systems 10 ($z_{\rm spec}=2.91$) and 19 ($z_{\rm spec}=6.70$).

In our \ALP model, we use the exact same 29 MIS and galaxy catalog as \citet{Meena2026_swallowtails}, and foreground galaxies lying close to lensed images are included in our galaxy catalog, while their weights are left free to vary.
To model M0454, one DMH was placed at the position of the BCG, with the center left free to vary.
The best-fit model presents a \rms of $0.40''$.

\subsubsection{MACS J0553.4-3342 (M0553)} \label{sec:m0553}

MACS J0553.4-3342 ($z=0.41$) is a massive, post-collision, bimodal cluster merger, whose lens model was first described with \Lenstool by \citet{Ebeling2017_M0553}, numbering 10 MIS using HST observations. The cluster exhibits two spectacular giant arcs (systems 1 and 3), both extending over $\sim24''$, thus providing significant constraints on its mass distribution. Concurring \textit{Chandra} X-ray observations concluded to a striking displacement between the western DMH and the ICM, thus supporting a merger scenario with complete stripping of the western halo's cool core.
After being observed with the RELICS survey, its JWST observations by the SLICE program expanded the constraints to 21 systems \citep{Cerny2026SLICE}. Building upon this legacy, with four spectroscopically confirmed systems, we use a total of 171 lensed images across 62 MIS \citep[including identifications by][]{Oguri2026_VENUS} to constrain the two-halo lens model, converging to $\rms = 0.79''$.

\subsubsection{RXC J0600.1-2007 (R0600)} \label{sec:r0600}

RXC~J0600.1-2007 ($z=0.43$, also known as MACS~J0600.1-2007), is one of the largest galaxy clusters observed to date \citep[][$\theta_{E, z_s=2} = 49.7''\pm5.0$]{Furtak2024_MACS0600}, possibly rivaling M0717 (see Section~\ref{sec:m0717}).
It was included in the RELICS survey prior to VENUS, and hosts the `Cosmic Grapes', a clumpy ALMA-detected spiral galaxy at $z=6.01$, which is quintuply-imaged \citep{Laporte2021_MACS0600,Fujimoto2021_MACS0600,Fujimoto2025_CosmicGrapes}. 
The cluster was initially modeled with \Lenstool, \glafic, and \LTM in \citet{Laporte2021_MACS0600} and \citet{Fujimoto2021_MACS0600} based on 10 MIS with spectroscopic redshifts from MUSE. 
However, models of the time had great difficulties reproducing the multiple image configurations without a significant external shear component.
Using dedicated \textit{Gemini}/GMOS observations (PID GN-2020B-Q-903, PI: Zitrin), this led to the discovery of massive, extended structures to the east of the cluster, right beyond the HST field of view, which are potentially more massive than the part of R0600 observed with RELICS \citep{Furtak2024_MACS0600}. 
Using new deep MUSE observations to derive 13 additional spectroscopic redshifts, as well as early JWST/NIRCam observations of the Cosmic Grapes by \citet{Fujimoto2025_CosmicGrapes}, \citet{Furtak2024_MACS0600} constructed a new \ZA SL model of R0600 which included the eastern structure, though these had no confirmed MIS at the time.

Now with VENUS, the strong lensing properties of these extended eastern structures of R0600 are confirmed in the NIRCam imaging, showing numerous previously unknown bright arcs in that part of the cluster (see Fig.~\ref{fig:clusters_5}). We construct a new model of R0600, building on the multiple images and spectroscopic redshifts compiled in \citet{Furtak2024_MACS0600}, adding new multiple images in the east of the cluster. Our model of R0600 has four DMHs, two each in the east and west, and is constrained with 173 multiple images of 58 sources, 23 of which have spectroscopic redshifts \citep{Furtak2024_MACS0600,Fujimoto2025_CosmicGrapes}, reaching a lens-plane $\rms=0.68''$ and $\theta_{E, z_s=2} = 46.3 \pm 4.6''$. This model of R0600 will be enhanced with new JWST/NIRSpec spectroscopic redshifts, scheduled to be observed in February~2027 (GO-6796), and published in a dedicated paper, Furtak et al.\ (in prep.).

\subsubsection{SPT-CL J0615-5746 (SPTCL0615)} \label{sec:sptcl0615}

Galaxy cluster SPT-CL J0615-5746 ($z=0.97$) is the most distant cluster of the VENUS sample. It was first imaged with the HST RELICS survey, and consequently modeled by \cite{PaternoMahler2018_SPTCL0615}. Later spectroscopy by \cite{Connor2019_SPTCL0615} suggested the cluster was in its relaxing phase, however later intra-cluster light (ICL) and ICM analyses suggest it is an ongoing merger \citep{JimenezTeja2023_SPTCL0615}.
\cite{Salmon2018} discovered the `Cosmic Gems' arc (SPT0615-JD1, $z=9.625$, systems 9-11 here), a group of five young star clusters, further studied in \citet{Salmon2020HighzRelics,Welch2023,Bradley2025_CosmicGems, Adamo2024_CosmicGems, Messa2026_SPTCL0615, Vanzella2026_CosmicGems}.

After including the new VENUS MIS discovered by \Oguri and 
recent JWST spectroscopy by \cite{Messa2026_SPTCL0615}, we supplemented the 20 existing MIS with three new systems. Using 2 DMHs with free central positions, our model converged to a \rms of $0.39''$.

\subsubsection{Bullet (CL0658)} \label{sec:cl0658}

The \textit{Bullet} cluster ($z=0.30$, also catalogued as CL J0658-5657 or 1E 0657-55.8) was first discovered as an extended X-ray source in \cite{Tucker1995_Bullet}. Later \textit{Chandra} observations revealed two distinct ICMs with a characteristic bullet shape, witnessing a prominent bow shock \citep{Markevitch2002_Bullet}. Its configuration, of on-sky collision with two distinctive collisionless components (galaxies, DM) and shocked ICMs lagging behind made it an ideal laboratory to constrain DM physics, and especially the self-interacting DM (SIDM) cross-section \citep{Markevitch2004_Bullet,Clowe2006Bullet,Bradac2006Bullet, Randall2008,Kahlhoefer2014, Robertson2017_Bullet}.
Using JWST lensing data, \citet{Cha2025_Bullet} reported a stronger limit of $\sigma/m_\chi<0.2\,\mathrm{cm^2\,g^{-1}}$, however such constraints are necessarily model-dependent \citep{Kahlhoefer2014,Robertson2017_Bullet}. 
We moreover note the recent discovery of the $z\sim 11$ `BulletArc-z11', a triply-imaged galaxy, allowing for a $\sim 10$\,pc resolution of a galaxy in the Reionization era \citep{Bradac2025_BulletArc}.
Past SL models, multiple images identification and spectroscopy here used to model this cluster include \cite{Mehlert2001_Bullet, Bradac2006Bullet, Bradac2009_Bullet, Gonzalez2010_Bullet, Paraficz2016Bullet, Richard2021_MUSEatlas, Cha2025_Bullet, Rihtarsic2026_Bullet}.

As with the other clusters already deeply surveyed in the past by JWST, we simply compiled the largest catalog of MIS, reaching 72 systems, including 52 spectroscopically confirmed. Following e.g.\ \cite{Rihtarsic2026_Bullet}, we used three halo components, centered on the main BCGs, although their positions are left free. A population of foreground galaxies, a background group at $z\sim 1$, as well as the ICM's gravitational potentials are not modeled in our \ALP optimization, constituting a limit to our model, which converges to $\rms = 0.62''$.

\subsubsection{MACS J0717.5+3745 (M0717)} \label{sec:m0717}

MACS~J0717.5+3745 ($z=0.55$) has for a long time been the most massive known SL cluster, with an Einstein radius of $\theta_{E,z_s=2}=55 \pm 3''$ \citep[e.g.][]{Zitrin2009_macs0717}. It was extensively and deeply observed with HST as part of the CLASH and HFF surveys, as well as in deep ground-based spectroscopy with \textit{Keck} \citep[e.g.][]{Limousin2012_M0717,Limousin2016MACS0717,Ebeling2014_HFFspectro}. M0717 is the last of the HFF targets to be observed with JWST, as part of the VENUS survey. 
As a HFF cluster, M0717 has been extensively studied and modeled in SL, from the first \LTM \citep{Zitrin2009_macs0717} and \Lenstool \citep{Limousin2012_M0717, Jauzac2012WL0717} models, to full CLASH and HFF models with ground-based spectroscopy \citep[e.g.][]{Medezinski2013M0717,Richard2014FF,Diego2015_0717,Zitrin2015CLASH25,Kawamata2016,Limousin2016MACS0717,Jauzac2018Beast}.

With VENUS providing the first JWST observations of M0717, we here model the cluster with six cluster-scale DMHs, as well as an external shear component. We constrain our model with 237 multiple images of 81 sources, 7 of which have spectroscopic redshifts \citep{Ebeling2014_HFFspectro,Limousin2016MACS0717} and has a lens-plane average image reproduction error of $\rms=1.04''$. We find M0717 to have an exceptionally large Einstein radius, $\theta_{E,z_s=2}=49.1''\pm4.9''$, in agreement with previous measurements, and $\theta_{E,z_s=10}=67.3''\pm6.7''$ for a source $z_s=10$, making it still the `largest' SL cluster observed to date (for $z_s\geq3$, see Fig.\,\ref{fig:figs_stats_3}).

\subsubsection{SMACS J0723.3-7327 (SM0723)} \label{sec:sm0723}

SMACS J0723.3-7327 ($z=0.39$) was discovered in the southern extension of the MAssive Cluster Survey \citep{ReppEbeling2018} and is a massive, X-ray luminous system \citep{Mahler2023}. 
The elongated mass, X-ray and galaxy distributions and the diffuse radio emission have led several studies to favor a merger scenario \citep{Mahler2023,Finner2023}, with \citet{Finner2023} placing the cluster near first pericenter passage. 
In prevision for its observation as part of the Early Release Observations (ERO), as the first strong lens observed by JWST, a first published model was presented by \citet{Golubchik2022} from HST imaging alone, followed by the RELICS-based \cite{Sharon2023}.
The first JWST imaging prompted several independent analyses \citep{Pascale2022SMACS0723,Caminha2022,Mahler2023}, which together increased the number of known multiply-imaged systems by more than a factor of four.
The comparison between these ERO and the VENUS JWST epoch revealed multiply-lensed SN Atalanta, which will be further described in Agrawal et al.\ (in prep.).

Our model is constrained by 16 MIS from \citet{Pascale2022SMACS0723}, comprising 45 multiple images, which we re-identify in the VENUS imaging, rooted in 7 spectroscopic redshifts secured through MUSE and NIRSpec observations \citep{Golubchik2022,Mahler2023, Noirot2023}.
The mass distribution requires two halos and reaches a \rms of $0.34''$.

\subsubsection{RXC J0911.1+1746 (R0911)} \label{sec:rxc0911}

The cluster RXC~J0911.1+1746 ($z=0.51$), also known as MACS~J0911.1+1746, was discovered as an SL cluster by \citet{Zitrin2011_12macsclusters}, who modeled it with \LTM using 3 MIS without any spectroscopic redshifts. R0911 was later included in the RELICS HST survey, but no SL model of this cluster has been published since. Now for VENUS, we model R0911 with \ALP, parameterizing it with one DMH, and using 13 multiple images of 6 sources as constraints. Our model has a final lens-plane $\rms=0.22''$, but we note that since there are no spectroscopic redshifts of multiple images in this field, the model remains uncertain.

\subsubsection{MACS~J0940.9+0744 (M0940)} \label{sec:m0940}

MACS~J0940.9+0744 ($z=0.34$) is another relatively small cluster, which was not part of any large-scale space-based surveys prior to VENUS, but was included in the MUSE Atlas by \citet{Richard2021_MUSEatlas}. M0940 is host of a multiply-imaged \Lya arc at $z=4.03$ \citep{Claeyssens2019LAEs}, which appears as a spectacular string of small star-forming clumps in the VENUS JWST imaging.
A first SL model of M0940 was derived by \citet{Leethochawalit2016_MACS0940}, and subsequent \Lenstool models of the cluster achieved $\rms=0.23''$ based on 7 MIS \citep{Claeyssens2019LAEs,Richard2021_MUSEatlas}. We model the cluster using the VENUS JWST imaging with one DMH. Our model is constrained with 30 multiple images of 10 sources, 8 of which have spectroscopic redshifts, and achieves a final lens-plane average image reproduction error of $\rms=0.27''$.

\subsubsection{RXC J0949.8+1707 (M0949)} \label{sec:m0949}

Cluster RXC J0949.8+1707 ($z=0.38$) has not yet been observed with VENUS. We will describe its lens model in a future update.

\subsubsection{MACS J1115.8+0129 (M1115)} \label{sec:m1115}

The cluster MACS~J1115.8+0129 ($z=0.35$) was observed with HST in the framework of the CLASH program. It is mostly known for a prominent galaxy-galaxy lensing system by two cluster members \citep{Parry2016_MACS1115}. As a CLASH cluster, M1115 was modeled with \LTM and \piemdenfw by \citet{Zitrin2015CLASH25} and their results were used by \citet{Chiu2018_MACS1115} to derive the triaxial shape if the cluster. As for other CLASH clusters, e.g.\ M0429 or M1311 (see Sections~\ref{sec:m0429} and~\ref{sec:m1311}), \citet{Caminha2019} obtained spectroscopic redshifts in M1115 with MUSE and used 3 MIS to constrain a \Lenstool SL model of the cluster. With the VENUS data, we now construct a new model of M1115, parameterizing the cluster with one DMH, and using 22 images of 7 sources as constraints, one of which has a spectroscopic redshift \citep{Caminha2019}. Our model has a final lens-plane average image reproduction error of $\rms=0.85''$.

\subsubsection{PLCK G165+67.0 (PLCKG165)} \label{sec:plckg165}

PLCK G165.7+67.0 ($z=0.35$) is a massive SL cluster, whose bimodal mass distribution and prominent $\sim20''$-long giant arcs were first revealed through HST imaging (\citealt{Frye2019}, subsequently modeled by \citealt{Pascale2022_PLCKG165} using new spectroscopic observations). 
It was included in the JWST PEARLS survey, which discovered a triply-lensed SN type Ia, SN H0pe in 2023 \citep{Frye2024SN_H0pe}.
SN H0pe's redshift was then spectroscopically confirmed to be $z=1.78$ \citep{Polletta2023_SNH0pe}.
Combined with a follow-up DDT proposal to measure the SN time delays, PEARLS data enabled to perform the first $H_0$ measurement from a multiply-imaged SN Ia, by comparing multiple independent lens models. As a standard-candle, knowledge of its magnification further assisted breaking lens model degeneracies \citep{Pascale2025_H0pe}.

Combining these HST and JWST detections with the earlier infrared and spectroscopic detections by \citet{Canameras2015, Canameras2018b}, we assembled a catalog of 41 MIS, 25 of which are spectroscopically confirmed. Our model left the two DMHs' centroids free to vary, and converged with a $\rms = 0.11''$.

\subsubsection{MACS J1149.5+2223 (M1149)} \label{sec:m1149}

MACS J1149.5+2223 ($z=0.54$) is a massive, dynamically complex merging cluster. It is one of the most extensively studied gravitational lenses, owing to its multiply-imaged systems wealth, extensive spectroscopic coverage, multiply-lensed SN and exceptional magnification of background sources. It was therefore the subject of numerous strong-lensing surveys (CLASH, HFF, CANUCS) and mass reconstructions \citep[][]{Zitrin2009_macs11495,Smith2009M1149,Zitrin2011_12macsclusters,Richard2014FF,Rau2014M1149,Johnson2014_HFF,Oguri2015_Refsdal,Sharon2015_M1149,Diego2016M1149,Jauzac2016_M1149,Kawamata2016,Finney2018_M1149,Zitrin2021_M1149,Schuldt2024_M1149,Schuldt2026_M1149}.
Its exceptional lensing configuration includes the first observed multiply-imaged supernova, SN Refsdal \citep[$z=1.4888$,][]{Refsdal1964MNRAS,Kelly2015Sci}, whose five images provide valuable constraints on image parity, relative magnification, and time delays \citep{Rodney2016_SN_Refsdal,Kelly2023_Refsdal}.
Extensive spectroscopic observations established the redshifts of numerous MIS \citep[][Rihtar\v{s}i{\v c} et al., in prep.]{Ebeling2014_HFFspectro,Treu2016Refsdal,Grillo2016_Refsdal, Sarrouh2026_CANUCS}, while the SN Refsdal system has also enabled independent cosmographic investigations \citep{Kelly2023_Refsdal_H0, Grillo2024_Refsdal}. 

Using this abundant literature, we compile a total of 269 individual multiple images, distributed across 93 MIS, 82 of which were spectroscopically identified. In particular, we here make use of the CANUCS catalogs of MIS and spectroscopic identifications (private communication), which will be detailed alongside a new \Lenstool model in Rihtar\v{s}i{\v c} et al.\ (in prep.). 
Our model uses the flux ratios reported in \cite{Kelly2023_Refsdal} as additional constraints. 
With three halos, all with free centroids, our best-fit model converges to $\rms = 0.86''$.

\subsubsection{PLCK G287.0+32.9 (PLCKG287)} \label{sec:plckg287}

PLCK~G287.0+32.9 ($z=0.39$, also catalogued as SPT-CL J1150-2805) is the second largest SZ-selected galaxy cluster from the \textit{Planck} catalog \citep{Planck2011ClustersV1}, and one of the most massive SL clusters observed to date \citep{Zitrin2017_PLCKG287}. PLCKG287 was observed with HST in the RELICS program, and with JWST in SLICE and VENUS. The first SL model of PLCKG287 was derived by \citet{Zitrin2017_PLCKG287} using the \LTM method, finding it to have an exceptionally large Einstein radius of $\theta_{E, z_s=2}=42 \pm 4''$ for a source at $z_s=2$. The cluster was then observed with three MUSE pointings, which enabled \citet{DAddona2024_PLCKG287} to measure numerous spectroscopic redshifts in the field and derive a \Lenstool model based on 47 multiple images of 12 sources, which confirmed the large Einstein radius measured by \citet{Zitrin2017_PLCKG287}.
\citet{Betts2026_SPTCL1150} recently published a model using the SLICE observations and \textit{Chandra} X-ray data to model the cluster together with its ICM, finding $\theta_{E, z_s=2} = 42.41_{-0.04}^{+0.05}\,''$.

With the VENUS data, we construct a new \ALP model of PLCKG287. Our model comprises 4 DMHs, and is constrained with 101 multiple images of 33 sources, 14 of which have spectroscopic redshifts \citep{DAddona2024_PLCKG287}. Our model achieves a final lens-plane average image reproduction error $\rms=0.75''$, and we find PLCKG287 to have an Einstein radius of $\theta_{E, z_s=2}=45.4''\pm4.5''$ at $z_s=2$ and $\theta_{E, z_s=10} = 60.5\pm6.1''$ at $z_s=10$, making it one of the largest SL clusters ever observed, alongside e.g.\ M0717 (see Section~\ref{sec:m0717}).

\subsubsection{MACS J1206.2-0847 (M1206)} \label{sec:m1206}

Galaxy cluster MACS J1206.2-0847 ($z=0.44$) was included in the CLASH and SLICE surveys, and modeled extensively in e.g.\ \cite{Zitrin2012CLASH1206, Caminha2017_M1206, Bergamini2019_3cl, Richard2021_MUSEatlas, Limousin2022}, which also present spectroscopic observations.
It hosts the famous Snake arc, spanning $15''$ across the cluster western field, located at $z=1.037$ \citep[see][system 2 here]{Ebeling2009}. Due to its large magnified area, it is very CCE-sensitive, thus yielding an impressive star detection, M1206-LS1, by comparing VENUS and SLICE imaging. This detection, more brilliant by at least 1 mag than the A370 Dragon arc detections (see Section \ref{sec:a370}), was followed up with JWST \href{https://www.stsci.edu/jwst-program-info/download/jwst/pdf/12775/}{DDT-12775} program for low-resolution spectroscopy, which will be presented in a forthcoming paper from the collaboration.

We compiled a catalog of 37 pre-existing MIS, including one with a hyperbolic-umbilic geometry.
Based on our lens model, we suspect that the field west to the Snake arc ($10-30''$) is hosting more MIS, which would benefit from MUSE coverage. As these could not be confirmed, they were not used here to constrain the model.
Our final model uses 44 MIS to constrain the lens potential, with 3 position-free DMHs, yielding $\rms = 0.48''$.

\subsubsection{MACS J1311.0-0310 (M1311)} \label{sec:m1311}

MACS~J1311.0-0310 ($z=0.49$) is a CLASH cluster that was previously lens-modeled with \LTM and \piemdenfw in \citet{Zitrin2015CLASH25}, based on two photometric MIS. A subsequent \Lenstool model was derived by \citet{Caminha2019} using new MUSE data, which however only yielded one source redshift in M1311. Our VENUS model of the cluster is similarly built, with one DMH, and constrained with 13 multiple images of 4 sources, 1 of which has a spectroscopic redshift \citep{Caminha2019}, reaching a final lens-plane $\rms=0.23''$.

\subsubsection{Abell 1689 (A1689)} \label{sec:a1689}

Cluster Abell 1689 ($z=0.18$) is a classic strong-lensing cluster, with a very large Einstein radius and an extensive set of multiply-imaged sources identified in deep HST imaging \citep{Broadhurst2005_A1689a,Limousin2007_A1689,Coe2010_A1689}.
Later free-form reconstructions extended the constraints \citep{Diego2014A1689, Alavi2016_A1689, Ghosh2023_A1689}, while MUSE spectroscopy confirmed multiply-imaged sources' redshifts \citep{Bina2016}.
We here note that the VENUS epoch led to the detection of CCEs (Palencia et al.\ in prep.).
Our literature review yielded a first catalog of MIS, which had to be cleaned of spurious spectral detections, thus yielding 50 MIS, then supplemented by our addition of 21 new systems, for a total of 213 images. 
In the present \ALP model, we characterized the relaxed cluster mass distribution through 2 DMH, both with a free center, yielding a $\rms = 0.74''$ precision, and an Einstein radius $\theta_{E, z_s=2} = 50.4 \pm 5.0''$ for a source at $z_s = 2$.

\subsubsection{RX J1347.5-1145 (R1347)}

Cluster RX J1347.5-1145 ($z=0.45$) has not yet been observed with VENUS. We will describe its lens model in a future update.

\subsubsection{Abell 1835 (A1835)}

Cluster Abell 1835 ($z=0.25$) has not yet been observed with VENUS. We will describe its lens model in a future update.

\subsubsection{MACS J1423.8+2404 (M1423)}

MACS~J1423.8+2404 ($z=0.55$) is a relaxed cluster that was observed with HST in CLASH and JWST in CANUCS. It is known for the `Firefly Sparkle', a strongly sheared $z=8.3$ arc composed of multiple star clusters \citep{Mowla2024_Firefly}, as well as a very highly magnified LRD with a resolved host galaxy, also at $z=8.3$ \citep{Judez2024_BestFriend}. First HST-based SL models of this cluster were derived with \Lenstool \citep{Limousin2010M1423}, and \LTM \citep{Zitrin2011_12macsclusters}, using three MIS. M1423 was also included in CLASH-wide sample SL model papers with \LTM and \piemdenfw \citep{Zitrin2015CLASH25} and \glafic \citep{Umetsu2014CLASH_WL}. The most complete, HST-based, SL model of M1423 published to date was constructed with \Lenstool by \citet{Patel_2024_MACS1423} with new deep MUSE observations from the KALEIDOSCOPE survey (PID 0104.A-0801; PI: A. Edge). A CANUCS \Lenstool model of M1423 will be published in Desprez et al.\ (in prep.).

For VENUS, we model M1423 using two DMHs, and constraining the model with 32 multiple images of 9 sources, 7 of which have spectroscopic redshifts \citep{Limousin2010M1423,Patel_2024_MACS1423,Sarrouh2026_CANUCS}. Our model achieves a final lens-plane average image reproduction error of $\rms=0.41''$.

\subsubsection{RXC J1532.9+3021 (R1532)} \label{sec:r1532}

RXC J1532.9+3021 ($z=0.35$) is a massive, X-ray luminous cluster with an exceptionally cool core, in which AGN feedback supplies enough mechanical power to balance radiative cooling \citep{HlavacekLarrondo2013}. 
The cluster HST field also hosts SN CLN12Did, a type Ia supernova at $z=0.85$ discovered by the CLASH survey, whose magnification provided an early consistency test of the cluster mass maps \citep{Patel2014}.
As a lens, however, R1532 had remained poorly constrained: \citet{Zitrin2015CLASH25} identified only one candidate system and presented a correspondingly preliminary mass model, and the cluster was subsequently excluded from the weak-lensing component of the CLASH lensing analyses \citep{Umetsu2016}. 

The VENUS imaging transforms this picture. We identify 5 new MIS, which together with the candidate system of \citet[][system 1 in our model]{Zitrin2015CLASH25} bring the total to 6 systems and 16 multiple images, providing the first secure strong-lensing constraints for this cluster. Systems 2--5 are also presented in \Oguri, and system 6 was identified independently in both analyses. None of the systems has a spectroscopic redshift; we therefore adopt photometric redshifts from the VENUS catalog as priors and leave all source redshifts free in the optimization. The mass distribution is described by a single DMH, and the model reaches a \rms of $0.26''$.

\subsubsection{PLCK G004.5-19.5 (PLCKG004)} \label{sec:plckg004}

PLCK G004.5-19.5 ($z=0.54$) is a massive, SZ-selected \textit{Planck} cluster \citep{Sifon2014}. The cluster is dynamically disturbed and hosts both a peripheral radio relic and a central radio halo, consistent with an ongoing merger \citep{Sifon2014,Albert2017}. Its strong-lensing potential was first explored by \citet{Sifon2014}, whose model was constrained by a single giant arc at $z=1.601$ and yielded an effective Einstein radius of $\theta_{E} = 30_{-4}^{+1}\,''$ for $z_s=1.6$. The cluster was subsequently imaged with HST as part of RELICS \citep{Coe2019RELICS}.
We further make use of archival VLT/MUSE observations of the cluster core, from which we measure new spectroscopic redshifts for three systems identified in the VENUS imaging: system 1 at $z=1.097$, system 3 at $z=1.484$ and system 4 at $z=1.471$. A fourth system (system 2) corresponds to the arc reported by \citet{Sifon2014}, for which we adopt their spectroscopic redshift of $z=1.601$. 
Our \ALP model is constrained by 9 MIS, two of which are distinct clumps within the same lensed galaxy, comprising 23 multiple images in total. In line with the disturbed state of the cluster, two DMHs are required to describe the mass distribution. The model reaches a \rms of $0.74''$.

Two of the constrained systems host LRDs. The first (system 9), a pair of LRDs embedded in a lensed star-forming galaxy dubbed `Red Eyes', is magnified by $\mu\sim15-20$ and separated by only $\sim70$\,pc in the source plane \citep[in line with our previous \ZA model, see][]{Yanagisawa2026_LRDs}. 
The second (system 10) is a further LRD system, for which the model predicts a fifth image that falls outside the NIRCam field of view and therefore cannot be confirmed. Spectroscopic redshifts of $z=7.49$ and $z=5.60$ have since been measured for these two systems, however these postdate the model presented here and were not used as constraints.

\subsubsection{MACS J1931.8-2635 (M1931)} \label{sec:m1931}

MACS J1931.8-2635 ($z=0.35$) hosts both a cool core with a strongly disturbed ICM and one of the most active AGN known in the BCG \citep{Ehlert2011MACS1931,Santos2016_M1931_Herschel}. This relaxed CLASH cluster was modeled in \cite{Zitrin2015CLASH25, Caminha2019} using HST and MUSE detections, and subsequently in \cite{Allingham2026_M1931} with VENUS observations, after the discovery of SN Eos. This multiply-lensed SN IIP motivated NIRSpec/PRISM (CLEAR) spectroscopic observations (PID: 9493-DDT, PI: Coulter) for both observed images (system 101), amounting to a total 10,504 s of exposure each. 
The analysis of this SN (system 101) is detailed in \cite{Coulter2026_SN_Eos}; here we only make use of the NIRSpec-confirmed spectroscopic redshift of SN Eos at $z=5.133$, in agreement with the four images detected in MUSE for its host galaxy (system 100), an ultra-faint ($M_{\rm UV}= -14.4 \pm 0.3$), metal-poor ($Z < 0.01 Z_{\odot}$) source presented in \citet{Asada2026_GalEos}. 
\cite{Allingham2026_M1931} moreover report the $z_{\rm phot}\simeq 3.5 \pm 0.2$ triply-lensed galaxy Charybdis (system 14), the most distant multiply-lensed grand-design spiral galaxy observed to date.

We here slightly update the \cite{Allingham2026_M1931} model, with one centroid-free DMH and a BCG very free to vary in weight, reaching $\rms = 0.65''$.
This allows to derive systematic errors on the Eos system magnification, our updated model yielding $\mu_{100} \simeq 15-16$ on both galaxy images and $\mu_{101.1} \simeq 21$ and $\mu_{101.2} \simeq 27$ on the observed SN Eos images. Comparing these figures with our past model enables to derive systematic errors (dominant here) of $20\%$ on the magnification of these images, fully consistent with \cite{Allingham2026_M1931}.

\subsubsection{SMACS J2031.8-4036 (SM2031)} \label{sec:sm2031}

SMACS~J2031.8-4036 ($z=0.34$) was previously imaged in various programs with HST (GO-12166 and GO-12884), as well as with JWST in SLICE. It hosts a famous multiply-imaged extreme N\,\textsc{iv}]$\lambda1486$\AA\ and \Lya emitter at $z=3.5$ \citep{Christensen2012Specs,Christensen2012LAEs,Patricio2016LAE,Claeyssens2019LAEs}, which was recently re-visited by \citet{Marques-Chaves2024SMS} as a candidate for forming stellar clusters hosting supermassive stars.

The cluster has been extensively modeled with \Lenstool, first by \citet{Christensen2012Specs} using one MIS, whose model was later improved with 13 spectroscopically confirmed MIS \citep{Richard2015SMACS2031,Richard2021_MUSEatlas}. The most recent \Lenstool model of SM2031 is based on SLICE JWST data and uses 15 MIS as constraints, including for the first time two MIS in the western extension of the cluster \citep{Cerny2026SLICE}. We model SM2031 based on VENUS data with \ALP using two cluster-scale DMHs and constrain it with 68 multiple images belonging to 20 sources, 17 of which have spectroscopic redshifts. Our model achieves a lens-plane $\rms=0.32''$.

\subsubsection{MACS J2129.4-0741 (M2129)} \label{sec:m2129}

Cluster MACS J2129.4-0741 ($z=0.59$) was first modeled in \cite{Zitrin2011_12macsclusters}, before its observations by the CLASH and SLICE surveys, thus leading to their lensing mass  models in \cite{Zitrin2015CLASH25, Monna2017_M2129, Caminha2019, Jauzac21_3SLcl, Cerny2026SLICE}, while additional spec-$z$ identification include 
\cite{Christensen2012Specs, Treu15, Huang2016_M2129_LAE}. 
Building on previous models \citep[including][]{Oguri2026_VENUS}, we constrain the cluster's potential using 21 MIS, 13 of which are spectroscopically confirmed.
Contrarily to previous models, we introduce a second DMH in the south where several bright galaxies are located, on top of the BCG-centered one (in line with the ICL, displaying merger-typical excess fraction; Jiménez-Teja et al., in prep.). Although this region is not constrained by SL, this inclusion, with DMH centroids let free, decreases the overall $\rms$ from $0.99''$ (1 DMH model) to $0.83''$.

\subsubsection{RX J2129.7+0005 (R2129)} \label{sec:r2129}

Galaxy cluster RX J2129.7+0005 ($z=0.23$) was first modeled by \citet{Richard2010locuss20}, before being imaged by the CLASH survey, leading to new lens models \citep{Zitrin2015CLASH25, Caminha2019}, supported by spectroscopic identifications (\citealt{Belli2013Clusters}, and the $z=9.4$ system 8 by \citealt{Williams2022z9p5}).
In 2022, the HST Snapshot (HST-SNAP) program GO-16729 (Cycle 29, PI Kelly) discovered the multiply-imaged lensed supernova type Ia SN 2022riv behind cluster R2129 ($z=1.5220$, triply-imaged). 
\cite{Dhanasingham2026_SN} summarizes multiple different lens models, with 14 MIS, and presents this discovery in depth. We here adopt the same MIS, with one DMH to model the relaxed geometry, thus reaching $\rms = 0.18''$.

\subsubsection{SMACS J2131.1-4019 (SM2131)} \label{sec:sm2131}

Cluster SMACS J2131.1-4019 ($z=0.44$) was initially observed with HST snapshot imaging, motivating a first mass reconstruction \citep{ReppEbeling2018}. \citet{Richard2021_MUSEatlas} leveraged a new combination of HST space-imaging with MUSE spectroscopy to present a new detailed \Lenstool mass model, with an Einstein radius $\theta_{E, z_s=4} = 24.4''$ for a source at $z_s=4$.  With complementary MUSE spectroscopy by \citet{Werle2022}, used here for cluster member identification, we assembled a new mass model with a new system, thus amounting to a total of 9 MIS, 8 of which have a spectroscopic observation. 
We here removed a system identified by \citet{Richard2021_MUSEatlas}, apparently incompatible with other systems, and lacking an optical or IR identification. The model includes two centroid-free DMHs, and a weighed dPIE potential to represent the foreground galaxy appearing near the cluster center, thus yielding a best-fit \rms of $0.55''$. We find $\theta_{E,z_s=4} = 25.8 \pm 2.6''$, in good agreement with \citet{Richard2021_MUSEatlas}.

\subsubsection{MACS J2137.3–2353 (M2137)} \label{sec:m2137}

MACS J2137.3–2353 ($z=0.31$) is a dynamically relaxed CLASH cluster \citep{SchmidtAllen2007} and a well-studied lens, whose core hosts a prominent radial arc. Its mass distribution has been modeled extensively \citep[e.g.,][]{Gavazzi2003_M2137,Newman2013,Zitrin2015CLASH25}, most recently by \citet{Cerny2025}, who obtained spectroscopic redshifts for all of the 4 MIS in their model, including one that had previously been identified from photometry alone. 
We adopt the systems of \citet{Cerny2025}, together with their measured MUSE spectroscopic redshifts, and complement them with 4 new systems identified in the VENUS imaging. The resulting 22 multiple images are well reproduced by a single DM halo, with a \rms of $0.40''$.

\subsubsection{RXC J2211.7-0349 (R2211)} \label{sec:r2211}

SL cluster RXC J2211.7-0349 ($z=0.40$) was observed as part of the RELICS and SLICE surveys and consequently modeled by \cite{Cerny2018, Cerny2026SLICE}. 
At the occasion of the discovery, with VENUS, of two multiply-lensed LRD candidates, forming two Einstein crosses \citep{Zhang2025LRD} with up to $130$ years predicted time delays, various models (\glafic, \Lenstool, \texttt{WSLAP+}) were assembled, including the \ALP model here presented.
With 2 DMHs, both centered on the BCG, with free centroids, the model converged to a relaxed geometry with a precision $\rms = 0.85''$. We refer to \cite{Zhang2025LRD} for more details.

\subsubsection{MACS J2214.9-1359 (M2214)} \label{sec:m2214}

Cluster MACS J2214.9-1359 ($z = 0.50$) was first modeled in \cite{Zitrin2011_12macsclusters}, vastly completed by the MUSE atlas with 15 systems with spectroscopic redshifts \citep{Richard2021_MUSEatlas}, and 3 more systems in \Oguri. 
We here compile a total of 23 MIS, for 67 multiple images (including 18 systems with a spectroscopically confirmed redshift). This seemingly dynamically relaxed cluster was modeled with three DMHs with free centroids, reducing the typical \rms to $0.62''$.

\subsubsection{Abell S1063 (AS1063)} \label{sec:as1063}

Abell~S1063 ($z=0.35$), also known as RXC~J2248.7-4431 and SPT-CL~J2248-4431, is a famous massive X-ray cluster \citep[e.g.][]{Gomez2012X-rayAS1063} which was extensively imaged with HST in the CLASH, HFF and Flashlights surveys, and more recently with JWST surveys GO-1840 and GLIMPSE. As a result, AS1063 is the deepest field observed to date, achieving 31\,magnitude depths in the NIRCam broad bands and HST ultra long-pass bands. 
It hosts a famous quintuply-imaged object at $z=6.111$ \citep{Monna2014RXC2248}, which was recently re-visited with JWST for its N-enhancement and Wolf-Rayet star signatures \citep{Topping2024N-enhancement,Berg2026WR-stars}, as well as numerous lensed \Lya emitters \citep[e.g.][]{Caminha2016LAE,Karman2017AS1063LAEs,Bergamini2019_3cl}, a particularly luminous low-redshift LRD \citep[$z=3.501$,][]{Kokorev2026LRD}, and several extremely lensed stars at $z=1-4$ \citep[][]{Diego2026Hedorah,Furtak2026GLIMPSE}.

As a CLASH and HFF cluster, AS1063 has a long history of SL mass modeling with numerous models \citep[e.g.][]{Monna2014RXC2248,Johnson2014_HFF,Richard2014FF,Zitrin2015CLASH25,Karman2015AS1063}, which were later complemented with deep ground-based spectra, confirming dozens of MIS \citep[e.g.][]{Caminha2016AS1063,Diego2016AS1063,Karman2017AS1063LAEs,Bergamini2019_3cl,Beauchesne2024AS1063}. JWST-based models of AS1063 to date include a free-from model by \citet{Diego2026Hedorah}, and the GLIMPSE \ALP model by \citet{Furtak2026GLIMPSE}. Since the VENUS data of AS1063 are much shallower than the previously obtained GLIMPSE observations, we do not derive a dedicated VENUS SL model for this cluster but rather here use the GLIMPSE \ALP model, which comprises two DMHs and is constrained with 91 multiple images of 34 MIS, 31 of which have spectroscopic redshifts from MUSE and JWST/NIRSpec \citep{Furtak2026GLIMPSE}, achieving a lens-plane $\rms=0.52''$. This model was published in \citet{Furtak2026GLIMPSE} and we refer the reader to that work for more details on this model.

\subsubsection{Abell 2537 (A2537)}

Abell~2537 ($z=0.30$) is a relaxed, X-ray luminous cluster which was included in the RELICS and SLICE surveys. First lensing analyses by \citet{Newman2013} combined HST, Keck, and X-ray observations. A RELICS-based \Lenstool model identified 4 MIS in A2537, one of which was spectroscopically confirmed \citep{Cerny2018}, and later improved with SLICE by adding seven newly identified MIS detected with JWST \citep{Cerny2026SLICE}. For VENUS, we model A2537  using one DMH and constraining our model with 57 multiple images of 15 sources, one of which has a spectroscopic redshift. Our model has a lens-plane $\rms=0.65''$.

\subsubsection{SPT-CL J2325-4111 (AS1121)} \label{sec:as1121}

Cluster SPT-CL J2325-4111 ($z = 0.36$, also known as Abell S1121) was first studied and SL-modeled by \cite{Mahler2025_twobigSLcl} on the base of HST (GO-15937) observations, detecting 6 systems with a combination of GMOS and FIRE spectroscopy and including earlier detections from \cite{Bayliss2016_GMOS}. Two of these MIS, systems 1 and 2, are knots of a $z=1.579$, $18''$-long giant arc. 
Numerous additional knots are distinguishable in this arc, and monitoring it may allow to discover CCEs, such as in the Dragon arc of Abell 370 (see Section \ref{sec:a370}). 
Our investigation added 24 new systems, none of which has spectroscopic detections, provided the lack of IFU spectroscopy on this cluster field. Among these, two systems (13 and 33-34) appear to be multiply-lensed LRDs, the second with a blue companion.
We note that a $z_{\rm phot} = 0.18$ foreground spiral galaxy may add lensing deflection to the cluster field. Moreover, the cluster appears to extend out of the main NIRCam module, with at least another BCG in the west (observed with HST/ACS, redshift not confirmed), with possible multiply-lensed images too.
This formidable cluster appears to be mostly of relaxed geometry, and although models with one DMH centered yielded a good $0.77''$ \rms, two centroid-free DMHs were retained in the final \ALP model, yielding a $\rms = 0.65''$.

\begin{figure*}
    \centering
    \begin{minipage}[t]{0.48\textwidth}
    \includegraphics[width=\linewidth]{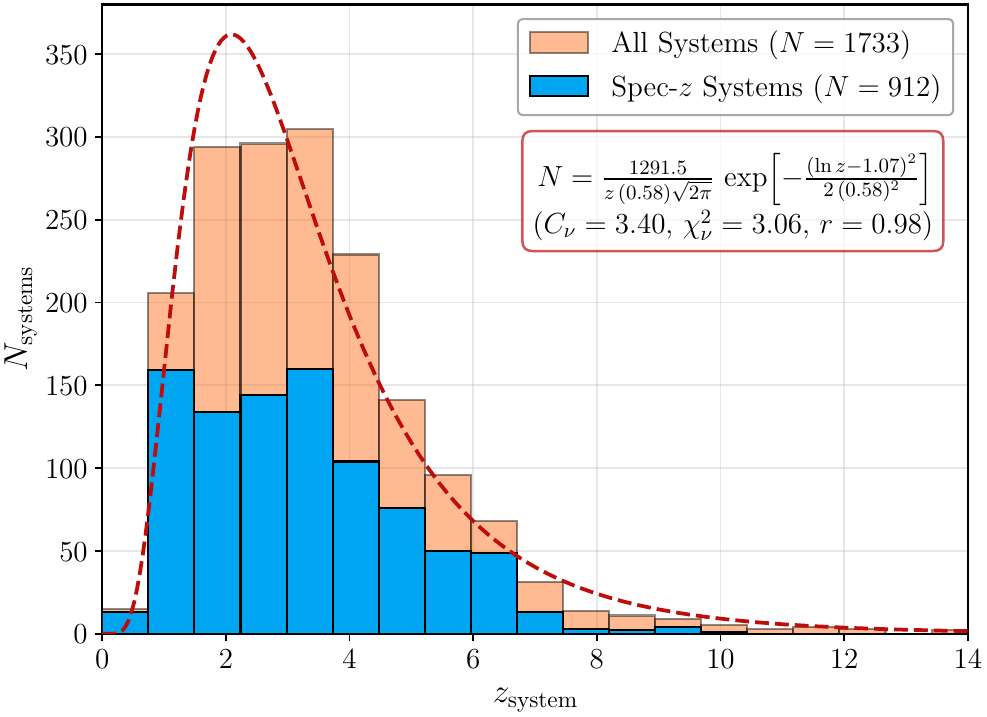}
    \end{minipage}
    \hfill
    \begin{minipage}[t]{0.48\textwidth}
    \includegraphics[width=\linewidth]{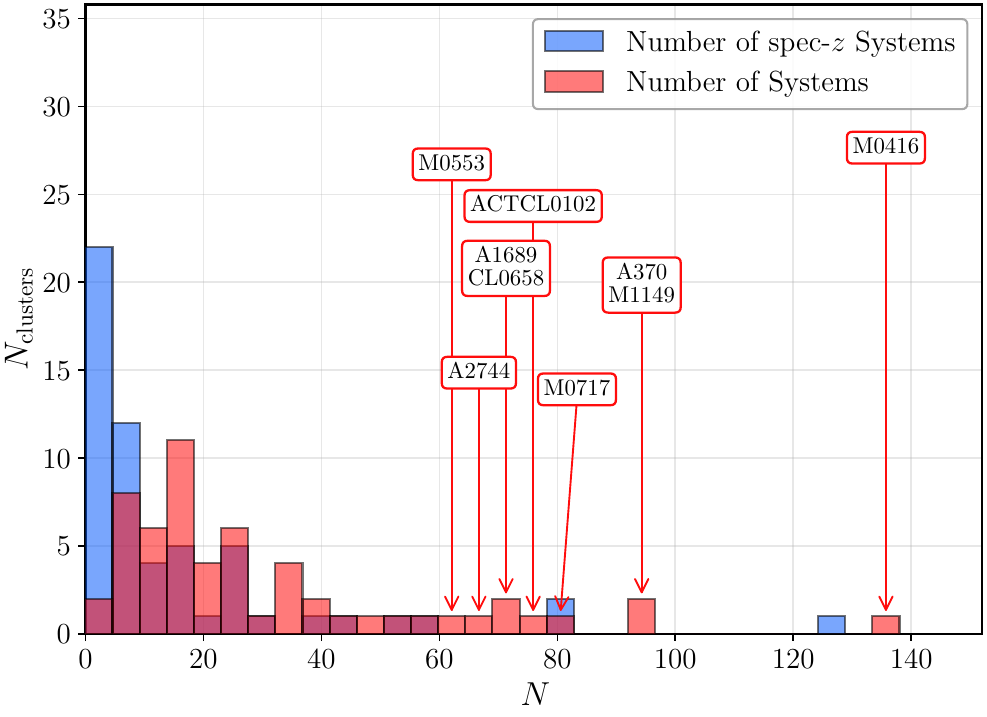}
    \end{minipage}
    \caption{\emph{Left}: Redshift distribution of the multiply-imaged galaxies used in this work, for both spectroscopic redshift measurements (blue) and for all systems (orange). The latter distribution is fit with a log-normal distribution, with the best-fit formula indicated on the figure. \emph{Right}: Histogram of the total number of MIS, and number of spectroscopically confirmed MIS, per cluster. We specify by name all clusters which show more than 60 known MIS. The most famous HFF (except M0717), UNCOVER, CANUCS and GLIMPSE targets, as well as El Gordo and the Bullet cluster, have a significantly higher number of secured spec-$z$ MIS in comparison with the rest of the sample.}    
    \label{fig:figs_stats_1}
\end{figure*}

\section{Discussion} \label{sec:discussion}

\subsection{Statistical inference from the cumulated results} \label{sec:stats_VENUS_SL}

\subsubsection{Uncertainties}

Besides supplying lens models for each cluster to facilitate studies of the individual clusters and their magnified backgrounds, one key motivation for compiling this lensing catalog is to enable statistical studies over the VENUS cluster sample as a whole.
Such analyses require to gauge uncertainties associated with the derived lensing observables (Table \ref{tab:all_clusters_3}), which is essential for understanding the significance of cluster-to-cluster variations and the scatter in the scaling relations probed here. 
Since the systematic uncertainties, stemming from modelers' and software's biases, dominate over the statistical ones by one order of magnitude, we take the total uncertainty on each observable to be the systematic component alone, conservatively assumed to be $10\%$ on the Einstein radii ($\theta_E$), $15\%$ on the Einstein area enclosed mass ($M_{\rm 2D}(\in \mathcal{A})$), $20\%$ on the magnified area above a threshold ($\mathcal{A}(|\mu|\geq \mu_T)$) and $30\%$ on the integrated cylindrical mass ($M_{\rm 2D} (<200$\,kpc)), following previous studies \citep{Zitrin2015CLASH25}.
The companion paper, \Oguri, performs similar modeling of these clusters with the independent \glafic code. A thorough comparison of both works' results will allow this cross-pipeline systematic to be quantified, which is deferred to future work.
\begin{figure}
    \centering
\includegraphics[width=\columnwidth]{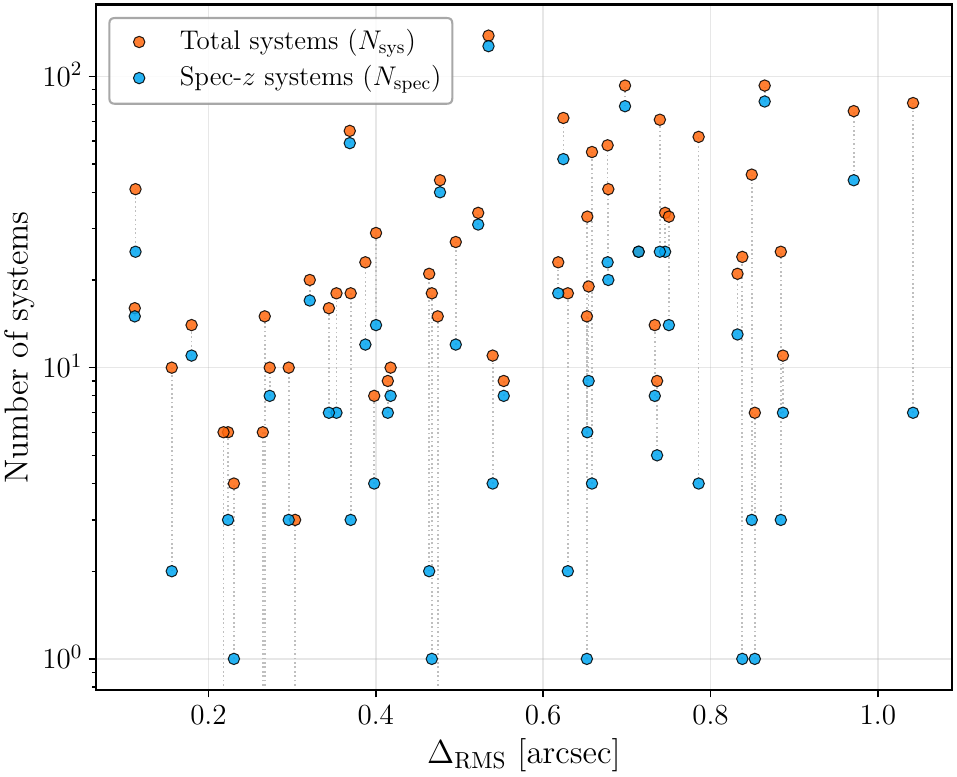} 
\caption{\rms vs Number of Systems (total in orange, spectroscopically confirmed in blue). There is no obvious trend of \rms increasing or decreasing with the number of systems, as more constraints also capture the physical subtlety of lensing in clusters.}
\label{fig:rms_nsys}
\end{figure}
\begin{figure*}
    \centering
    \begin{minipage}[t]{0.48\textwidth} 
    \centering
    \panel{(i)}{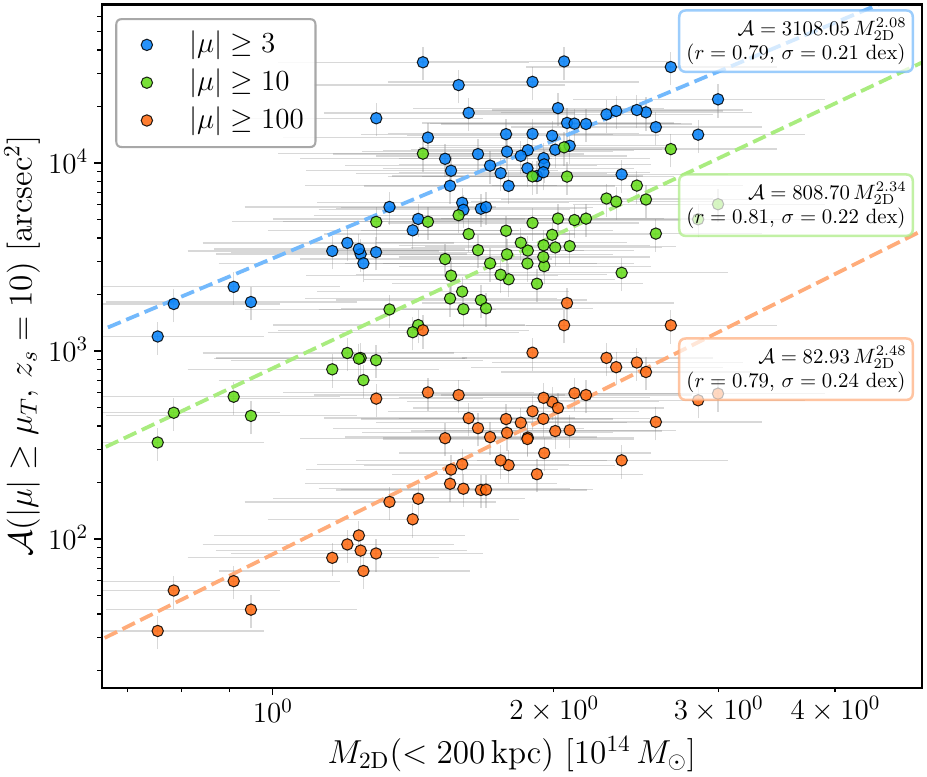}{\linewidth}
    \label{fig:area_M200kpc}
    \end{minipage}
    \hfill
    \begin{minipage}[t]{0.48\textwidth}
    \panel{(ii)}{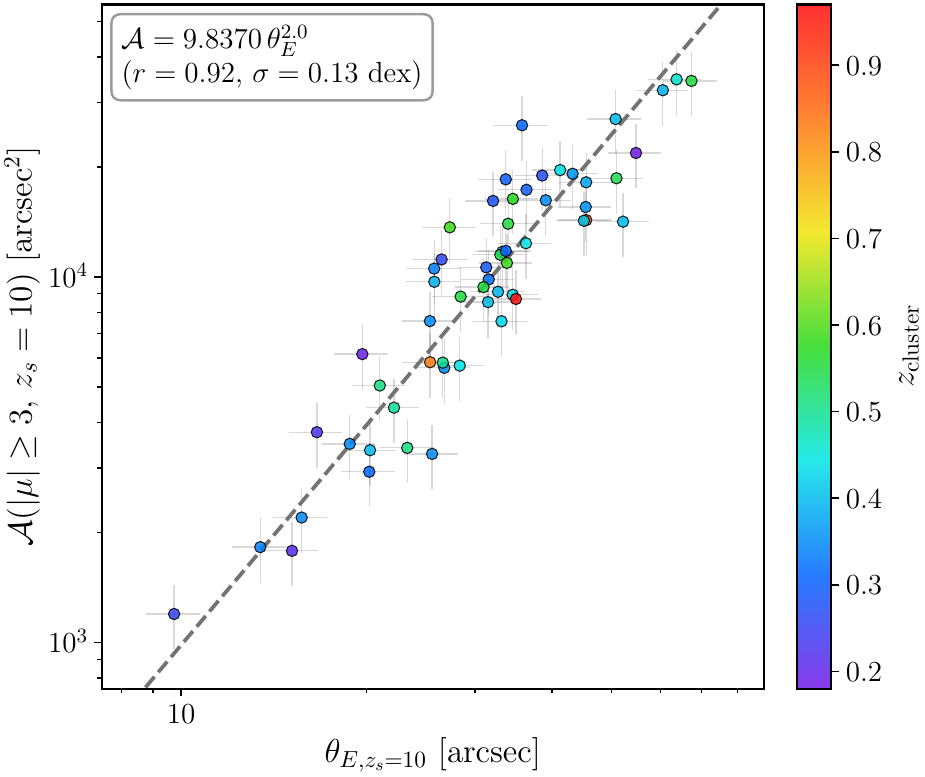}{\linewidth}
    \label{fig:thetaE_vs_area_zs10}
    \end{minipage}
    \begin{minipage}[t]{0.48\textwidth} 
    \centering
    \panel{(iii)}{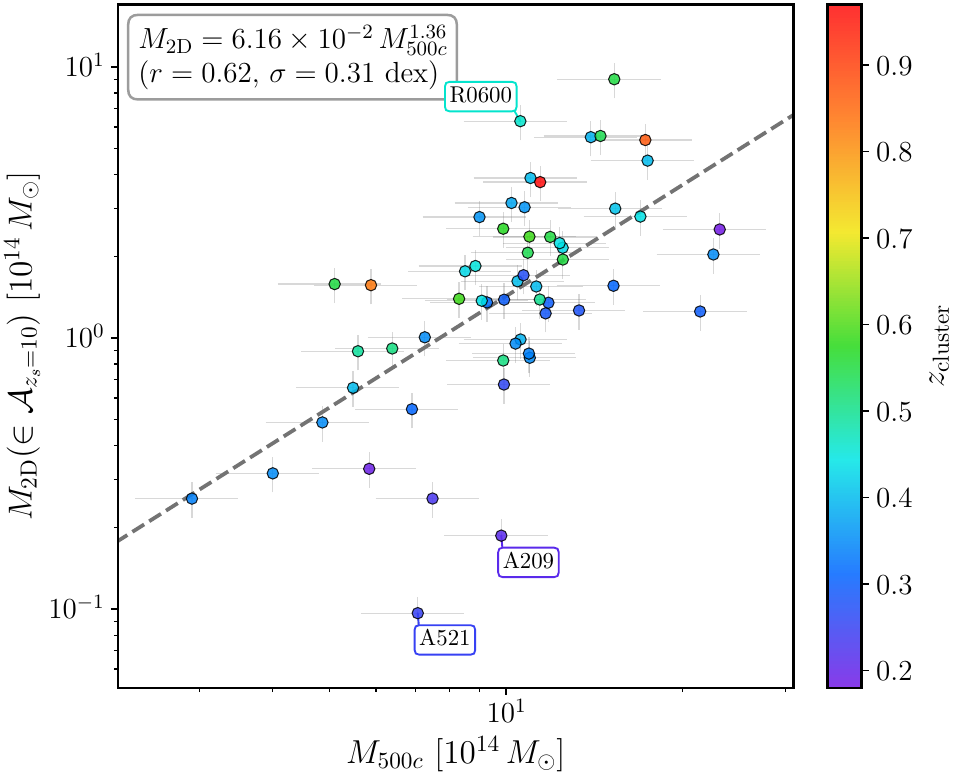}{\linewidth}
    \label{fig:Me_M500}
    \end{minipage}
    \hfill
    \begin{minipage}[t]{0.48\textwidth}
    \panel{(iv)}{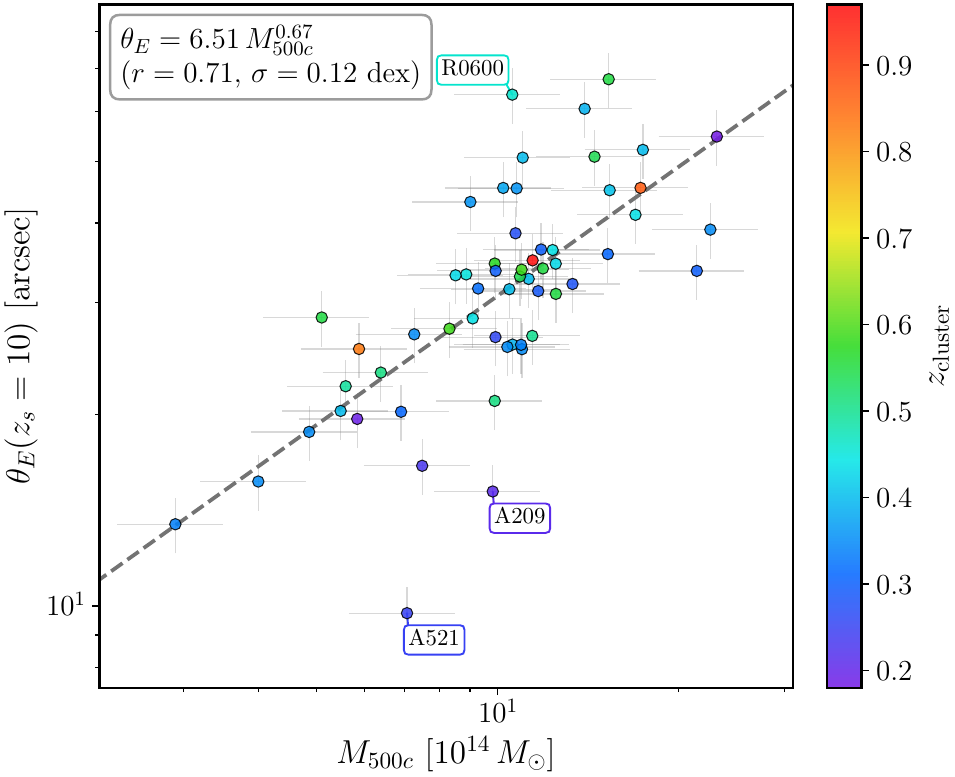}{\linewidth}
    \label{fig:thetaE_vs_M500}
    \end{minipage}
    \begin{minipage}[t]{0.48\textwidth}
    \centering
    \panel{(v)}{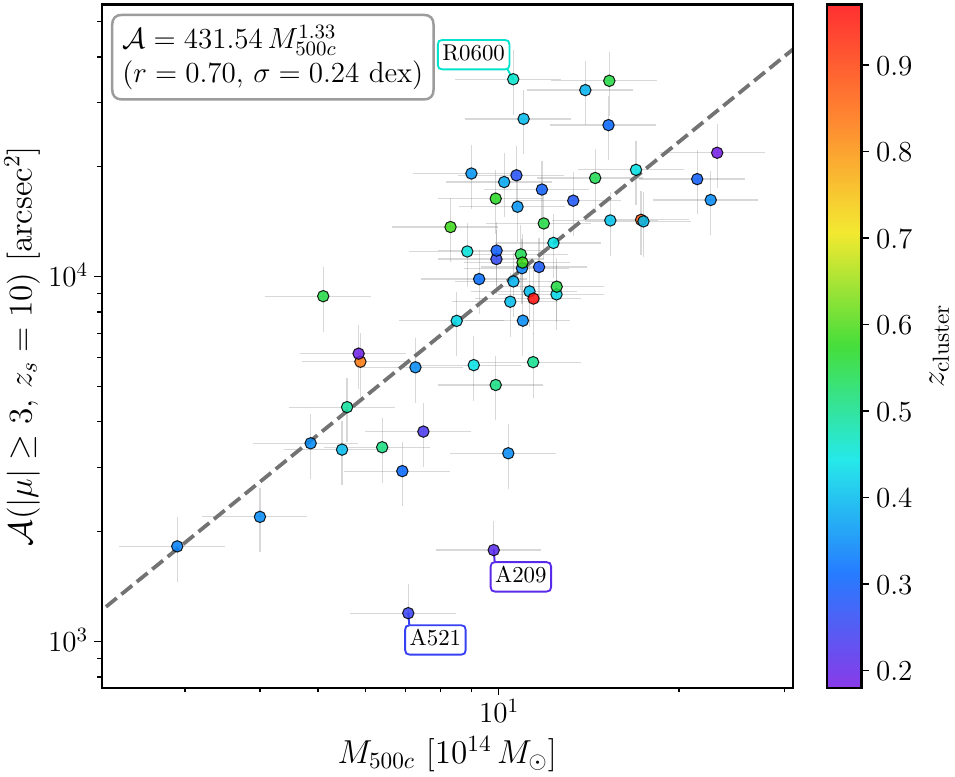}{\linewidth}
    \label{fig:area_vs_M500}
    \end{minipage}
    \hfill
    \begin{minipage}[t]{0.48\textwidth}
    \panel{(vi)}{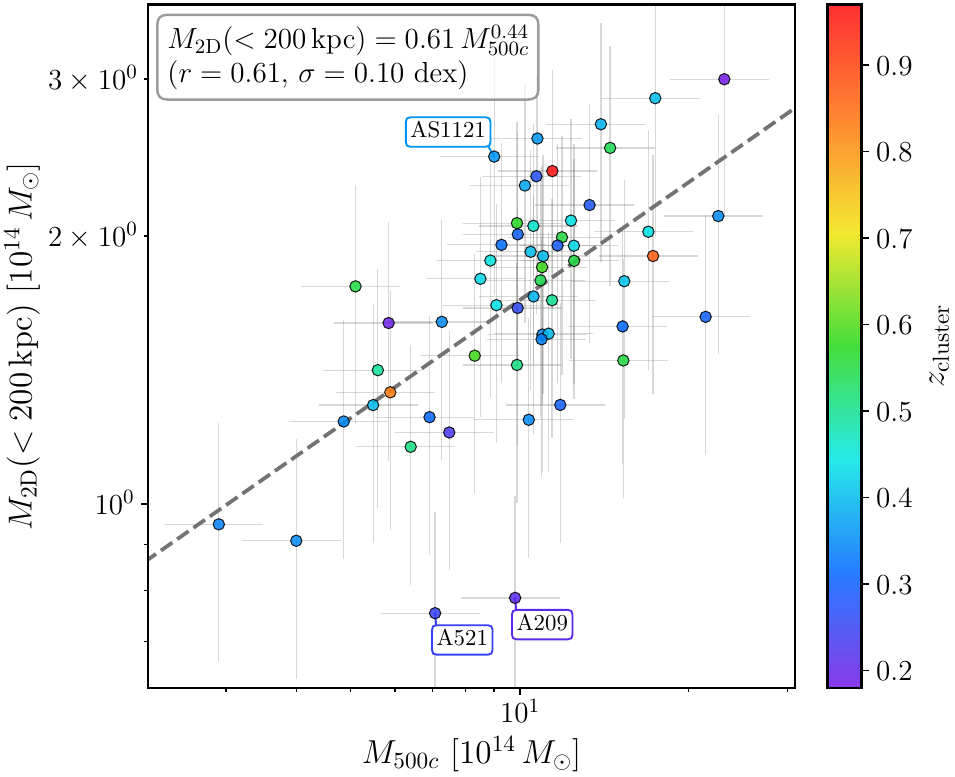}{\linewidth}
    \label{fig:M200kpc_vs_M500}
    \end{minipage}
    \caption{\textbf{Scaling relations:} (i) Magnified area above a magnification threshold $\mu_T$ vs $M(<200$\,kpc). (ii) Magnified area above a threshold, $\mathcal{A}_{z_s=10}(|\mu|\geq3)$ for a source at $z_s=10$, vs Einstein radius. (iii) Mass contained within an Einstein radius vs \Mfivehundred. (iv) Einstein radius vs \Mfivehundred. (v) Magnified area $|\mu| \geq 3$ at $z_s = 10$ vs \Mfivehundred.  (vi) Cylindrical mass within 200 kpc vs \Mfivehundred. For panels (iii)-(vi), we notice outliers, the most notable of which is cluster A521, whose lensing constraints only apply to the cluster core, thus yielding very small lensing statistics (see Section \ref{sec:a521}).}
    \label{fig:figs_stats_2}
\end{figure*}
We hereafter explore empirical correlations among the lensing observables across the VENUS sample. 
Like every other SL sample, we note that it is skewed by its selection function (see Section \ref{sec:VENUS_imaging}).

\subsubsection{Multiply-imaged systems distribution}

Provided the long standing interest in the redshift distribution of the sources enabled by lensing \citep[discussed e.g.\ in][]{Bayliss2011, Bayliss2012, Richard2021_MUSEatlas}, we display in the left panel of Figure \ref{fig:figs_stats_1} the distribution of all MIS, with and without spectroscopic identification, distributed by redshift.
As this work constitutes the largest compilation of multiply-lensed sources, this Figure offers a representative view of such sources in typical JWST and HST observations.
We provide an effective estimate of the redshift distribution expected for the \ncl clusters by fitting the log-normal likelihood, which we fit with Poisson-Cash maximum likelihood estimate:
\begin{equation}
    N_{\rm Sys}(z) = \frac{A}{\sigma \sqrt{2 \pi} z} \exp \left[ - \frac{\left(\ln z - M\right)^2}{2 \sigma^2} \right],
\end{equation}
where $A = 1291.5 \pm 31.0$, $\sigma = 0.58 \pm 0.01$ and $M = 1.07 \pm 0.01$, with a reduced Cash statistics $C_{\nu} = 3.4$, goodness of fit reduced $\chi^2_{\nu} = 3.1$ and Pearson coefficient $r=0.98$.
We note that this is an effective fit, representing a specific SL sample of clusters, for a given observational depth (28 AB mag here).
Nonetheless, in such cases, it provides a rough estimate \textit{a priori} for future surveys.

Figure \ref{fig:figs_stats_1}, right panel presents the distribution of MIS (both total and spectroscopically identified) per cluster.
A total of $53\%$ of all \totmis MIS have a spectroscopic identification, with large variations between target clusters. Some notable clusters stand out, such as e.g.\ M0553, R0600, M0717 as priorities for follow-up spectroscopy, e.g.\ with NIRSpec or NIRISS.
More than 20 clusters have $\leq 5$ spectroscopically identified systems, thus also witnessing the need for more spectroscopic observations and studies in order to anchor lens models for a large number of clusters.

Figure \ref{fig:rms_nsys} presents the final image-reproduction $\rms$ of our models for all \ncl clusters, distributed by number of multiply-imaged systems (both spec-$z$ and total). We find an average $\rms$ of \avrms, with a standard deviation $0.24''$, and a median $0.54''$.
We note that a larger number of spectroscopic systems should generally yield a better quality model, however this is not necessarily manifested in terms of \rms.
Indeed, a larger set of MIS constraints may witness complexities in the underlying mass distribution (such as substructures, multi-plane lensing, etc.).
These intricate properties can increase the image-plane \rms even when the additional constraints lead to a more accurate and physically representative model.
Conversely, a low level of constraints ($\lesssim 5$ MIS) can yield extremely low \rms, due sometimes to overfitting.

\subsubsection{Scaling relations: lensing and ICM}

Figure \ref{fig:figs_stats_2} displays effective scaling relations between various lens observables and the ICM-inferred mass \Mfivehundred.
In detail, panel (i) displays the correlation between the magnified area at redshift $z_s=10$ ($\mathcal{A}_{z_s=10}(|\mu|\geq \mu_T)$ above different $\mu_T$ thresholds and the 2D mass within 200 kpc:
\begin{equation}
\begin{split}
\mathcal{A}_{z_s=10}^{|\mu|\geq3} &= 3108 \times M_{\rm 2D, <200\,\mathrm{kpc}}^{2.08}~(r=0.79; \sigma = 0.21~\mathrm{dex}),\\
\mathcal{A}_{z_s=10}^{|\mu|\geq10} &= 809 \times M_{\rm 2D, <200\,\mathrm{kpc}}^{2.34}~(r=0.81; \sigma = 0.22~\mathrm{dex}),\\
\mathcal{A}_{z_s=10}^{|\mu|\geq100} &= 82.9 \times M_{\rm 2D, <200\,\mathrm{kpc}}^{2.48}~(r=0.79; \sigma = 0.24~\mathrm{dex}),\\
\end{split}
\end{equation}
where $r$ is the Pearson coefficient and $\sigma$ the scatter, and where areas are in arcsec$^2$ and masses in $10^{14} M_{\odot}$.
Provided the biases on the measurement of $M_{\rm 2D} (<200$\,kpc), i.e.\ a measurement centered on the main BCG, the strong correlation between both observables is notable. In particular, the increase of the slope with larger magnification thresholds signifies that high-magnification regions are disproportionately sensitive to mass concentration.
\citet{FoxMahlerSharon2022} reached the same qualitative conclusion for lens models of HFF, RELICS and SGAS clusters: the projected core's mass is a good predictor of the $|\mu|\geq3$ area.

Panel (ii) on Figure \ref{fig:figs_stats_2} shows the correlation between the magnified area $\mathcal{A}_{z_s}(|\mu|\geq3)$ for $z_s \in \left\{2, 5, 10\right\}$ and the Einstein radius. We first linearly reduce this relationship in the log-log plane, and find a general estimate:
\begin{equation}
\begin{split}
\mathcal{A}_{z_s=2}^{|\mu|\geq3} &= 49.6 \times \theta_{E,z_s=2}^{1.57}~(r=0.81; \sigma = 0.20~\mathrm{dex}),\\
\mathcal{A}_{z_s=5}^{|\mu|\geq3} &= 18.3 \times \theta_{E,z_s=5}^{1.83}~(r=0.91; \sigma = 0.14~\mathrm{dex}),\\
\mathcal{A}_{z_s=10}^{|\mu|\geq3} &= 17.4 \times \theta_{E,z_s=10}^{1.83}~(r=0.92; \sigma = 0.13~\mathrm{dex}),\\
\end{split}
\end{equation}
where areas are in arcsec$^2$ and Einstein radii in arcsec.
The higher source redshifts correspond to a better quality fit and to the power index approaching $2$, agreeing with the findings of \citet{FoxMahlerSharon2022}.
This square power law corresponds to a correlation of the magnified area with the effective Einstein area, in line with the expected relationship presented in \citet{Vega-Ferrero2019probHFF}, for HFF clusters; and seemingly matching the trend displayed in \citet{Cerny2026SLICE}, Fig.\ 7.
Indeed, assuming a homologous lensing potential profile, the magnified area across the cluster field ($\mathcal{A}_{z_s}(|\mu|)$) is self-similar to the Einstein critical area.
Expecting this natural correlation, we also fit the same relationship keeping the Einstein radius power index constant, and found the following results (displayed on Fig.\,\ref{fig:figs_stats_2} panel (ii)):
\begin{equation}
\begin{split}
\mathcal{A}_{z_s=2}^{|\mu|\geq3} &= 13.0 \times \theta_{E,z_s=2}^{2}~(r=0.81; \sigma = 0.21~\mathrm{dex}),\\
\mathcal{A}_{z_s=5}^{|\mu|\geq3} &= 10.5 \times \theta_{E,z_s=5}^{2}~(r=0.91; \sigma = 0.14~\mathrm{dex}),\\
\mathcal{A}_{z_s=10}^{|\mu|\geq3} &= 9.8 \times \theta_{E,z_s=10}^{2}~(r=0.92; \sigma = 0.13~\mathrm{dex}),\\
\end{split}
\end{equation}
which does not degrade the fit or the scatter, while presenting strong physical motivations, thus confirming the aforementioned pattern.

Figure \ref{fig:figs_stats_2} panel (iii) presents a general relationship between the 2D mass in the lens plane contained within the Einstein radius at $z_s=10$ ($M_{\rm 2D}(\in \mathcal{A}_{z_s=10})$) and the ICM-inferred mass \Mfivehundred, while panel (iv) presents the relationship between the Einstein radius itself and \Mfivehundred. We note that, by definition, both relationships are not independent. We linearly reduce these relationships in the log-log plane and found a general estimate:
\begin{equation}
\begin{split}
M_{\rm 2D} (\in \mathcal{A}_{z_s=2}) &= 4.4 \times 10^{-2} \Mfivehundred^{1.31}\\&~~~~(r=0.62; \sigma = 0.30~\mathrm{dex}),\\
M_{\rm 2D} (\in \mathcal{A}_{z_s=5}) &= 5.6 \times 10^{-2} \Mfivehundred^{1.36}\\&~~~~(r=0.63; \sigma = 0.30~\mathrm{dex}),\\
M_{\rm 2D} (\in \mathcal{A}_{z_s=10}) &= 6.2 \times 10^{-2} \Mfivehundred^{1.36}\\&~~~~(r=0.62; \sigma = 0.31~\mathrm{dex}),\\
\theta_{E,z_s=2} =&~5.0 \times \Mfivehundred^{0.65}~(r=0.65; \sigma = 0.14~\mathrm{dex}),\\
\theta_{E,z_s=5} =&~6.1 \times \Mfivehundred^{0.68}~(r=0.71; \sigma = 0.12~\mathrm{dex}),\\
\theta_{E,z_s=10} =&~6.5 \times \Mfivehundred^{0.67}~(r=0.71; \sigma = 0.12~\mathrm{dex}),\\
\label{eq:ThetaE_M500}
\end{split}
\end{equation}
where masses are in $10^{14} M_{\odot}$ and Einstein radii in arcsec.
The relatively large scatter on both relations shows that \Mfivehundred constitutes an imperfect prediction of the SL region. This result could be expected provided that \Mfivehundred is a 3D measurement of the mass which encapsulates the total cluster mass, while the Einstein radius is inherently a lensing, and thus a projected observable, which probes the densest central regions of the clusters. Moreover, \Mfivehundred being ICM-measured, it is sensitive to baryonic processes, further increasing possible scatter.
These results are in line with previous works, displaying an important scatter on the Einstein radius -- cluster mass relation \citep[e.g.][]{Killedar2018}. 
We note that a rapid inspection of the clusters' dynamical state using our lens models does not display an obvious trend, although a more thorough analysis is left to future studies.
Despite their significant scatter, such relationships are useful to estimate the inner lensing regions' extent and mass compared to the ICM-inferred \Mfivehundred, e.g.\ when designing a SL survey from ICM-inferred data.
Nonetheless, such relations should be used with caution. Indeed, the VENUS clusters were selected for a large magnified area, so the sample is biased toward lensing-efficient clusters.
\begin{figure}
    \centering
    \includegraphics[width=\columnwidth]{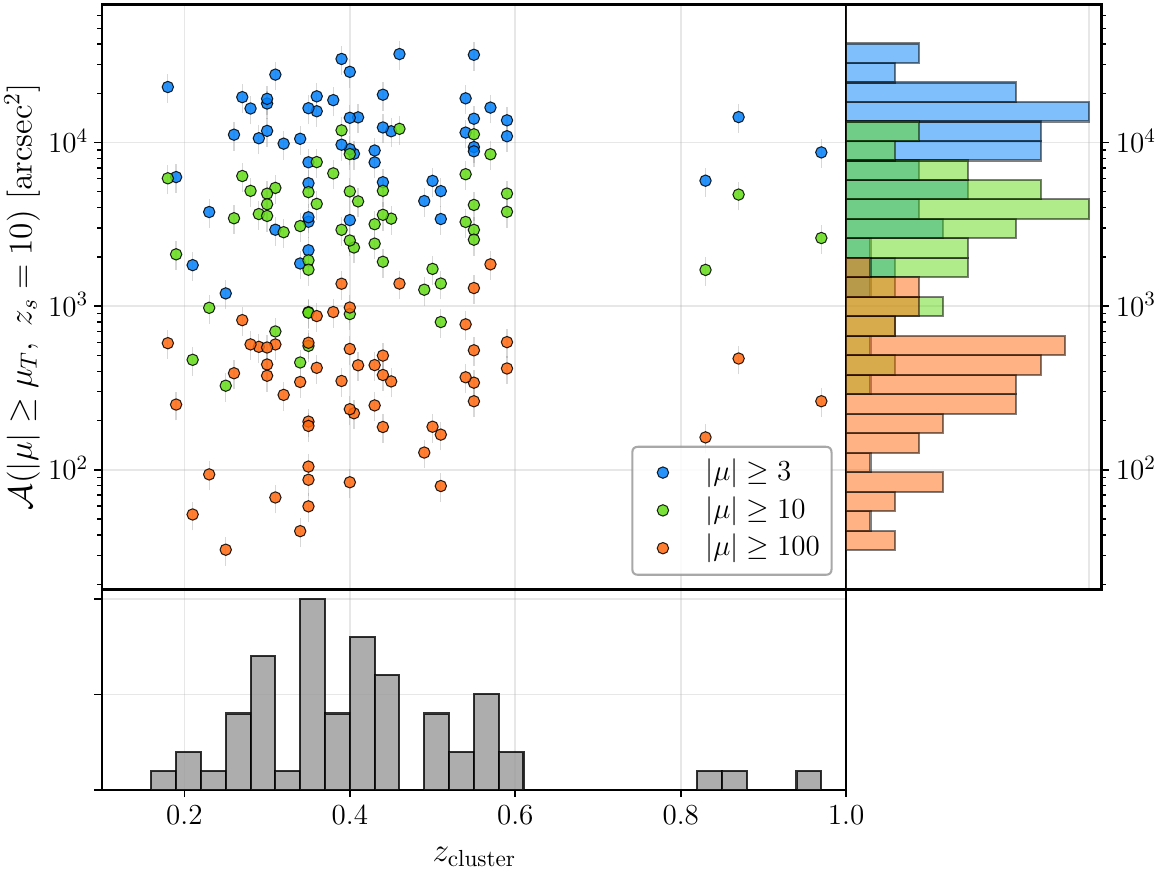}
    \caption{Magnified area at $z_s=10$ above magnification thresholds $\mu_T \in \left\{3, 10, 100\right\}$ compared to the cluster redshift.}
    \label{fig:area_vs_z}
\end{figure}

\begin{figure*}
    \centering
    \begin{minipage}[t]{0.48\textwidth}
    \panel{(i)}{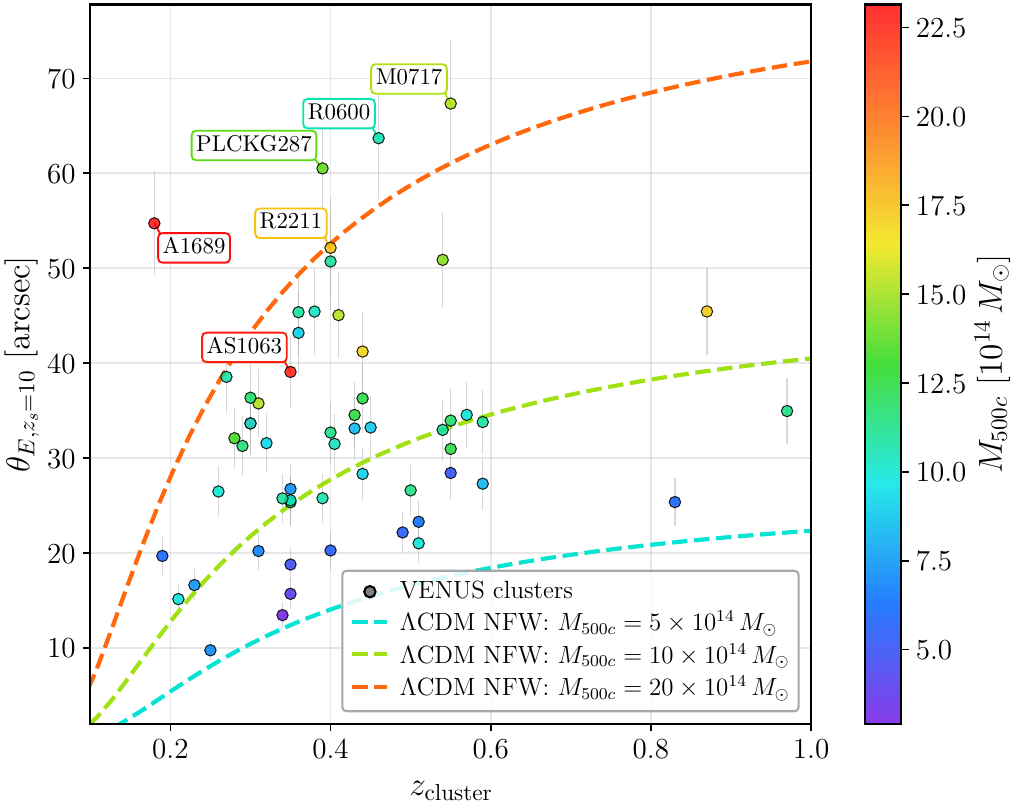}{\linewidth}
    \label{fig:thetaE_vs_z_LCDM}
    \end{minipage}
    \hfill
    \begin{minipage}[t]{0.48\textwidth}
    \panel{(ii)}{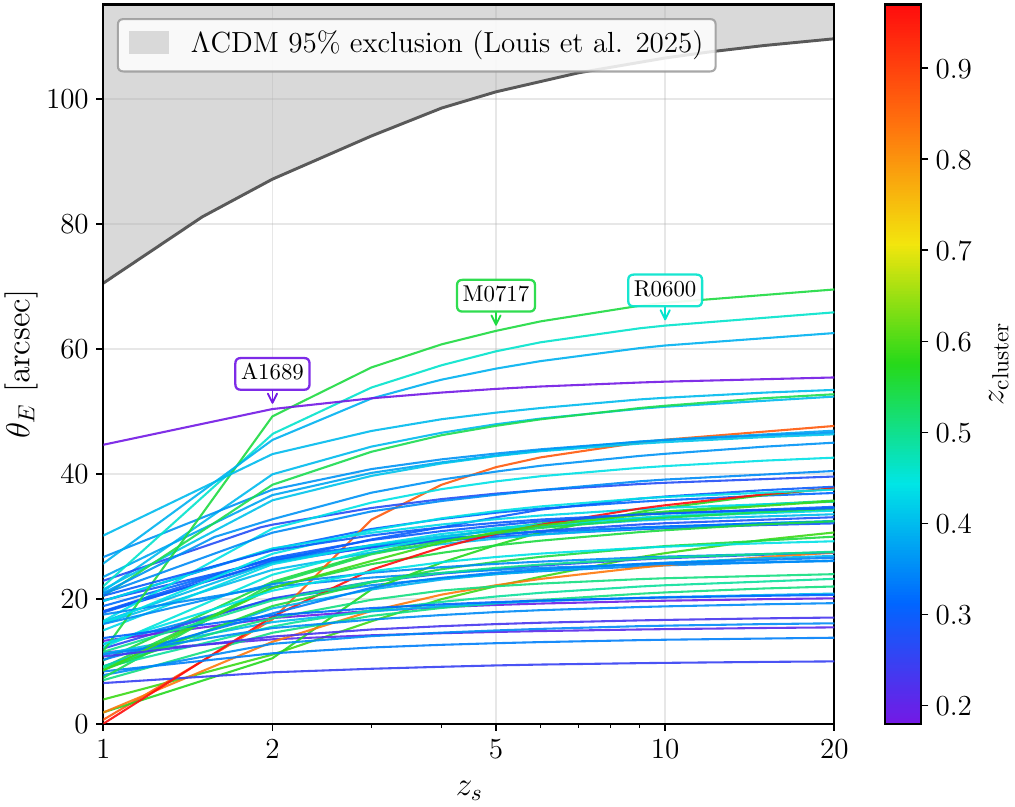}{\linewidth}
    \label{fig:thetaE_zs}
    \end{minipage}
    \caption{(i) Cluster Einstein radius for a source $z_s=10$ compared to the cluster redshift. The marker color represents \Mfivehundred, while the colored dashed lines represent the evolution of the expected $\theta_{E, z_s=10} (z_{\rm cluster})$ values for a NFW \citep[][]{Navarro1996} spherical potential of various \Mfivehundred values \citep[scaled using the mass-concentration relationship from][]{Duffy2008}. 
    (ii) Einstein radius evolution with source redshift for all \ncl modeled VENUS clusters. \LCDM limits were computed using \cite{OguriBlandford2009}, with cosmological parameters from \citet{Louis2025_ACTDR6}.}
    \label{fig:figs_stats_3}
\end{figure*}
Similarly, for panel (v) in Fig.\,\ref{fig:figs_stats_2}, we display a comparable the relationship for the magnified area above a threshold ($\mathcal{A}_{z_{s}}^{|\mu| \geq \mu_T}$) as a function of \Mfivehundred, fitted to:
\begin{equation}
\begin{split}
\mathcal{A}_{z_s=2}^{|\mu|\geq3} &= 278 \times \Mfivehundred^{1.38}~(r=0.70; \sigma = 0.25~\mathrm{dex}),\\
\mathcal{A}_{z_s=5}^{|\mu|\geq3} &= 389 \times \Mfivehundred^{1.35}~(r=0.71; \sigma = 0.24~\mathrm{dex}),\\
\mathcal{A}_{z_s=10}^{|\mu|\geq3} &= 432 \times \Mfivehundred^{1.33}~(r=0.70; \sigma = 0.24~\mathrm{dex}),\\
\label{eq:MagA_M500}
\end{split}
\end{equation}
where areas are in arcsec$^2$ and masses in $10^{14} M_{\odot}$.
With a comparable scatter to eq.\,\ref{eq:ThetaE_M500}, this relationship quantitatively assesses the lensing efficiency as a function of the total mass, in line with previous studies \citep{Johnson2014_HFF,FoxMahlerSharon2022}. Although the correlation found here still displays a significant scatter, the Pearson coefficient ($r \simeq 0.70$) is witnessing a much stronger correlation than the one found in \citet{FoxMahlerSharon2022}\footnote{We however note that if we fit with the $\mathcal{A}_{0.5, z_s}^{|\mu|\geq3}$ observable used in \citet[][eq. 2]{FoxMahlerSharon2022}, the resulting regression presents a much lower correlation ($r<0.63$) than eq.\,\ref{eq:MagA_M500}.}.

Panel (vi) in Fig.\,\ref{fig:figs_stats_2} displays the cylindrical mass within $200$\,kpc ($M_{\rm 2D} (<200$\,kpc)) as a function of \Mfivehundred, fitted to:
\begin{equation}
\begin{split}
    M_{\rm 2D}(<200\,\mathrm{kpc}) =~&0.61 \times \Mfivehundred^{0.44}\\ 
    &(r=0.61; \sigma = 0.10~\mathrm{dex}),
\end{split}
\end{equation}
where masses are in $10^{14} M_{\odot}$.
This relation, limited in its correlation, can serve as a qualitative indicator of variations in central mass concentration for a fixed \Mfivehundred.
Clusters lying above (below) this best-fit relation have a higher (lower) projected central mass than typically expected. Such an interpretation must however be interpreted conservatively, provided the biases of the lensing sample and caveats of $M_{\rm 2D} (<200$\,kpc) measurements. For instance, the clusters are prolate, and may be aligned with the line-of-sight.
Although no equivalent relationship may be found in the literature, \cite{Bonamigo2018, Caminha2019} show relationships for $M_{\rm 2D} (<0.1 R_{200,c}) / M_{200,c} \simeq 0.13$ with a relatively low scatter ($\simeq5\%$ for respectively 3 and 8 clusters), i.e.\ lower than obtained here, albeit with different observables.

\subsubsection{Cosmic evolution}

Figure \ref{fig:area_vs_z} shows the distribution of magnified area (at $z_s=10$, $\mathcal{A}_{z_s=10}(|\mu|\geq \mu_T$) above different thresholds $\mu_T$ as a function of cluster redshift. 
We do not detect any redshift evolution, in line with e.g.\ Fig.\,1 in \citet{FoxMahlerSharon2022}. 
This contradicts expectations for higher-$z$ clusters, which, being younger, and would be expected to be less massive on average, and thus less efficient lenses. Nonetheless, the VENUS sample does not allow to draw any statistically significant conclusion, due to the number of $z>0.6$ clusters, limited to three targets, as well as by the sample selection bias, favoring the largest lensing strengths. Future surveys with higher-redshift cluster lenses may more efficiently probe such a redshift evolution.

Figure \ref{fig:figs_stats_3} panel (i) compares the redshift evolution of the $z_s=10$ Einstein radius, $\theta_{E,z_s=10}$, for different Navarro-Frenk-White \citep[NFW,][]{Navarro1996} spherical halo \Mfivehundred masses \citep[scaled using the mass-concentration relationship from][]{Duffy2008}. 
These predicted typical Einstein radii are not meant here to constrain \LCDM cosmology, but rather to see which clusters are exceptionally lensing-efficient for their mass, such as e.g.\ A1689, PLCKG287, R0600 or M0717. 
Overall, the vast majority of VENUS clusters exceed their expected lensing efficiency (based on NFW halo predictions), which is not surprising given the selection bias toward potent lenses, the lenses' ellipticity, possible merger state, and line-of-sight alignment. These effects all concur to make our cluster sample more lensing-efficient \citep{Meneghetti2007}.
This result is in agreement with Einstein radii seemingly much larger than SZ-inferred masses predicted \citep{Gralla2011}.

\begin{figure}
    \centering
    \includegraphics[width=\columnwidth]{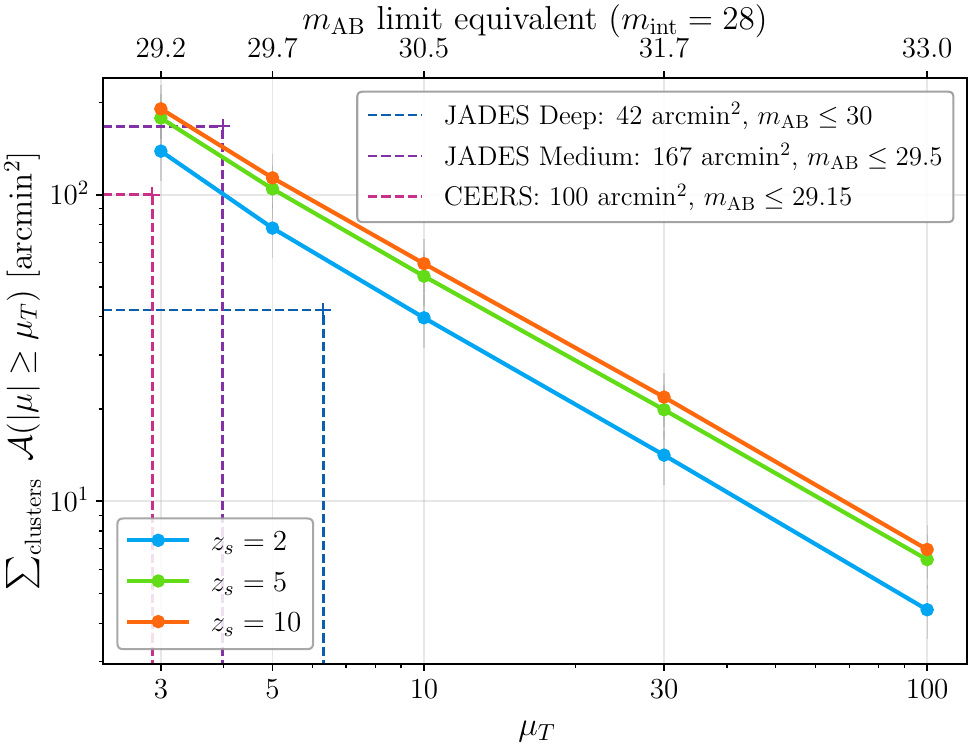}
    \caption{Cumulative total magnified area above a given threshold $\mu_T$ for different source redshifts, contrasted with blank fields JADES and CEERS. The top $x$-axis provides equivalent magnitudes to magnifications.}
    \label{fig:cumulative_magnified_area_tot}
\end{figure}
Panel (ii) in Fig.\,\ref{fig:figs_stats_3} displays the Einstein radius for $z_s\in[1,20]$, for all DR1 VENUS lens models. The magenta line with the largest $\theta_{E, z_s =1} = 44.6\pm 4.5''$ is A1689, while the green one, with highest $\theta_E$ for $z_s>3$, is M0717.
The \LCDM 95\% exclusion limits are computed following \citet[][adapted to the last ACT DR6 cosmological parameters, see \citealt{Louis2025_ACTDR6}]{OguriBlandford2009}, thus allowing a comparison with our companion article \Oguri. All our computed Einstein radii are well within the \LCDM predicted exclusion limits, and therefore do not contradict the standard cosmology paradigm.

\subsubsection{Survey breadth and depth}

Finally, Figure \ref{fig:cumulative_magnified_area_tot} displays the total image-plane magnified area for all \ncl VENUS clusters of this DR1, above different magnification thresholds. The corresponding magnification of $\mu_T=1$ represents the nominal minimum $5\sigma$ VENUS $28$ AB magnitude. For instance, $\sum_{\rm clusters} \mathcal{A}_{z_s=10}(|\mu|\geq 3) = 190.6$\,arcmin$^2$ and $\mathcal{A}_{z_s=10}(|\mu|\geq 10) = 59.6$\,arcmin$^2$, at respective equivalent AB depth 29.2 and 30.5. 
In the context of blank fields, these figures respectively exceed the areas of CEERS \citep[${\sim}100$\,arcmin$^2$, $m_{\rm AB}\simeq29.15$,][]{Finkelstein2025_CEERS} and JADES Deep \citep[$42$\,arcmin$^2$, $m_{\rm AB}\lesssim30$,][]{Eisenstein2026_JADES, Johnson2026_JADES_DR5}.
As for the JADES Medium field, its probed area ($167$\,\arcminsq) is comparable to the VENUS image-plane area at equivalent depth ($m_{\rm AB} \leq29.5$).

\subsection{Opportunity of observation} \label{sec:opportunities}
 
The Einstein radius or the magnified area alone do not constitute a perfect ranking of clusters for a given science goal. The interest of SL clusters resides in the combination of lensing area and high magnifications, and model accuracy, driven by larger numbers of spectroscopically confirmed systems, as well as the ancillary data availability. 
We highlight the clusters identified by our models as the most promising targets for the main VENUS science cases, although a more complete ranking is left to Furtak et al.\ (in prep.).

\paragraph{Dark-matter assembly and constraints.}
Dissociative mergers, in which the collisional ICM is spatially separated from collisionless galaxies and DM, represent pristine laboratories for DM physics, exploiting the offset between the lensing mass and the X-ray detected peak. While CL0658 (the Bullet cluster) is the natural target for self-interaction constraints \citep{Clowe2006Bullet}, clusters AS295, ACTCL0102, M0025, A3192, ACTCL0152 and SPTCL0615 display various merger configurations at different stages and redshifts.
Although their `bulleticity' is yet to be confirmed with X-rays for some, large SL surveys enable a deeper study of `Bullet-like' clusters, thus limiting degeneracies such as three-dimensional merger geometry, collision velocity and merger state.
While JWST has already enabled a detailed Bullet Cluster SL model through the CANUCS survey \citep{Rihtarsic2026_Bullet}, VENUS extends such constraints to a class of similar clusters.

\paragraph{Interplay with the intracluster medium.}
A joint study of lensing mass maps with X-ray, SZE and radio data, a necessary step toward understanding `Bullet-like' clusters, more broadly enables to probe the dynamical state of the ICM relative to the DM, and therefore to test the hydrostatic mass bias \citep{Nagai2007pressureProfile, Rasia2012_LensingXray}.
Conversely, joining ICM studies to SL allows to better isolate each component (stellar, ICM and DM), and to better understand the potential in the central regions of clusters \citep{Beauchesne2024AS1063} and in the outskirts \citep{Allingham2024}.
The whole VENUS sample is suited for such a systematic comparison, as it contains targets of all merger status, such as R1347, containing the most X-ray luminous cluster known \citep[][not yet observed by VENUS]{Gitti2007_R1347_Xray} or the extreme AGN feedback from the BCG of M1931 \citep{Ehlert2011MACS1931}, responsible for destroying the cool-core. 
The VENUS dataset could be split by merger status, and each category (dynamically relaxed, minor or major merger,  Bullet-like) could see its ICM and SL properties (centroid offset, inferred mass, etc.) systematically compared.
This would moreover allow to deepen our comparison between lensing-inferred observables and the ICM-inferred \Mfivehundred (Fig.\,\ref{fig:figs_stats_2} panels (iii-vi)).
An example of X-ray survey probing the ICM and supplying the relevant observables (\Mfivehundred, ICM temperature, dynamical state, etc.) could be the \textit{XMM-Newton} survey \cite{CHEXMATE2021}, overlapping with VENUS for 24 clusters.
The further addition of weak lensing, the intracluster light and stellar dynamics to this picture would constitute the deepest multi-probe understanding for any cluster sample, and sharpen the lensing maps at the subhalo level.

\paragraph{Multiply-lensed little red dots.}
Prior to VENUS, only one system of multiply-lensed LRDs was known \citep{Furtak2023_LRD}. This survey reveals a growing population of such sources, including the doubly-imaged A383-LRD1 with a blue companion \citep[$z=6.03$,][]{Golubchik2025_LRD_VENUS, Baggen2026_A383LRD}, two Einstein crosses LRD candidates behind R2211 \citep[$z_{\rm phot} \simeq 4.3$,][]{Zhang2025LRD} and the close `Red Eyes' pair lensed by PLCKG004 \citep[$z=7.49$,][]{Yanagisawa2026_LRDs}.
Such strongly-lensed LRD discoveries have enabled to precisely reconstruct source properties, such as the intrinsic luminosity, the separation between the LRD and its blue companion, and to calculate the time-delay predictions (predicted to be as high as $\sim 130$ years for the R2211 systems).
Such time delays provide a unique opportunity to investigate the long-term LRD variability, offering insights into the nature of their central engines and the structure of their emitting regions. 
Further refinement of the VENUS lens models, combined with  VENUS Cycle 5 and other spectroscopic follow-up (e.g.\ \href{https://www.stsci.edu/jwst-program-info/download/jwst/pdf/10075/}{GO-10075}) will enable better constraints on their intrinsic properties.

\paragraph{Giant arcs and caustic-crossing events.}
Giant arcs with extreme magnification are prime locations for CCEs of individual stars \citep{Kelly2018Icarus,Welch2022EarendelJWST,Meena2023_lensedstars}.
Their occurrence and properties are governed by a combination of chance alignments between extended lensed sources and critical curves, and the lensing cluster properties (concentration, ellipticity, line-of-sight alignment), with substructures and mergers providing enhanced magnifications.
The Dragon arc in A370 has yielded more than a hundred of such events
\citep{Li2024_A370_Dragon_Flashlights,Fudamoto2025_40lensedstars,Palencia2026_CCE_Dragonarc}, which also constrain DM models
\citep{Broadhurst2025_DM_Dragon,Palencia2026_CCE_Dragonarc}. 
The combination of SLICE and VENUS imaging allowed to detect lensed star M1206-LS1 in the Snake arc (hosted in M1206), brighter by $\geq 1$\,mag than A370 detections (publication in prep.). More giant arcs, such as the $18''$-long AS1121 arc at $z=1.579$ \citep{Mahler2025_twobigSLcl}, and more arcs hosted in e.g.\ clusters A2813, A383, M0553 and PLCKG165 offer such CCE opportunities, monitorable in future epochs.

\paragraph{First stars and first galaxies.}
With strong gravitational lensing, high magnification regions unveil the otherwise unobservable most distant and metal-poor stellar populations. M1931 provides the instance of SN Eos, a metal-poor supernova type IIP at $z=5.13$, with a combined magnification $\mu\simeq55$, occurring in a faint galaxy \citep{Coulter2026_SN_Eos,Asada2026_GalEos}.
Moreover, the Misty Moons, a multiply-lensed resolved galaxy in the background of M0257 at $z_{\rm phot}\sim11$--$12$ \citep{Nakane2025_VENUS_highzgal}, and VCQG-1, a multiply-lensed massive quenched galaxy at $z_{\rm phot} \sim 5.18$, were discovered by VENUS \citep[][]{Pan2026_CL0152}.
Other examples of highly-lensed galaxies re-observed with VENUS include, in clusters CL0658, M1423, M0308 and SPTCL0615 respectively, the Bullet Arc, a highly-magnified galaxy at Cosmic Dawn \citep[$z\sim 11.1$;][]{Bradac2025_BulletArc}, the Firefly Sparkle, a spiral galaxy in the early stages of its assembly \citep[$z\sim8.3$;][]{Mowla2024_Firefly}, the Cosmic Spear, a multiply-imaged galaxy \citep[$z\sim 6$;][]{Fudamoto2024_CosmicSpear}, and the Cosmic Gems, young star clusters \citep[$z\sim 10$;][]{Salmon2018,Adamo2024_CosmicGems}. 
The two most metal-poor known galaxies, the highly magnified LAP~1 \citep{Vanzella2023_LAP1} and AMORE-6 \citep{Morishita2025_AMORE6}, are also part of the VENUS sample, lensed by M0416 and A2744, respectively.
Such detections probe early structure formation and the interstellar medium, and their updated lens models now anchor error measurements in their intrinsic luminosity more precisely than ever.
Clusters with the largest magnified area at high source redshifts ($\mathcal{A}_{z_s=10}^{|\mu|\geq10}$, amounting to higher chances to detect high-redshift objects), such as R0600 (3.4 \arcminsq), PLCKG287 (3.3 \arcminsq), M0717 (3.1 \arcminsq), R0032 (2.4 \arcminsq) and W0137 (2.4 \arcminsq) are the best targets to detect such magnified sources at $z\gtrsim10$.

\paragraph{The faint end of the UV luminosity function.}
For UVLF statistics, the same figure of merit, i.e.\ the magnified area, applies with a lower threshold ($\mu_T=3$), and additionally requires reliable magnification maps \citep[e.g.][]{Atek2025_GLIMPSE}, anchored on a large number of spectroscopic MIS identifications.
A2744, ACTCL0102, A370, M0416, CL0658, M1149, M1206, AS1063 combine $\gtrsim30$ spectroscopic systems with magnified areas of $\mathcal{A}_{z_s=10}(|\mu|\geq3) \geq 2.5$\,\arcminsq, and therefore constitute robust candidates.
We note that, due to a lack in spectroscopic identifications, the four clusters with the largest lensing areas, R0600, M0717, PLCKG287 and R0032, are all excluded from this selection criterion despite their magnified areas $\mathcal{A}_{z_s=10}(|\mu|\geq3) \geq 7.5$\,\arcminsq,
and are therefore priority targets for further MUSE, KCWI, NIRSpec or other spectroscopic observations.
Lensed supernovae extend this reach to transient-hosted galaxies far below blank-field limits, with hosts at $M_{\rm UV}=-14.4$ for SN Eos \citep{Asada2026_GalEos} and $M_{\rm UV}\gtrsim-13$ for SN Helios \citep{Fujimoto2026_SNHelios}.

\paragraph{Time-delay cosmography.}
VENUS has revealed several multiply-imaged supernovae (Ares, Eos, Athena, Atalanta, Helios, Hydra, Typhon), whose reappearances (e.g.\ SN~Athena in M0417, predicted by early 2027; Pascale et al.\ in prep.) will not only further test the lens models, but also allow to constrain cosmology \citep[e.g.][]{Pierel2026_Encore}. 
While high-cadence monitoring on predicted reappearance will minimize the time-delay measurement uncertainties, the dominant systematic for time-delay cosmography is the lens model error.
Only a combination of lens models using different software and priors, anchored by a large number of spectroscopically confirmed MIS, may reduce the model systematic error. 
This work, in combination with our companion paper \Oguri and the upcoming Cycle 5 follow-up, provides the sample depth and model diversity to proceed toward precise cluster time-delay cosmography.
In turn, time delay measurements themselves represent significant constraints on lens models, making several multiply-lensed SNe in one cluster, such as PLCKG165, a rare cosmographic opportunity.

\section{Summary} \label{sec:summary}

We present a lens model for each of the \ncl VENUS clusters observed to date as part of the JWST cycle 4 (updated to 60 clusters once all observations complete), therefore creating the largest observed homogeneous set of cluster lens models to date, with state-of-the-art JWST era SL constraints.
This work is performed alongside our companion paper \Oguri, which will enable to derive robust systematic errors, dominating the error budget of lens models.
We summarize the main results of this article:
\begin{itemize}
    \item With a combination of newly acquired JWST/NIRCam observations from the VENUS survey, the VENUS-related DDTs accepted during the survey, ancillary HST and JWST imaging from previous surveys, and spectroscopic data from MUSE, NIRSpec, NIRISS and other observatories, we compiled a total of \textbf{\totmis multiply-lensed systems spanning \totmulim lens images}, including \textbf{\totnewmis new system detections} and \textbf{a total of \totmisspec spectroscopically-confirmed system detections}, across \textbf{\ncl SL clusters}.
    \item Leveraging this wealth of constraints, we constructed the cluster lens models with our \ALP software, thus yielding an average \avrms \rms between observations and model, thus witnessing the quality of our reconstruction.
    The VENUS DR1 will publish the lens models on the collaboration website at a later date.
    \item Clusters MACS J0717.5+3745, RXCJ0600.1-2007 and PLCKG287.0+32.9 are particularly notable for magnifying the largest Einstein radii at $z_s = 10$, with $\theta_{E, z_s=10} \geq 60''$ and the largest $|\mu|\geq3$ magnified areas ($\mathcal{A}_{z_s=10}^{|\mu|\geq3} \geq 9$ arcmin$^2$) and can therefore be considered priority targets to observe the high-redshift Universe.
    \item Leveraging the largest set of cluster SL models assembled for a unique survey, we derived empirical scaling relations between different lensing observables, and for their dependence on the ICM-measured \Mfivehundred (Fig.~\ref{fig:figs_stats_2}), as well as a fit for the redshift distribution of multiply-imaged sources (Fig.~\ref{fig:figs_stats_1}). Despite the inherent survey biases, we foresee these empirical relations informing target selection for future cluster lensing programs.
    \item Thanks to the magnification provided by cluster strong lensing, the VENUS survey covers a larger image-plane area at fainter depths than the main blank fields CEERS and JADES (see Fig.~\ref{fig:cumulative_magnified_area_tot}), for instance, demonstrating the observational capabilities of cluster strong-lensing surveys.
\end{itemize}
Foreshadowing the broader scientific impact of the VENUS DR1, the lens models presented here will provide the community with comprehensive resources for the statistical exploitation of the survey and a wide range of future investigations. 
These results demonstrate the impressive potential of VENUS to constrain the high-redshift UV luminosity function, inform the design of future surveys, and pave the way to the further studies of the faintest and most distant objects in the Universe, from early stars and galaxies to transients, LRDs, and AGNs.

\begin{contribution}

JA, LF, MG and AM developed the \ncl lens models here presented, with the advice of AZ.
AZ developed \ZA, which is the core algorithm based on which JA developed \ALP.
SF is the PI of VENUS.
GB and VK reduced the NIRCam VENUS data, as well as archival HST and JWST imaging.
VK provided the \eazy photometric catalogs.
\end{contribution}

\begin{acknowledgments} 
The authors want to thank Guillaume Mahler, David Lagattuta, Keren Sharon, Catherine Cerny, Benjamin Beauchesne and Mathilde Jauzac, from the SLICE collaboration, for useful discussions and advice.

This work is supported by the Israel Science Foundation Grant No.\ 864/23. 
LJF acknowledges support from the University of Texas at Austin Cosmic Frontier Center.
SF acknowledges support from the Dunlap Institute, funded through an endowment established by the David Dunlap family and the University of Toronto, and from an NSERC discovery grant (RGPIN-2026-07931; DGECR-2026-00232).
CJC acknowledges support from the European Research Council (ERC) Advanced Investigator Grant EPOCHS (788113).
CMC is supported by the University of Toronto Faculty of Arts \& Science Research Excellence Postdoctoral Fellowship, co-funded by the Department of Astronomy \& Astrophysics.
FEB acknowledges support from ANID-Chile BASAL CATA FB210003 and FONDECYT Regular 1241005.
PAAL thanks the support from CNPq, grants 310260/2025-6 and 404160/2025-5.
GEM acknowledges the Villum Fonden research grant 13160 “Gas to stars, stars to dust: tracing star formation across cosmic time,” grant 37440, “The Hidden Cosmos,” and the Cosmic Dawn Center of Excellence funded by the Danish National Research Foundation under the grant No. 140.
MB, GR, and NM acknowledge support from the ERC Grant FIRSTLIGHT \#101053208 and Slovenian national research agency ARIS through grants N1-0238 and P1-0188. We acknowledge the support of the Canadian Space Agency (CSA) [25JWGO4A18].
MO acknowledges the supports from the World Premier International Research Center Initiative (WPI Initiative), MEXT, Japan, the joint research program of the Institute for Cosmic Ray Research (ICRR), the University of Tokyo, and KAKENHI (25H00674) through JSPS.
PD warmly acknowledges support from an NSERC discovery grant (RGPIN-2025-06182).
RAW acknowledges support from NASA JWST Interdisciplinary Scientist grants NAG5-12460, NNX14AN10G and 80NSSC18K0200 from GSFC.
TK acknowledges support by KAKENHI (26KJ1232) through Japan Society for the Promotion of Science (JSPS).
YJ-T acknowledges financial support from the State Agency for Research of the Spanish MCIU through Center of Excellence Severo Ochoa award to the Instituto de Astrofísica de Andalucía CEX2021-001131-S funded by MCIN/AEI/10.13039/501100011033, the grant PID2022-136598NB-C32 "Estallidos8" funded by MCIN/AEI/10.13039/501100011033, and the project ref. AST22-00001-Subp-15 funded by the EU-NextGenerationEU.
RA acknowledges support of Grant PID2023-147386NB-I00 funded by MICIU/AEI/10.13039/501100011033 and by ERDF/EU, and  the Severo Ochoa award to the IAA-CSIC CEX2021-001131-S and from grant PID2022- 136598NB-C32 “Estallidos8”.

The authors acknowledge the use of the Canadian Advanced Network for Astronomy Research (CANFAR) Science Platform operated by the Canadian Astronomy Data Center (CADC) and the Digital Research Alliance of Canada (DRAC), with support from the National Research Council of Canada (NRC), the Canadian Space Agency (CSA), CANARIE, and the Canadian Foundation for Innovation (CFI).
This work is based on observations made with the NASA/ESA/CSA James Webb Space Telescope. The data were obtained from the \texttt{Mikulski Archive for Space Telescopes} (\texttt{MAST}) at the \textit{Space Telescope Science Institute} (STScI), which is operated by the Association of Universities for Research in Astronomy, Inc., under NASA contract NAS 5-03127 for JWST. These observations are associated with program \# 6882 (VENUS). The VENUS JWST observations are accessible via \dataset[doi: 10.17909/6myr-m048]{https://doi.org/10.17909/6myr-m048}. Support for program \# 6882 was provided by NASA through a grant from the Space Telescope Science Institute, which is operated by the Association of Universities for Research in Astronomy, Inc., under NASA contract NAS 5-03127.
Some of the data presented herein were retrieved from the Dawn JWST Archive (DJA). DJA is an initiative of the Cosmic Dawn Center (DAWN), which is funded by the Danish National Research Foundation under grant DNRF140.

We acknowledge the use of a generative AI tool to refine the text, figures and the code used in this work. All figures and quoted values were reproduced and verified by the authors.

This work made use of the following software packages: \texttt{Astropy} \citep{astropy:2013,astropy:2022}, \texttt{Matplotlib} \citep{Hunter:2007}, \texttt{NumPy} \citep{numpy}, \texttt{pandas} \citep{mckinney-proc-scipy-2010,pandas_17229934}, \texttt{Python} \citep{python}, \texttt{SciPy} \citep{2020SciPy-NMeth,scipy_17467817}, \texttt{Photutils} \citep{Bradley2025photutils}, \texttt{Cython} \citep{cython:2011}, \texttt{emcee} \citep{emcee-Foreman-Mackey-2013,emcee_10996751}, \texttt{Numba} \citep{numba:2015,Numba_17874185}, \texttt{h5py} \citep{collette_python_hdf5_2014,h5py_7560547}  and the \texttt{MAAT} Astronomy and Astrophysics tools for \texttt{MATLAB} \citep[][]{maat14}.  
Software citation information aggregated using \texttt{\href{https://www.tomwagg.com/software-citation-station/}{The Software Citation Station}} \citep{software-citation-station-paper}.
\end{acknowledgments}


\bibliographystyle{aasjournalv7.1}
\bibliography{MyBiblio}

\begin{toggleblock}
\appendix
\vspace{-1cm}
\begin{figure}[!htbp]
    \centering
    \includegraphics[width=1\textwidth, height=0.90\textheight, keepaspectratio]{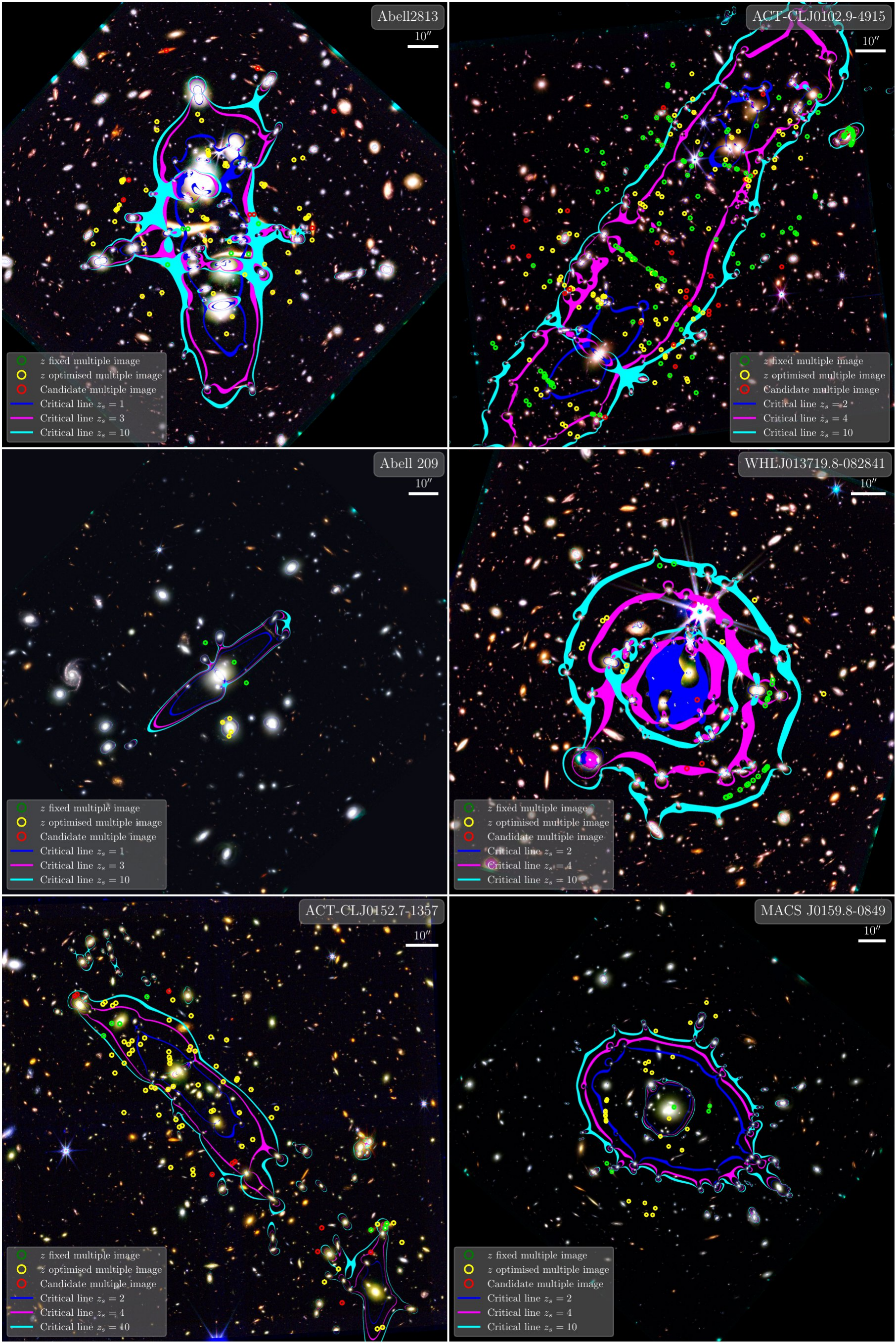}
    \caption{Similar to Fig.\,\ref{fig:clusters_1}. Galaxy clusters A2813, ACTCL0102 (El Gordo), A209, W0137, ACTCL0152, M0159, represented with critical curves and MIS. For W0137, the critical lines are represented for $|\mu|\geq300$, instead of 100 for other clusters.}
    \label{fig:clusters_2}
\end{figure}
\end{toggleblock}

\begin{toggleblock}
\begin{figure*}
    \centering
    \includegraphics[width=0.85\linewidth]{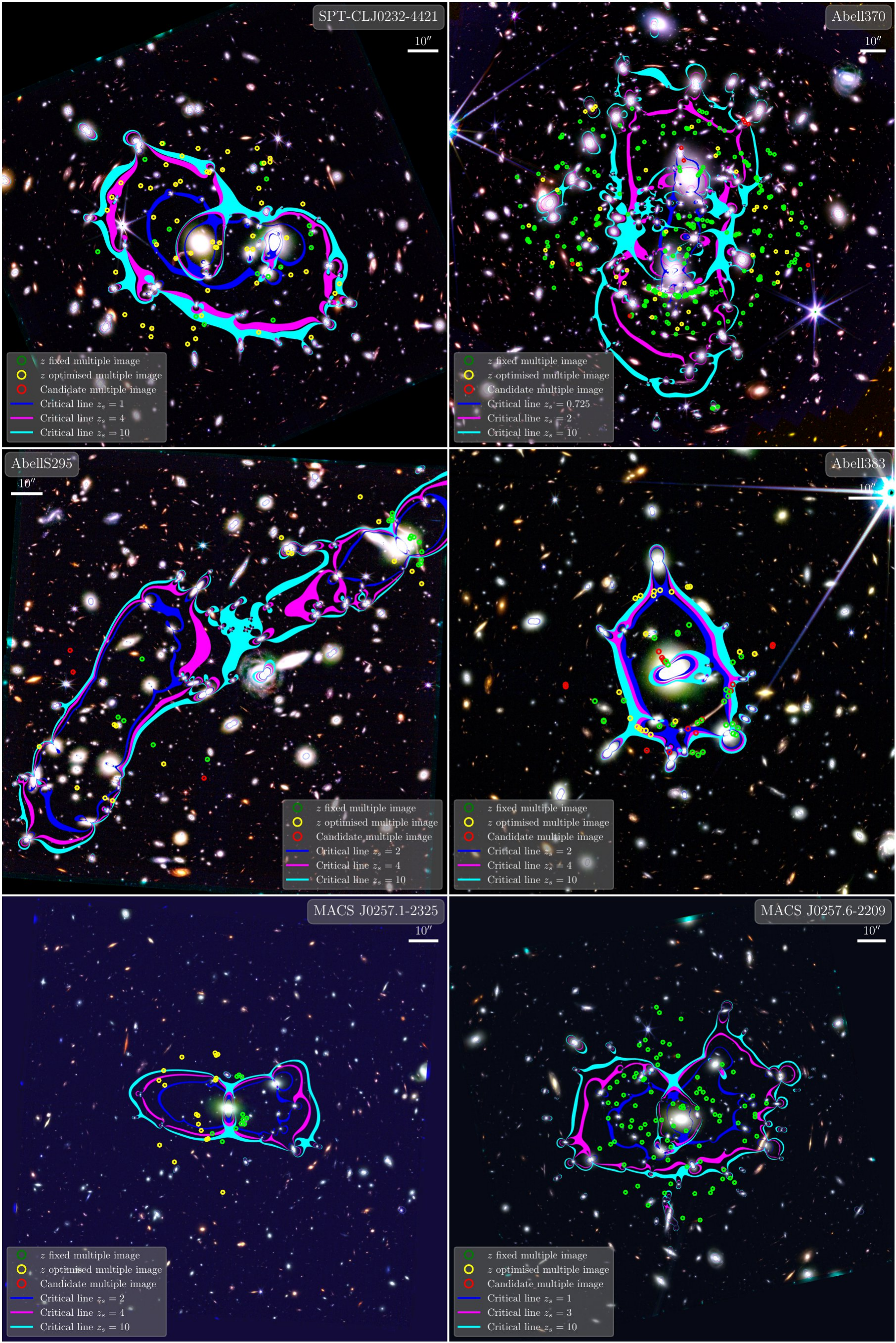}
    \caption{Galaxy clusters SPTCL0232, A370, AS295, A383, M0257, A402, represented with critical curves and MIS.}
    \label{fig:clusters_3}
\end{figure*}

\begin{figure*}
    \centering
    \includegraphics[width=0.8\linewidth]{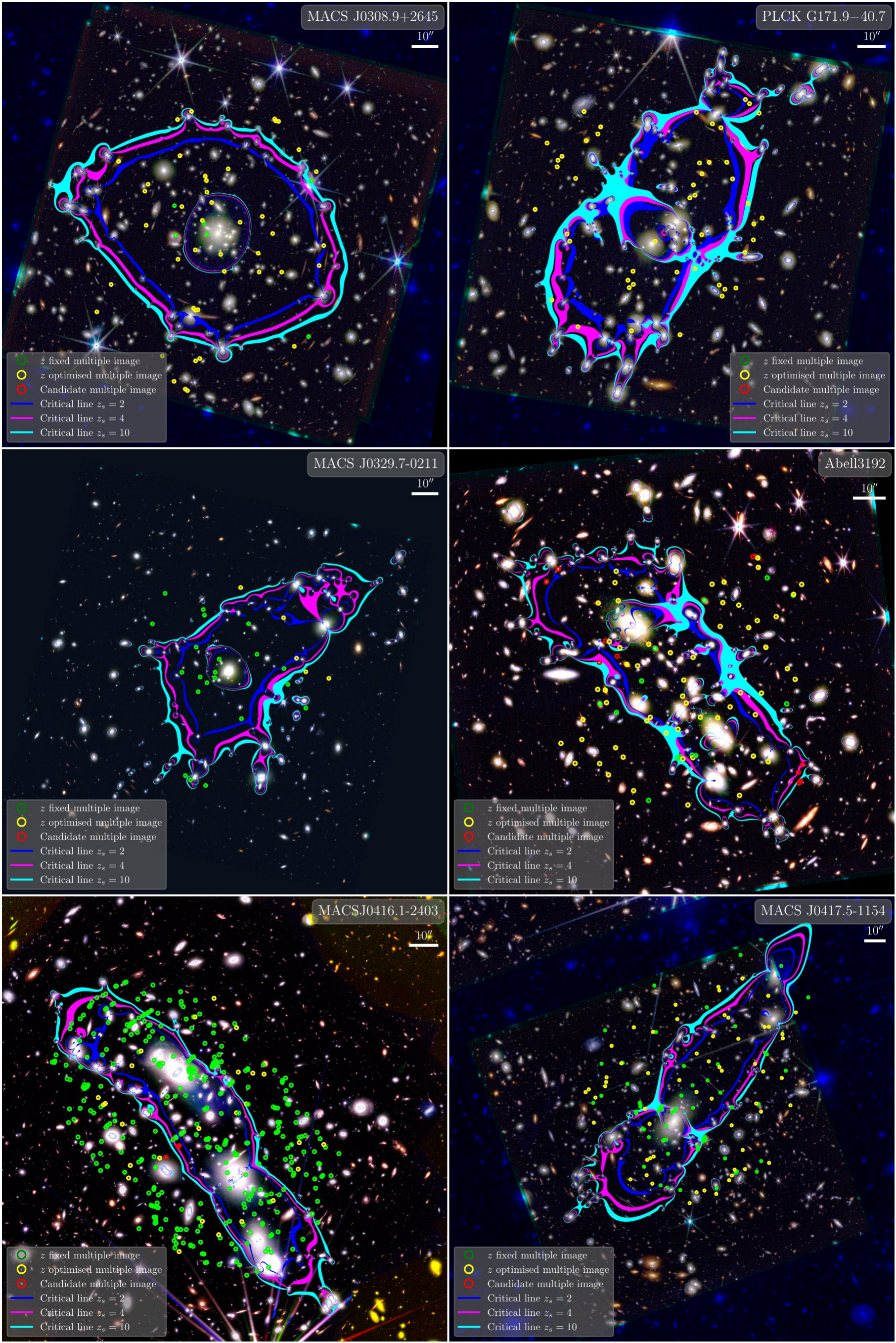}
    \caption{Galaxy clusters M0308, PLCKG171, M0329, A3192, M0416, M0417, represented with critical curves and MIS.}
    \label{fig:clusters_4}
\end{figure*}

\begin{figure*}
    \centering
    \includegraphics[width=0.85\linewidth]{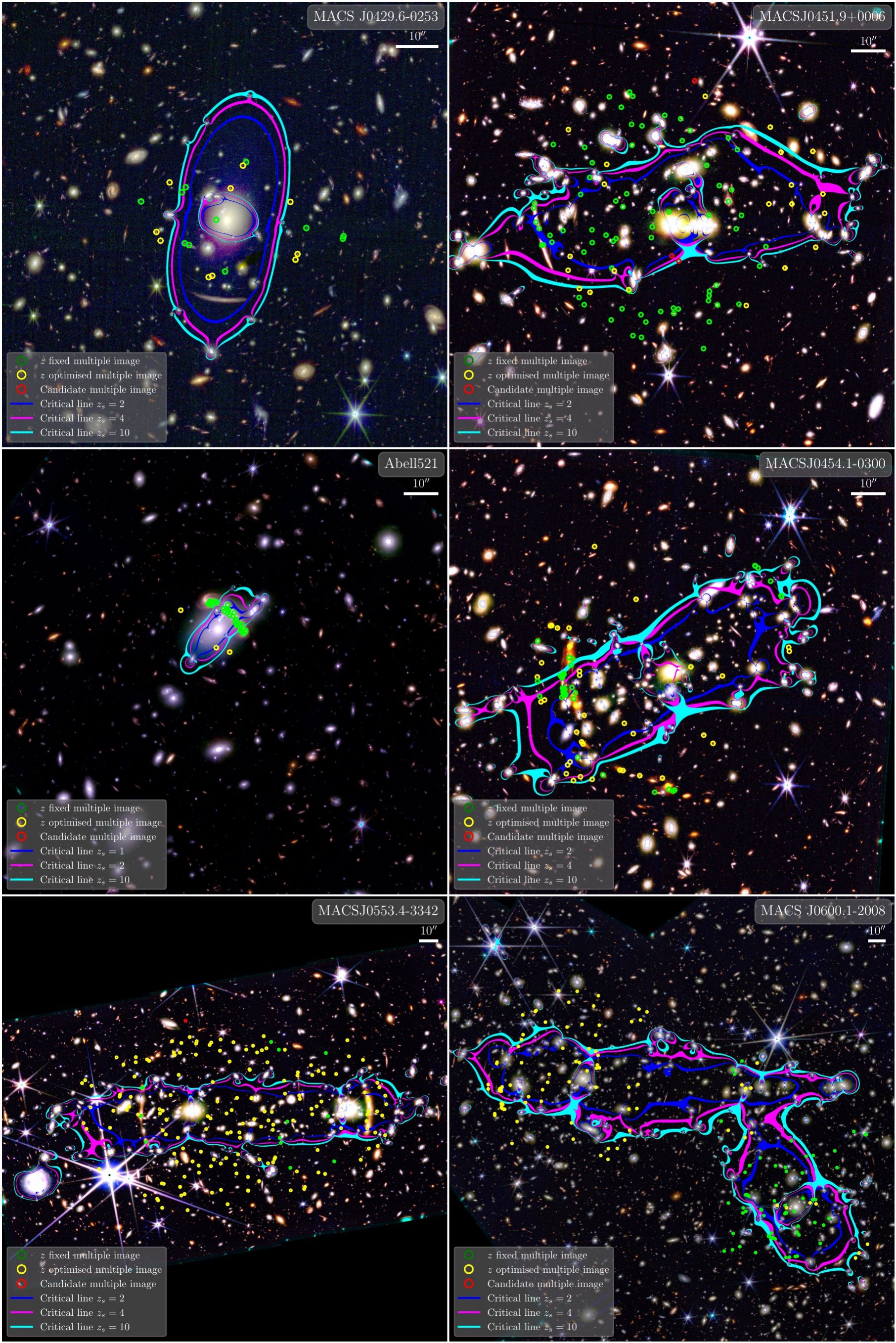}
    \caption{Galaxy clusters M0429, M0451, A521, M0454, M0553, R0600, represented with critical curves and MIS.}
    \label{fig:clusters_5}
\end{figure*}

\begin{figure*}
    \centering
    \includegraphics[width=0.85\linewidth]{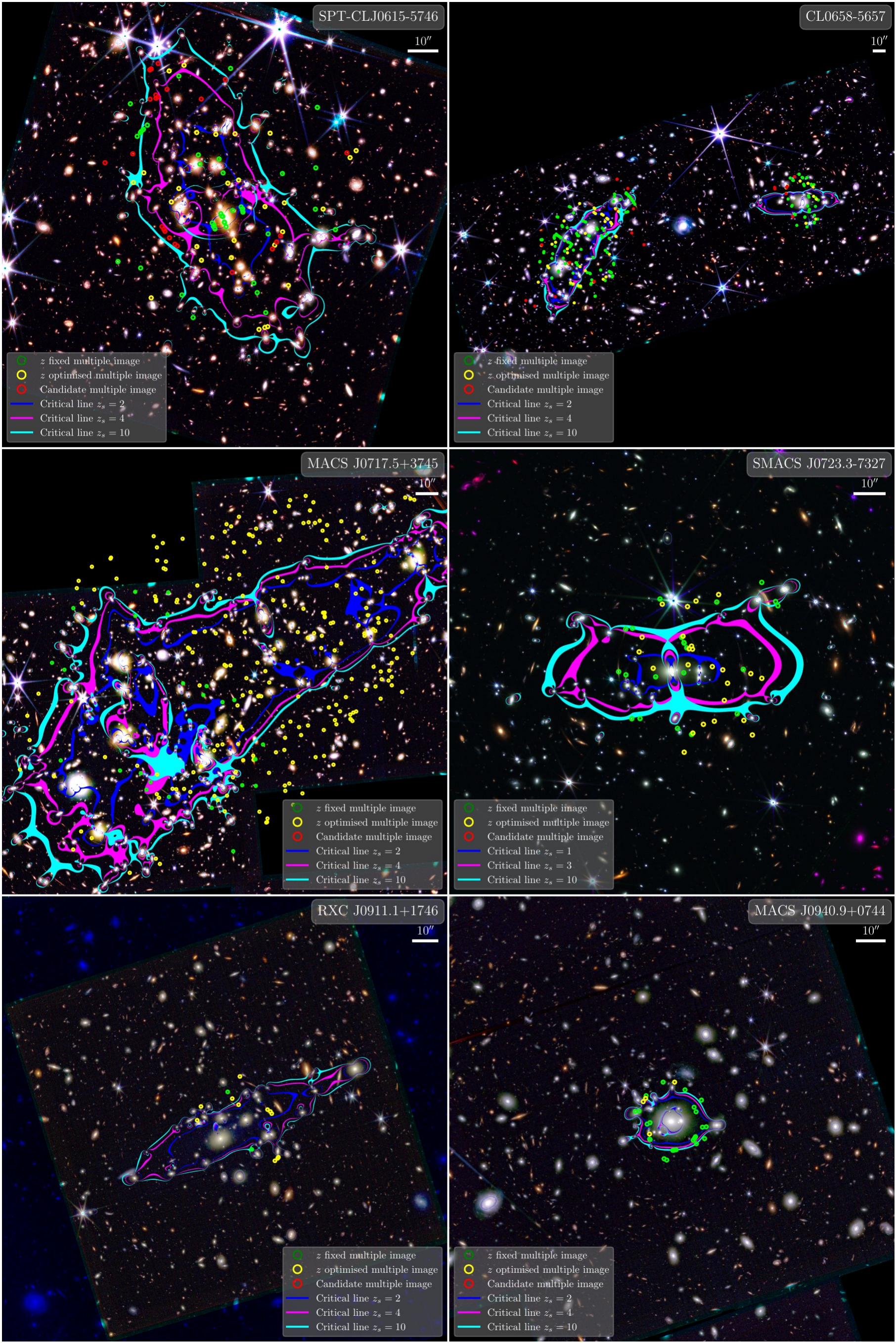}
    \caption{Galaxy clusters SPTCL0615, CL0658 (Bullet), M0717, SM0723, R0911, M0940, represented with critical curves and MIS.}
    \label{fig:clusters_6}
\end{figure*}

\begin{figure*}
    \centering
    \includegraphics[width=0.85\linewidth]{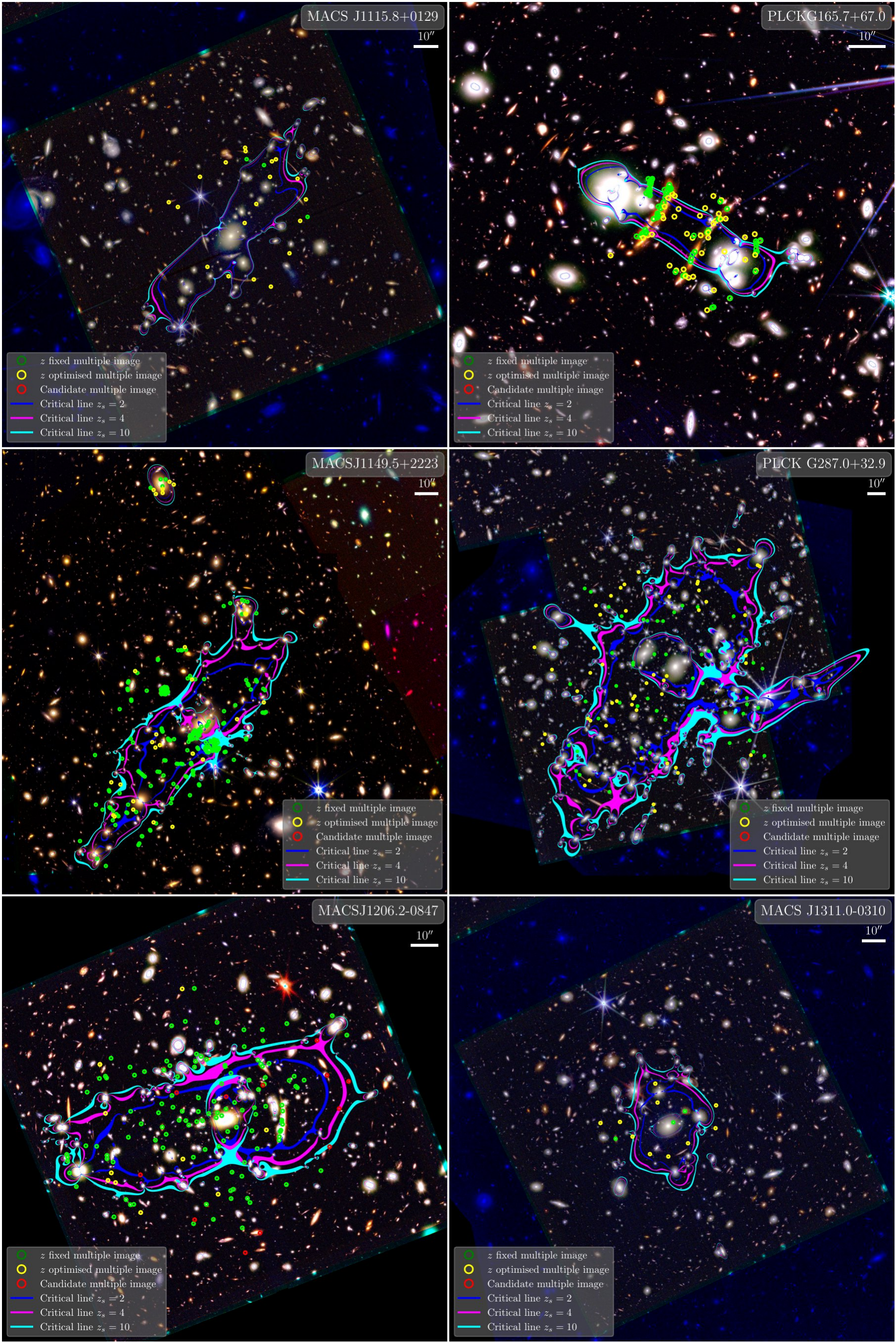}
    \caption{Galaxy clusters M1115, PLCKG165, M1149, PLCKG287, M1206, M1311, represented with critical curves and MIS.}
    \label{fig:clusters_7}
\end{figure*}

\begin{figure*}
    \centering
    \includegraphics[width=0.85\linewidth]{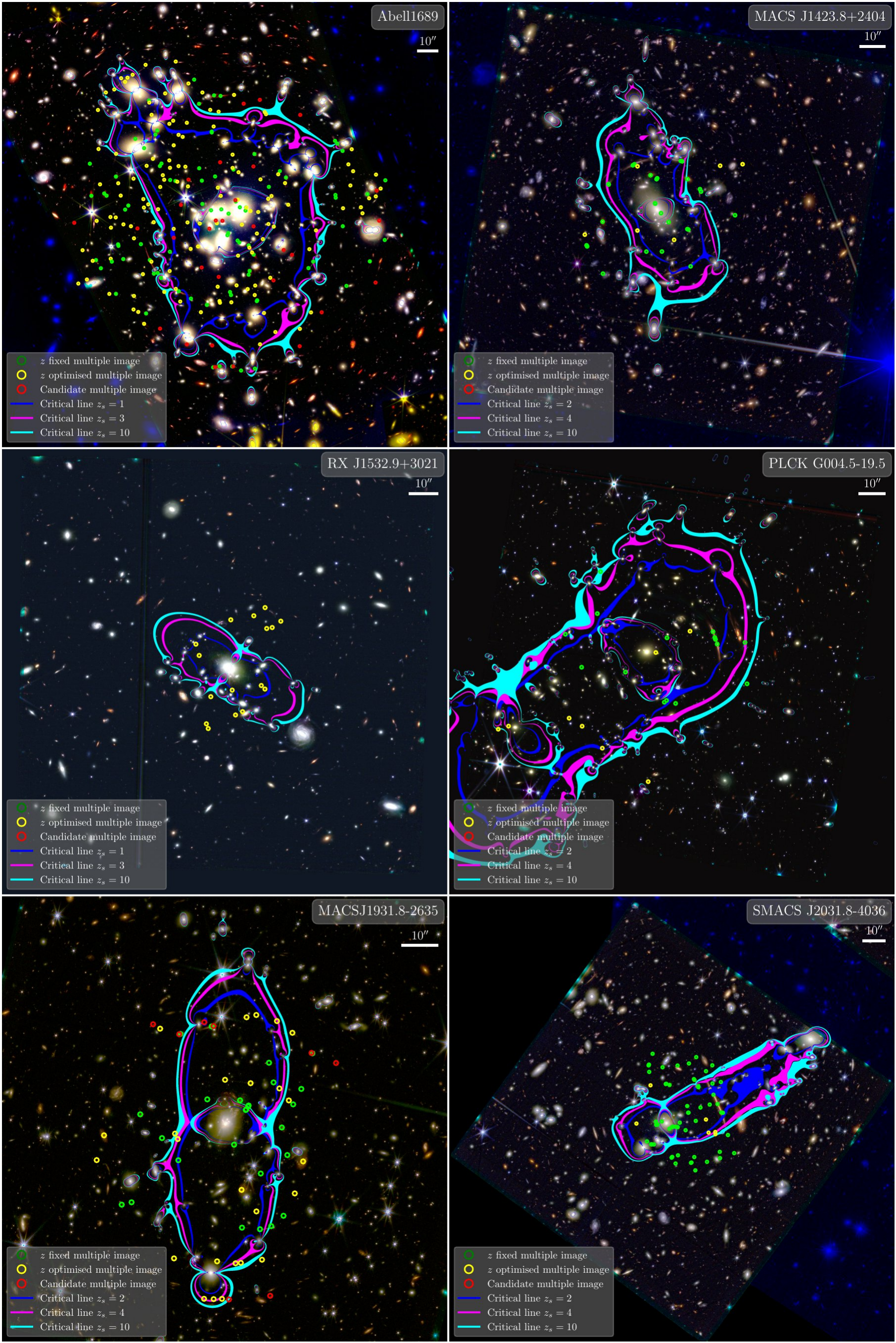}
    \caption{Galaxy clusters A1689, M1423, R1532, PLCKG004, M1931, SM2031, represented with critical curves and MIS.}
    \label{fig:clusters_8}
\end{figure*}

\begin{figure*}
    \centering
    \includegraphics[width=0.85\linewidth]{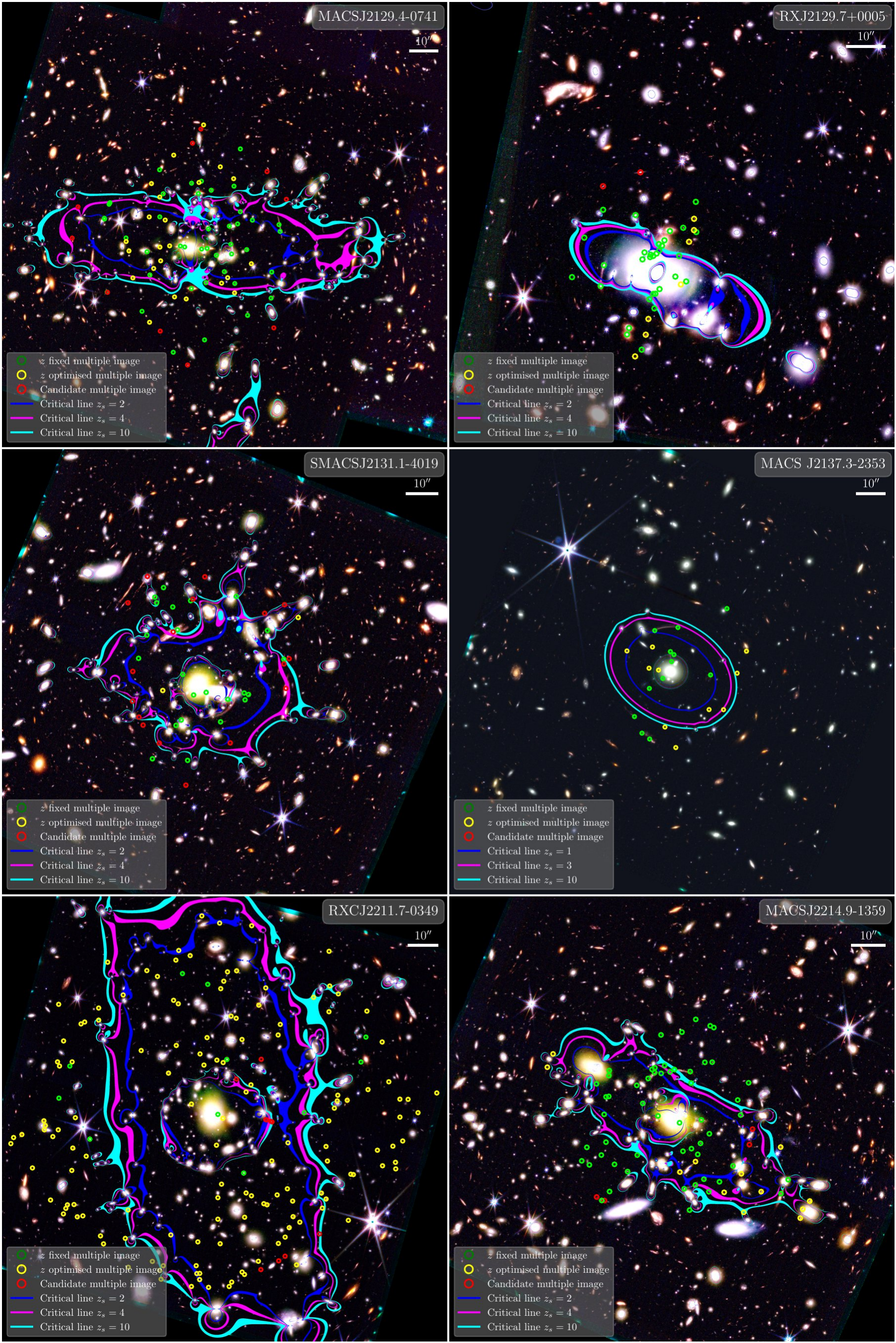}
    \caption{Galaxy clusters M2129, R2129, SM2131, M2137, R2211, M2214, represented with critical curves and MIS.}
    \label{fig:clusters_9}
\end{figure*}

\begin{figure*}
    \centering
    \includegraphics[width=0.85\linewidth]{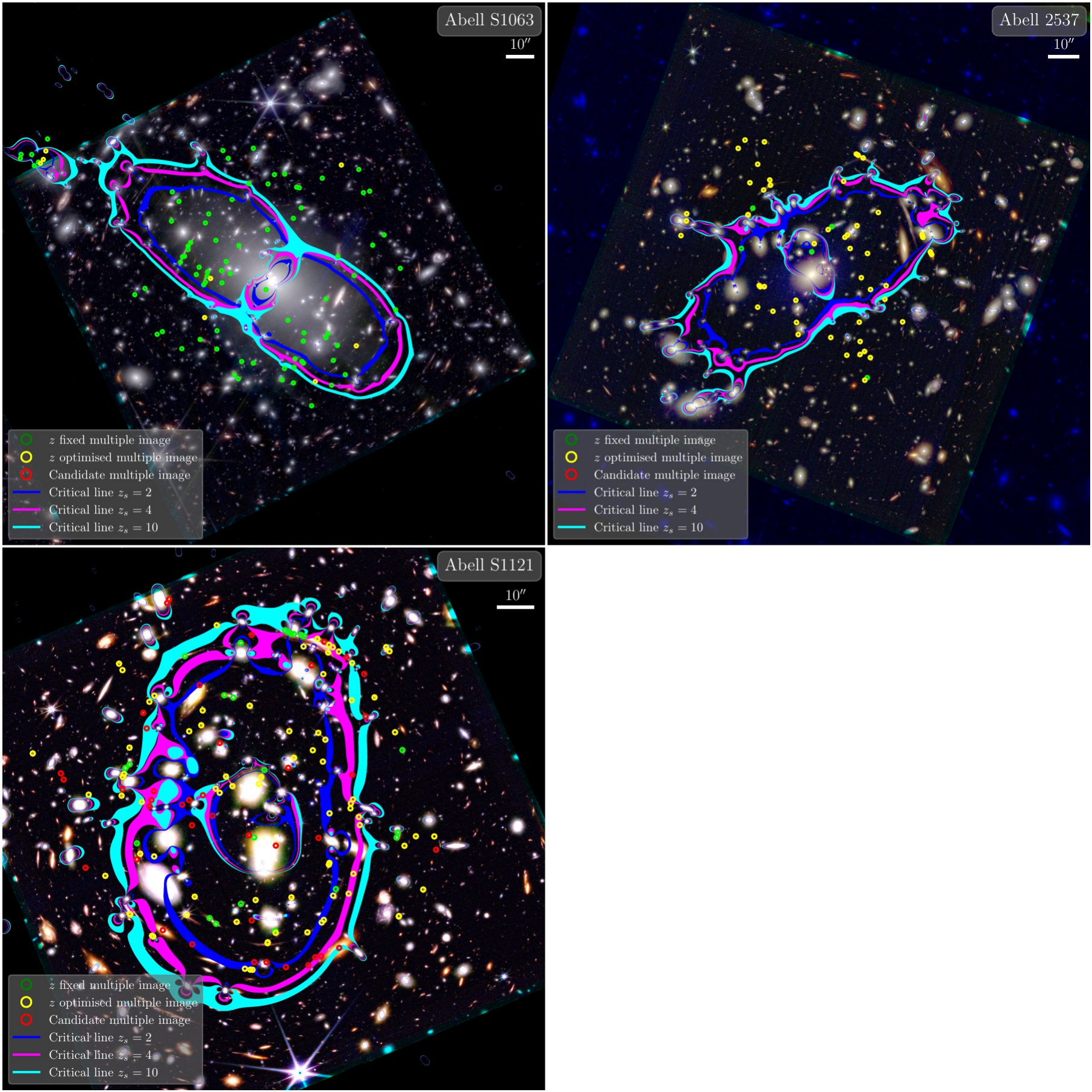}
    \caption{Galaxy clusters AS1063, A2537, AS1121, represented with critical curves and MIS.}
    \label{fig:clusters_10}
\end{figure*}
\end{toggleblock}

\end{document}